\documentclass[ppnp-LaTeX]{elsarticle}

\usepackage{amsmath,amssymb,pdflscape,rotating,xcolor,mathtools}
\usepackage{bm}
\usepackage{epsfig}
\biboptions{numbers,sort&compress}

\newcommand{\be}{\begin{equation}}
\newcommand{\ee}{\end{equation}}
\newcommand{\bea}{\begin{eqnarray}}
\newcommand{\eea}{\end{eqnarray}}

\newcommand{\bfvec}[1]{\bm{\mathrm{#1}}} 
\newcommand{\fref}[1]{Fig.~\ref{fig:#1}} 
\newcommand{\eref}[1]{Eq.~\eqref{eq:#1}}

\newcommand{\sref}[1]{Section~\ref{sec:#1}}
\newcommand{\cref}[1]{Chapter~\ref{ch:.#1}}
\newcommand{\tref}[1]{Table~\ref{tab:#1}}

\newcommand{\nnl}{\nonumber \\}

\newcommand{\beq}{\begin{equation}} 
\newcommand{\eeq}{\end{equation}} 
\newcommand{\ba}{\begin{array}}  
\newcommand{\ea}{\end{array}} 
\newcommand{\bi}{\begin{itemize}}  
\newcommand{\ei}{\end{itemize}}

\newcommand{\cO}{{\mathcal O}} 
 
\newcommand{\cL}{{\mathcal L}}

\newcommand{\ket}[1]{\:|\,{#1}\rangle\:}                      
\newcommand{\br}[1]{\:\langle{#1}|\:}

\newcommand{\eps}{\epsilon}
\newcommand{\eS}{\epsilon_S}
\newcommand{\eT}{\epsilon_T}
\newcommand{\eP}{\epsilon_P}
\newcommand{\eL}{\epsilon_L}
\newcommand{\eR}{\epsilon_R}
\newcommand{\teps}{{\tilde{\epsilon}}}
\newcommand{\teS}{{\tilde{\epsilon}_S}}
\newcommand{\teT}{{\tilde{\epsilon}_T}}
\newcommand{\teP}{{\tilde{\epsilon}_P}}
\newcommand{\teL}{{\tilde{\epsilon}_L}}
\newcommand{\teR}{{\tilde{\epsilon}_R}}
\newcommand{\eLc}{{\epsilon_L^{(c)}}}
\newcommand{\eLv}{{\epsilon_L^{(v)}}}
\newcommand{\eSP}{\epsilon_{S,P}}
\newcommand{\teSP}{{\tilde{\epsilon}_{S,P}}}

\begin{document}

\title{New physics searches in nuclear and neutron $\beta$ decay}

\author[label1]{M.\ Gonz\'alez-Alonso~\fnref{myfootnote1}}
\author[label2]{O.\ Naviliat-Cuncic~\fnref{myfootnote2}}
\author[label3]{N.\ Severijns~\fnref{myfootnote3}}

% Email
\fntext[myfootnote1]{martin.gonzalez.alonso@cern.ch}
\fntext[myfootnote2]{naviliat@nscl.msu.edu}
\fntext[myfootnote3]{nathal.severijns@kuleuven.be (Corresponding author)}

\address[label1]{
Theoretical Physics Department, CERN, 1211 Geneva 23, Switzerland
}
\address[label2]{National Superconducting Cyclotron Laboratory and Department of Physics and Astronomy,
Michigan State University, East Lansing, Michigan 48824, USA}
\address[label3]{Instituut voor Kern- en Stralingsfysica, Katholieke Universiteit Leuven, Belgium}

%%%%%%%%%%%%%%%%%%%%%%%%%%%%%%%%%%%%%%%%%%%%%%%%%%%%%%%%%%%%%%%%%%%%%%%%%%%%%%%%

\begin{abstract}
%%%%%%%%%%%%%%%%%%%%%%%%%%%%%%%%%%%%%%%%%%%%%%%%%%%%%%%%%%%%%%%%%%%%%%%%%%%%%%%%
The status of tests of the standard electroweak model and of searches for new
physics in allowed nuclear $\beta$ decay and neutron decay is reviewed 
including both theoretical and experimental developments. 
The sensitivity and complementarity of recent and ongoing experiments are
discussed with emphasis on their potential to look for new physics. 
Measurements are interpreted using a model-independent effective field
theory approach enabling to recast the outcome of the analysis in many
specific new physics models. Special attention is given to the connection
that this approach establishes with high-energy physics. 
A new global fit of available $\beta$-decay data is performed incorporating,
for the first time in a consistent way, superallowed $0^+\to 0^+$
transitions, neutron decay and nuclear decays. The constraints on exotic
scalar and tensor couplings involving left- or right-handed neutrinos
are determined while a constraint on the pseudoscalar coupling from
neutron decay data is obtained for the first time as well. 
The values of the vector and axial-vector couplings, which are associated within the standard model to $V_{ud}$ and $g_A$ respectively, are also updated.
The ratio between the axial and vector couplings obtained from the fit
under standard model assumptions is $C_A/C_V = -1.27510(66)$.
The relevance of the various experimental inputs and error sources is critically
discussed and the impact of ongoing measurements is studied. 
The complementarity of the obtained bounds with other low- and high-energy
probes is presented including ongoing searches at the Large Hadron Collider. 
%%%%%%%%%%%%%%%%%%%%%%%%%%%%%%%%%%%%%%%%%%%%%%%%%%%%%%%%%%%%%%%%%%%%%%%%%%%%%%%%

\end{abstract}

\maketitle

%\eject
%\vspace*{-2cm}
\begin{flushright}
CERN-TH-2018-050
\vspace*{2mm}
%\today
\end{flushright}

\tableofcontents

%%%%%%%%%%%%%%%%%%%%%%%%%%%%%%%%%%%%%%%%%%%%%%%%%%%%%%%%%%%%%%%%%%%%%%%%%%%%%%%%
%%%%%%%%%%%%%%%%%%%%%%%%%%%%%%%%%%%%%%%%%%%%%%%%%%%%%%%%%%%%%%%%%%%%%%%%%%%%%%%%
\section{Introduction}
\label{sec:intro}

The standard electroweak model (SM)~\cite{Glashow1959, Salam1959, Weinberg1967}
has been remarkably successful for the description of processes at the most
elementary level. The celebrated culmination of the
existence of the long-predicted Higgs boson confirmed its internal consistency 
along with global fits between experimental data and SM predictions
\cite{Patrignani:2016xqp}. Nevertheless, many questions remain unanswered
such as the origin of parity violation, the mechanism behind the matter-antimatter asymmetry, 
the nature of dark matter, the large number of parameters of the theory, etc.
These are expected to find explanations in extended and unified theoretical
frameworks involving New Physics (NP). 
No new particle other than the Higgs boson has been found at the Large Hadron Collider
(LHC) and this has renewed much interest into low- and intermediate-energy probes that
could indicate the energy scale of NP. 
 
Experimental and theoretical developments in nuclear and neutron $\beta$ decay have
played~\cite{Lee:1956qn,Wu1957,Weinberg:2009zz} and continue to play a major role in
our understanding of the electroweak sector of the SM~\cite{Severijns:2006dr, Abele:2008zz,
Dubbers2011, Cirigliano:2013xha, Gonzalez-Alonso:2013uqa, Vos:2015eba}.
Precision measurements probe possible contributions of non-SM, so-called exotic,
currents such as scalar, tensor, pseudoscalar or right-handed vector currents, that
couple to hypothetical new particles heavier than the known $W$ and $Z$ vector bosons.
They can also test the properties of these exotic interactions as well as of the
$V-A$ interactions under the discrete symmetries of parity ($\mathcal{P}$), charge
conjugation ($\mathcal{C}$), and time-reversal ($\mathcal{T}$) or equivalently
the combined $\mathcal{CP}$ symmetry.
$\beta$ decay is a semileptonic strangeness-conserving process involving, to lowest
order, only the lightest leptons ($e, \nu_e$) and quarks ($u, d$) interacting via the
exchange of charged vector bosons $W^{\pm}$. The possibilities of nuclear and neutron
$\beta$ decay experiments to constrain the SM are therefore limited.
Such limitations are however partly compensated by the large variety of nuclear states
and observables available, allowing the selection of the most sensitive ones to a given
NP search or symmetry test.
Other important experimental assets are the high intensity with which many $\beta$
emitters and neutrons can presently be available at existing nuclear and neutron
facilities. Several $\beta$-decay parameters can nowadays be measured with high
precision and access has been gained to parameters that were previously very difficult
or impossible to address. These developments were also accompanied by new and advanced
detection
techniques, new polarization methods, the use of atom and ion traps in nuclear
physics experiments, or the use of magnetic traps in experiments with ultra-cold
neutrons.

This review describes recent theoretical and experimental advances in the study
of nuclear and neutron $\beta$ decay and it is structured as follows.
\sref{theo} provides an overview of the theoretical formalism. An Effective-Field-Theory (EFT) framework
is constructed gradually from the quark level, over the nucleon level up to the level
of the nucleus. Special attention is paid to the hadronic charges, the calculation
of which has enjoyed great progress over the last decade mainly due to lattice
techniques. These calculations are important to translate current and future precision
measurements into stringent NP bounds without losing sensitivity due to theoretical
limitations. The relevant observables and their theoretical expressions are discussed
as well, including SM as well as non-SM sub-leading corrections. The same EFT approach
is then used to make the connection all the way up to the electroweak scale or even 
the LHC scales, with special attention to recent developments. This theoretical
framework, introduced decades ago, receives currently a renewed interest motivated
in part by the absence of NP signals in direct LHC searches. Finally, the connection
with model-dependent studies is shortly discussed.

In~\sref{exp} we review recent experimental achievements as well as ongoing and
planned developments. The observables include: (i) the corrected $\mathcal{F}t$ values of
superallowed Fermi transitions and of mirror $\beta$ transitions, including the
neutron; (ii) $\beta$-spectrum shape measurements, which enjoy a renewed interest
both as a probe to search for exotic scalar and tensor currents and to investigate
in more detail the contribution of weak magnetism to $\beta$ decay; and (iii)
correlation measurements between the spins and momenta of the different particles
involved in the $\beta$ decay. 
For each experiment, the important technical developments are
pointed out, as well as the results already obtained and the precision goals that are
being aimed at.

\sref{fit} presents a new global fit to the existing $\beta$-decay data.
The analysis incorporates, for the first time in a consistent way, superallowed $0^+\to 0^+$
transitions, neutron decay and nuclear decay data.
Experimental errors are rescaled using standard prescriptions to reflect internal
inconsistencies in some of the neutron measurements. The data are added step by step to illustrate the impact of each data set on the constraints.
Other improvements and differences with respect to previous global analyses available in the
literature are discussed. 
First, the bounds on interactions involving left-handed 
(LH) neutrinos are thoroughly studied, including current benchmark sensitivities
and projected ones based on claimed precision goals. 
The constraints on exotic scalar and tensor couplings are updated and a constraint
on the pseudoscalar 
coupling is obtained from neutron data for the first time. New values are provided for
the vector and axial-vector couplings, associated to $V_{ud}$ and $g_A$ in the SM.
Fits with scalar and tensor
currents involving right-handed (RH) neutrinos are then carried out, and the differences with the previous case are analyzed.

In \sref{alternativeprobes} we discuss the interplay between these bounds
with other low- and high-energy searches using the EFT framework. The probes considered
are pion decay, LEP precision measurements, LHC searches and neutrino mass. 
Finally, we conclude in~\sref{conclusion} with a summary of our main results, and a discussion about future opportunities for $\beta$-decay studies.

%%%%%%%%%%%%%%%%%%%%%%%%%%%%%%%%%%%%%%%%%%%%%%%%%%%%%%%%%%%%%%%%%%%%%%%%%%%%%%%%

\section{Theoretical formalism}
\label{sec:theo}
Every experiment that explores new levels of sensitivity
has a discovery potential. In order to prioritize and to interpret the results, a theoretical framework is needed, which unavoidable entails specific assumptions. It is also important to remain as general as possible. We will follow therefore a model-independent effective field theory approach, which we introduce in this section and whose basic assumptions are Lorentz symmetry and the absence of nonstandard light states, with the only possible exception of a right-handed neutrino.

The advantage of this approach is not only its generality but also its efficiency. The analysis of experimental data is done once and for all, and its outcome can be applied to a plethora of specific beyond the standard model (BSM) scenarios in a straightforward way. In this spirit, all effective operators of a given order in the EFT counting must be kept in the analysis simultaneously, and numerical results must be provided including correlations. A NP model can easily generate several operators simultaneously (see e.g. Ref.~\cite{Profumo:2006yu}), and even when only one effective operator is generated at tree-level, the others will unavoidably be generated through radiative corrections.
Moreover, this EFT theoretical framework allows us to compare in a general way the NP sensitivity among different $\beta$-decay experiments as well as with respect to other searches, such as those performed at the LHC at CERN. It makes also possible to write the bounds in a language that is more convenient for model builders.

\subsection{Quark-level EFT ($\eps_i,\tilde{\epsilon}_i$)}
\label{sec:quark-level-EFT}
The partonic process that underlies nuclear and neutron $\beta$ decays is $d\to u\, e^- \bar{\nu}_e$ or its crossed version. Simply assuming that the SM and BSM underlying physics mediating this process have a characteristic scale much higher than those involved in $\beta$ decays, one can write the following generic low-energy effective Lagrangian
	\bea
	\cL = \cL^{\rm eff}_{ude\nu}+\cL_{QED}+\cL_{QCD}~,
	\eea
where QED and QCD effects have of course not been integrated out. The first term, which represents the central element of this review, is given by~\cite{Herczeg:2001vk,Cirigliano:2012ab}\footnote{For the sake of simplicity we do not consider operators involving $\nu_\mu$ or $\nu_\tau$. The generalization is straightforward and the general formulae can be found in Ref.~\cite{Cirigliano:2012ab}.}$^,$\footnote{Various equivalent notations have been used throughout the literature, see e.g. Refs.~\cite{Herczeg:2001vk,Profumo:2006yu,Cirigliano:2009wk,Vos:2015eba}. None of them denotes the Wilson coefficients with the Greek letter $\epsilon$ though, and thus there should be no confusion with the notation of Ref.~\cite{Cirigliano:2012ab} followed in this work.}
\bea
\cL^{\rm eff}_{ude\nu}  =
- \frac{G^0_F V_{ud}}{\sqrt{2}} \Big[ \
&&\!\!\!\!\!\!\!\!\!\!\left( 1 +  \eL \right) \
\bar{e}  \gamma_\mu  (1 - \gamma_5)   \nu_e  \cdot \bar{u}   \gamma^\mu  (1 - \gamma_5)  d \Big.
~+~
\teL  \,\bar{e}  \gamma_\mu  (1 + \gamma_5)   \nu_e  \cdot \bar{u}   \gamma^\mu  (1 - \gamma_5)  d
\nnl
&&\!\!\!\!\!\!\!\!\!\!\!\!\!\!\!+\,   \eR   \,  \bar{e}  \gamma_\mu  (1 - \gamma_5)   \nu_e
\cdot \bar{u}   \gamma^\mu  (1 + \gamma_5)  d
~+~
\teR   \,  \bar{e}  \gamma_\mu  (1 +  \gamma_5)   \nu_e
\cdot \bar{u}   \gamma^\mu  (1 + \gamma_5)  d
\nnl
&&\!\!\!\!\!\!\!\!\!\!\!\!\!\!\!+\,
\eT    \   \bar{e}   \sigma_{\mu \nu} (1 - \gamma_5) \nu_e    \cdot  \bar{u}   \sigma^{\mu \nu} (1 - \gamma_5) d
~+~
\Big.
~\teT      \   \bar{e}   \sigma_{\mu \nu} (1 + \gamma_5) \nu_e    \cdot  \bar{u}
 \sigma^{\mu \nu} (1 + \gamma_5) d  \nnl
&&\!\!\!\!\!\!\!\!\!\!\!\!\!\!\!+\,  \eS  \, \bar{e}  (1 - \gamma_5) \nu_e  \cdot  \bar{u} d
~+~
\teS  \, \bar{e}  (1 +  \gamma_5) \nu_e  \cdot  \bar{u} d
\nnl
&&\!\!\!\!\!\!\!\!\!\!\!\!\!\!\!-\, \eP  \,  \bar{e}  (1 - \gamma_5) \nu_e  \cdot  \bar{u} \gamma_5 d
~-~
\teP  \,  \bar{e}  (1 + \gamma_5) \nu_e  \cdot  \bar{u} \gamma_5 d
~+~ \ldots
\Big] + {\rm h.c.}~,
\label{eq:leff-lowE}
\eea
where the dots refer to subleading derivative higher-dimensional operators. We work with the nowadays usual conventions, $\gamma^5\equiv i\gamma^0\gamma^1\gamma^2\gamma^3$ and $\sigma_{\mu\nu} \equiv \frac{i}{2}\left[ \gamma_\mu,\gamma_\nu \right]$, and in natural units ($\hbar=c=1$). The Lagrangian in \eref{leff-lowE} displays explicitly the SM tree-level contribution, namely the $(V-A)\times (V-A)$ Fermi interaction~\cite{Feynman1958,Sudarshan:1958vf} generated by the exchange of a $W$ boson. $V_{ud}$ is the $ud$ entry of the Cabibbo-Kobayashi-Maskawa (CKM) matrix. Up to this factor, the Wilson coefficient in front of the leading term is the famous Fermi constant, which is connected with the underlying SM physics at tree level by
\bea
G^{0,\rm{tree}}_F = \frac{g^2}{4\sqrt{2}m_W^2}~,
\eea
where $g$ is the gauge coupling of the $SU(2)_L$ group, and $m_W$ is the $W$ mass. Likewise, the $\eps_i$ and $\teps_i$ complex coefficients are model-dependent functions of the masses and couplings of the new particles. Since they were defined with respect to the SM contribution, one expects them, on pure dimensional grounds, to scale as
\bea
\eps_i, \teps_i \propto \left( \frac{m_W}{\Lambda} \right)^n~,
\eea
where $n\geq 2$, and $\Lambda$ denotes the characteristic NP scale. In the simplest case $n=2$ and the new interaction is generated \emph{\`a la} Fermi, which gives $\eps_i, \teps_i \sim 10^{-3}$ for NP scales at or above the TeV. The non-standard Wilson coefficients can also be suppressed (enhanced) by NP couplings smaller (larger) than the SM coupling $g$, or by loop factors. In~\sref{specificmodels} we discuss the specific form of the coefficients in various NP models.

The phenomenological extraction of the Fermi constant from muon decay gives $G^\mu_F / (\hbar c)^3 = 1.1663787(6) \times 10^{-5} \,\rm{GeV}^{-2}$ \cite{Tishchenko2013}. Since this process can also be affected by NP effects, we write $G_F^\mu = G_F^0 + \delta G_F$, but for the sake of simplicity we will omit the $\mu$ superindex hereafter. 
Up to an overall phase, there are ten real couplings and nine phases that are, at least in principle, phenomenologically accessible. 
It is to be stressed that the SM piece comes together with the $\eL$ and $\delta G_F$ coefficients and cannot be separated using only $\beta$-decay data. 
For this reason, we define
\bea
\label{eq:Vtilde}
\tilde{V}_{ud} \equiv V_{ud} \left(1 +  \eL+\eR\right)\left(1-\frac{\delta G_F}{G_F}\right) ~,
\eea
where we also included $\eR$ for later convenience, since the most precise $V_{ud}$ determination comes from vector-mediated transitions. 

Given the expected smallness of the NP couplings it is useful to work at linear order in them to identify their main effect on the different observables. In this approximation we can neglect the $\teps_i$ terms, since they involve right-handed neutrinos and thus their interference with the SM piece is suppressed by the smallness of the neutrino mass. The ``linearized" low-energy effective Lagrangian can be written as
\begin{eqnarray}
\label{eq:leff-lowE-linear}
\cL_{\rm eff}  &=&
- \frac{G_F \tilde{V}_{ud}}{\sqrt{2}} \
\Big\{
\bar{e}  \gamma_\mu (1\! -\! \gamma_5) \nu_{e}  \cdot \bar{u}
\gamma^\mu \left[ 1 - \left(1-2\eR \right) \gamma_5 \right]   d \Big.
\nonumber  \\
&&+~  \eS  \, \bar{e}  (1\! -\! \gamma_5) \nu_{e}  \cdot  \bar{u} d
 -~ \eP  \,  \bar{e}  (1\! -\! \gamma_5) \nu_{e}  \cdot  \bar{u} \gamma_5 d
+~
\Big.
\eT    \   \bar{e}   \sigma_{\mu \nu} (1\! -\! \gamma_5) \nu_{e}    \cdot  \bar{u}
\sigma^{\mu \nu} (1\! -\! \gamma_5) d \Big\}
+{\rm h.c.}~,
\end{eqnarray}
All in all, there are nine couplings left in this approximation:
\begin{itemize}
\item The overall normalization, given by $V_{ud}$ in the SM and now replaced by $\tilde{V}_{ud} \approx V_{ud} \left( 1 + \eL+\eR - \frac{\delta G_F}{G_F} \right)$. 
Its only consequence is the violation of the unitarity condition of the first row of the CKM quark mixing matrix;
\item The relative size of the axial current with respect to the vector one is modified by the presence of $\mathcal{CP}$-conserving non-standard right-handed currents Re$(\eR)$.
To probe this coupling requires however an accurate theoretical knowledge of the non-perturbative hadronization (and nuclearization) of this current;
\item The real parts of the (pseudo)scalar $\eSP$ and tensor $\eT$ couplings that modify the energy distributions and $\mathcal{CP}$-even correlation coefficients in $\beta$ decay. Moreover, $\eT$ and $\eP$ also modify at tree-level the radiative and non-radiative leptonic pion decay. These interactions are sometimes called chirality-flipping because the chiralities of the two fermions in each bilinear are different. That is, a left-handed neutrino comes with a right-handed electron;
\item The imaginary parts of $\eR,\eS,\eP$ and $\eT$, which modify $\mathcal{CP}$-odd observables. More precisely these are sensitive to the relative phase between these coefficients and the vector one $(1+\eL+\eR)$.
\end{itemize}
It is clear that the quark-level effective Lagrangian,~\eref{leff-lowE}, makes possible to compare model-independently nuclear and neutron $\beta$-decay searches with other semileptonic hadron decays that are governed by the same dynamics, like for example $\pi^{\pm} \to \pi^0 e^{\pm} \nu$.
The details of the hadronization are obviously different, with different form factors needed in each process, but the underlying partonic process is the same. 
On the other hand it facilitates the connection with particle physics. As a matter of fact, the effective Lagrangian in~\eref{leff-lowE} describes any flavor transition of the type $d^j\to u^i \ell^- \bar{\nu}_\ell$ once the necessary flavor indices are added. Finally, the non-trivial hadronization is performed once and for all in this model-independent framework, instead of doing it for every NP model.

It is important to keep in mind that this approach, as general as it is, does not capture more exotic scenarios with non-standard light particles, which could be emitted in $\beta$ decays, or with violations of Lorentz symmetry. A recent example of the former can be found in Ref.~\cite{Fornal:2018idm}, whereas the latter was discussed in great detail in Ref.~\cite{Vos:2015eba}.

\subsection{Nucleon-level EFT ($C_i$, $C'_i$)}
\label{sec:nucleon-level-EFT}
We start by discussing the simple case of neutron decay. The connection between the quark-level effective Lagrangian, Eq.~(\ref{eq:leff-lowE}), with the observables requires the calculation of a series of neutron-to-proton matrix elements. They can be parametrized in terms of Lorentz-invariant form factors as follows~\cite{Weinberg:1958ut,Bhattacharya:2011qm}:
\begin{subequations}
\label{eq:nucleonmatching}
\bea
\label{eq:nucleonmatchingV}
\br{p (p_p) } \bar{u} \gamma_\mu d \ket{n (p_n)} &=& \bar{u}_p (p_p)  \left[
g_V(q^2)  \,  \gamma_\mu
+ \frac{\tilde{g}_{T(V)} (q^2)}{2 M_N}   \, \sigma_{\mu \nu}   q^\nu
+ \frac{\tilde{g}_{S} (q^2)}{2 M_N}   \,  q_\mu  \right]  \hspace{-0.1cm}
 u_n (p_n)~,
\\
\label{eq:nucleonmatchingA}
\br{p (p_p) } \bar{u} \gamma_\mu \gamma_5  d \ket{n (p_n)} &=&
\bar{u}_p (p_p)  \left[
g_A(q^2)    \gamma_\mu
\hspace{-0.1cm}
+ \frac{\tilde{g}_{T(A)} (q^2)}{2 M_N}   \sigma_{\mu \nu}   q^\nu
+
\hspace{-0.1cm}
\frac{\tilde{g}_{P} (q^2)}{2 M_N}   q_\mu
\right]  \hspace{-0.15cm}  \gamma_5  u_n (p_n) ~,
\label{eq:inducedgP}
\\
\br{p (p_p) } \bar{u} \,   d \ket{n (p_n)} &\approx&
g_S(0)  \ \bar{u}_p (p_p)  \, u_n (p_n)  + \cO(q^2/M_N^2)~,
\label{eq:defgS}
\\
\br{p (p_p) } \bar{u} \,  \gamma_5 \,  d \ket{n (p_n)} &\approx&
g_P(0)  \ \bar{u}_p (p_p)  \, \gamma_5 \, u_n (p_n)  + \cO(q^2/M_N^2)
\label{eq:defgP}~,
\\
\br{p (p_p) } \bar{u} \, \sigma_{\mu \nu}  \,  d \ket{n (p_n)} &\approx&
g_T(0) \, \bar{u}_p (p_p) \,\sigma_{\mu \nu}  u_n (p_n)+ \cO(q/M_N)
\label{eq:defgT}~,
\eea
\end{subequations}
where $u_{p,n}$ are the proton and neutron spinor amplitudes,  $q = p_n - p_p$ and $M_N= M_n = M_p$ denotes a common nucleon mass. Since the (pseudo)scalar and tensor matrix elements appear multiplied by small NP Wilson coefficients we neglected small corrections due to the momentum dependence of the leading form factors and to additional $\cO(q/M_N)$ form factors~\cite{Weinberg:1958ut,Bhattacharya:2011qm}. In addition, the pseudoscalar bilinear $\bar{u}_p \gamma_5 u_n$ is itself of order $q/M_N$, and thus it has been traditionally dropped from the subsequent analysis. As discussed below, this is however not fully justified since the suppression is compensated by the large value of the form factor at zero momentum, $g_P(0)$~\cite{Gonzalez-Alonso:2013ura}.

Subleading corrections in the (axial-)vector bilinear cannot be dropped so lightly, as they do not undergo any NP suppression. The size of these corrections is determined by their dependence on the ratios $(M_n - M_p)/M_N , q/M_N\sim \cO(10^{-3})$, which control isospin-symmetry breaking and momentum-dependent corrections, respectively. Taking into account current and planned experimental sensitivities, which are in the $10^{-4}-10^{-2}$ range, it is crucial to keep effects that are linear in these ratios, but one can neglect quadratic contributions. At this order, the expressions simplify significantly and a hierarchy of effects can be identified:
\bi
\item The $q^2$ dependence of all form factors can be neglected. We denote their value at $q^2=0$, usually called charge, omitting the $q^2$ argument, e.g., $g_A\equiv g_A(0)$;
\item The first contribution, at order one, is given by standard (axial-)vector charges $g_{V,A}$. Moreover $g_V=1$ up to quadratic corrections in isospin-symmetry breaking~\cite{Ademollo:1964sr,Donoghue:1990ti}, while the value of the axial-vector charge $g_A$ is not fixed by symmetry arguments;
\item At first order in the \lq\lq recoil\rq\rq~expansion we have the contribution from the so-called {\bf weak magnetism}, $\tilde{g}_{T(V)}$, which is related by the Conservation of the Vector Current (CVC) to the difference of proton and neutron magnetic moments in the isospin-symmetric limit~\cite{Gell-Mann1958,Holstein:1974zf}; 
\item The contribution associated with the induced-pseudoscalar charge, $\tilde{g}_P$, is quadratic in the counting used here (pseudoscalar bilinear and an explicit $q/M_N$ suppression), but it is also enhanced by the pion pole, $\tilde{g}_P \sim M_N/m_q \sim 100$. This can be derived using the Partial Conservation of the Axial Current (PCAC), making the contribution to the amplitude of order $10^{-4}$. 
The effect of this term on the differential decay rate is known though~\cite{Holstein:1974zf}, and so it can be included in the analysis of experiments with such precision.
\item Last, we have the so-called {\bf second-class currents}, which enter with the induced-scalar and -tensor charges, $\tilde{g}_S$ and $\tilde{g}_{T(A)}$. They can be safely neglected because they vanish in the isospin-symmetric limit~\cite{Weinberg:1958ut}, are multiplied by one power of $q_\mu/M_N$, and do not present any PCAC enhancement.
\ei
All in all, the SM currents introduce only one free parameter in the analysis, $g_A$, and they predict a hierarchy of calculable subleading effects\footnote{$g_A$ is not a free parameter from a conceptual point of view, but only from a practical one, because our ability to calculate its QCD value is smaller than our ability to measure it.}. These represent milestones for current and future experiments in their search for new phenomena~\cite{Holstein:1974zf,Wilkinson:1982hu,Gardner:2000nk}. The identification of these effects, and the agreement with the SM expectation, indicates that they are well under control. This is also true for radiative corrections, which enter at a similar level, $\alpha/\pi\sim 10^{-3}$, and for kinematic corrections, also of order $q/M_N$.

Equation~(\ref{eq:nucleonmatching}) can be viewed as the matching conditions from the quark-level effective theory, \eref{leff-lowE}, to the nucleon-level effective theory, whose leading term was written down by Lee and Yang in 1956~\cite{Lee:1956qn}:\footnote{
The original paper of Lee and Yang~\cite{Lee:1956qn} (and many works afterwards, such as Refs.~\cite{Jackson:1957zz,Severijns:2006dr}) defined $\gamma_5$ with the opposite sign to the convention followed here. This is however taken into account in the Lagrangian in~\eref{leffJTW}, so that the definition of the $C^{(\prime)}_i$ couplings is here the same as theirs. Ref.~\cite{Herczeg:2001vk} defines the couplings $C_A, C'_{V,S,T}$ with an overall minus sign compared to the present one, whereas in Ref.~\cite{Vos:2015eba} there is an unphysical minus sign difference for all couplings.}
\bea
- {\cal L}_{n \to p e^- \bar{\nu}_e}
&=& ~~\bar{p}~n ~\left( C_S \bar{e} \nu_e - C'_S \bar{e} \gamma_5 \nu_e  \right) \nonumber\\
&&+~ \bar{p}\gamma^\mu n \left( C_V \bar{e} \gamma_\mu \nu_e - C'_V \bar{e} \gamma_\mu \gamma_5 \nu_e  \right)  \nonumber\\
&&+~\frac{1}{2}\bar{p}\sigma^{\mu\nu} n \left( C_T \bar{e} \sigma_{\mu\nu} \nu_e - C'_T \bar{e} \sigma_{\mu\nu} \gamma_5 \nu_e  \right)  \nonumber\\
&&\mathbf{-}~ \bar{p}\gamma^\mu \gamma_5 n \left( C_A \bar{e} \gamma_\mu \gamma_5 \nu_e - C'_A \bar{e} \gamma_\mu \nu_e  \right)  \nonumber\\
&&+~ \bar{p} \gamma_5 n ~ \left( C_P \bar{e} \gamma_5 \nu_e - C'_P \bar{e} \nu_e  \right)+\mbox{h.c.}~
\label{eq:leffJTW}
\eea
One should keep in mind that this Lagrangian does not contain the effect of weak magnetism, which cannot be neglected in general, as explained above. To include this effect, one has to work at one more order in the effective hadronic Lagrangian, where derivative terms are present. This is explicitly done in Ref.~\cite{Ando:2004rk}, where the complete NLO effective Lagrangian {\it generated by the SM} is built, and used to calculate neutron $\beta$-decay observables. This Lagrangian contains both the recoil and isospin-symmetry breaking corrections discussed above, as well as the electromagnetic effects, which we will discuss in \sref{corrections}. 
On the other hand, the use of the leading-order Lee-Yang Lagrangian to study NP effects is 
justified due to their small magnitude, although two-body corrections could induce a non-negligible uncertainty in the analysis (\sref{nuclearmatrixelements}).

The Lee-Yang effective couplings $C_i$, $C_i'$ can be expressed in terms of the quark-level parameters as
\bea
\label{eq:matchLeeYang}
\begin{aligned}
\overline{C}_V+\overline{C}_V' &= 2 \,g_V  \left(1 + \eL + \eR \right)  \\
\overline{C}_A+\overline{C}_A' &= -2 \,g_A  \left(1 + \eL - \eR \right)  \\
\overline{C}_S+\overline{C}'_S &=   2 \,g_S  \, \eS  \\
\overline{C}_P+\overline{C}'_P &=   2 \,g_P  \, \eP  \\
\overline{C}_T+\overline{C}'_T &=   8 \,g_T  \, \eT  \\
\end{aligned}
\qquad\qquad
\begin{aligned}
\overline{C}_V-\overline{C}_V' &= 2 \,g_V  \left( \teL + \teR \right)  \\
\overline{C}_A-\overline{C}_A' &= 2 \,g_A  \left(\teL - \teR \right)  \\
\overline{C}_S-\overline{C}'_S  &=   2  \,g_S  \, \teS \\
\overline{C}_P-\overline{C}'_P  &=   -2  \,g_P  \, \teP \\
\overline{C}_T-\overline{C}'_T  &=  8  \,g_T  \, \teT ~,
\end{aligned}
\eea
where $C_{i} = (G^0_F V_{ud} / \sqrt{2} )\,  \overline{C}_{i}$. We present explicitly the sums and differences $C_i\pm C_i'$ due to their very different theoretical and phenomenological nature. Namely, the former (latter) only involves left(right)-handed neutrinos, and thus the corresponding amplitude does (not) interfere with the SM leading contribution, which produces a linear (quadratic) sensitivity to them. Consequently we expect that much stronger bounds on $C_i+C_i'$ than on $C_i-C_i'$ will be set from experiments.

As mentioned in~\sref{quark-level-EFT}, the $\eR$ coupling changes the relative size of the axial current with respect to the vector one:
\bea
\label{eq:gAcontamination}
\frac{C_A+C_A'}{C_V+C_V'} = -\frac{g_A}{g_V} \frac{1+\eL-\eR}{1+\eL+\eR}\approx  -\frac{g_A}{g_V} (1-2\,{\rm Re}\,\eR )\,e^{i\phi_{AV}} \left[ 1 + \cO(\epsilon) \right]~,
\eea
where $\phi_{AV}\approx -2\, {\rm Im} \,\eR$ is the relative phase between the axial and the vector terms. It is clear that the access to ${\rm Re}(\eR)$ requires indeed a precise calculation of the hadronic axial-vector form factor $g_A$ (or the so-called mixing ratio $\rho$ in nuclear decays).

Analyzing nuclear and neutron $\beta$ decays using this nucleon-level effective Lagrangian has various advantages. First, it makes possible to compare and combine the plethora of existing results from nuclear and neutron $\beta$ decays without introducing the hadronic charges. Secondly, expressing the outcome of experimental analyses in terms of the nucleon $C_i^{(\prime)}$ coefficients ensures that these analyses do not become obsolete as soon as new values of the hadronic charges become available.

\subsubsection{Hadronic charges}
\label{sec:hadronic-charges}
As shown above, hadronic effects can be described with the necessary precision in terms of four charges, namely $g_{A,S,P,T}$. There has been recently an enormous progress concerning these charges that we briefly summarize in this section.

{\bf Axial-vector charge $g_A$.}
The nucleon axial-vector charge is a fundamental quantity that plays a central role in the theoretical understanding of hadron and nuclear physics. It controls the Gamow-Teller component of nuclear $\beta$ decay and several astrophysical and cosmological processes, which depend very sensitively on its value (see e.g. Ref.~\cite{Mathews:2004kc}). The axial charge is also a benchmark quantity for Lattice QCD (LQCD), since its value can be precisely extracted from the $\beta$-asymmetry parameter $A$ in neutron decay. The various experimental extractions are discussed in \sref{correlations}. As reference, the 2016 Particle Data Group (PDG) value is $g_A=1.2723(23)$~\cite{Patrignani:2016xqp} and is only valid in the absence of non-standard interactions.\footnote{
In the conventions followed here $g_A$ (and thus also $\lambda\equiv g_A/g_V$) is positive, consistent with the usual lattice QCD definition. On the other hand, the PDG and experimental works usually define $g_A$ with the opposite sign.}

Once NP terms are introduced, the comparison of the QCD calculations of $g_A$ and the experimental value makes possible to probe the right-handed coupling, 
\eref{gAcontamination}, or more explicitly we have~\cite{Herczeg:2001vk,Bhattacharya:2011qm,Gonzalez-Alonso:2013uqa}
\bea
\label{eq:gA}
\tilde{g}_A = g_A^{\scriptscriptstyle{QCD}}~\mbox{Re}\left( \frac{1+\eL-\eR}{1+\eL+\eR} \right)
\approx g_A^{\scriptscriptstyle{QCD}} \left( 1 - 2 \,\mbox{Re}\eR \right) \left[ 1 + \cO\left( \eps_i\right) \right]~,
\eea
where $\tilde{g}_A$ is the phenomenological value extracted e.g. from the $\beta$-asymmetry parameter $A$, with the subtraction (if necessary) of the Fierz term $b_n$ contribution, {\it cf.}~\eref{Xtilde} and \sref{correlations}.

Various systematic effects made $g_A$ a very difficult quantity to calculate in the lattice during decades. However, in the last couple of years the LQCD community has been able to produce the two first results with all systematics accounted for, with a precision at the few-percent level \cite{Bhattacharya:2016zcn,Berkowitz:2017gql}. We note that, adding errors in quadrature and neglecting correlations, one finds a $\sim 1.6\sigma$ tension between these two lattice determinations. This should however be taken with caution, since both determinations %were carried out using the same lattice ensembles. 
share a subset of the same lattice ensembles.

In the numerical analysis we will use the most precise of these results, $g_A=1.278(21)(26)$~\cite{Berkowitz:2017gql}. Ongoing efforts are expected to further improve this result, provide a more precise determination, and hopefully clarify the small tension between Refs.~\cite{Bhattacharya:2016zcn,Berkowitz:2017gql}. For instance, Ref.~\cite{Chang:2017oll} presented recently a preliminary value with $1.3\%$ relative uncertainty (mainly statistical), i.e.\ $g_A=1.285(17)$, using a new computational method inspired by the Feynman-Hellmann theorem. 
A summary of $g_A$ lattice results~\cite{Edwards:2005ym,Yamazaki:2008py,Capitani:2012gj,Horsley:2013ayv,Bali:2014nma,Abdel-Rehim:2015owa,Bhattacharya:2016zcn,Green:2017keo,Berkowitz:2017gql,Alexandrou:2017hac,Capitani:2017qpc} is shown in \fref{gAgT}.

{\bf \lq\lq Non-standard\rq\rq~charges $g_{S,P,T}$.}
These charges are well-defined non-zero QCD quantities that we call \lq\lq non-standard\rq\rq~simply to stress the fact that they appear in $\beta$-decay expressions only when the corresponding non-standard interactions $\eps_{S,P,T}$ are present. They are distinct to the {\it induced} charges $\tilde{g}_S,\tilde{g}_P, \tilde{g}_{T(V)}$ and $\tilde{g}_{T(A)}$ that appear in the SM description, Eqs.~(\ref{eq:nucleonmatchingV})-(\ref{eq:nucleonmatchingA}).

The non-zero values of the $g_{S,P,T}$ charges are not to claim a NP effect, but a decent knowledge of them is crucial to obtain strong NP bounds from current and future searches, so that the experimental effort is not ruined by large QCD uncertainties. They are also needed to connect with the underlying dynamics, and to be able to assess the NP sensitivity of any given experiment. This is clearly seen in the most extreme and obvious scenario where one of the charges is zero and any NP sensitivity to the corresponding interaction is lost. Under such conditions, a very strong bound on the hadronic coefficient $C$ becomes void at the underlying quark level.

In contrast to the (axial-)vector charges, the (pseudo-)scalar and tensor charges are renormalization-scale and -scheme dependent QCD quantities. Hereafter we quote their values in the $\overline{MS}$ scheme and at the $\mu=2$~GeV scale, unless otherwise stated.

Due to their \lq\lq non-standard\rq\rq~nature, these quantities received much less attention than the axial charge and their values were poorly known until very recently. A first lattice estimate for $g_S$ and an average of existing lattice $g_T$ results were provided in  Ref.~\cite{Bhattacharya:2011qm}, although with large errors ($g_S = 0.8\pm0.4,~g_T=1.05\pm0.35$) and with various systematic effects requiring more detailed studies.
The level of precision required for $g_{S,T}$ so that future bounds on the non-standard couplings $\epsilon_{S,T}$ would be dominated by experimental errors was also analyzed \cite{Bhattacharya:2011qm}, assuming a per-mil level limit on the neutron Fierz term (in addition to current bounds on the pure-Fermi Fierz term). 
It was concluded that an improvement by a factor of 2 in the error was necessary, which was considered to be a feasible goal if the appropriate efforts were made. This motivated a renewed effort in the QCD community to study these quantities, and several lattice collaborations have studied them using a variety of techniques~\cite{Green:2012ej,Bhattacharya:2013ehc,Bali:2014nma,Bhattacharya:2015esa,Bhattacharya:2015wna,Abdel-Rehim:2015owa,Bhattacharya:2016zcn,Alexandrou:2017qyt}.

The PNDME collaboration presented later a new determination of $g_T$ including,  for  the  first  time,  a simultaneous extrapolation in the lattice spacing, volume, and light quark masses to the physical point in the continuum limit~\cite{Bhattacharya:2015esa,Bhattacharya:2015wna}. The result was later further improved to obtain a value with a 6\% precision, namely $g_T=0.987(51)(20)$~\cite{Bhattacharya:2016zcn}. 
As shown in~\fref{gAgT}, this result is in agreement with other lattice calculations, which show little sensitivity to the number of flavors. Sum-rule, Dyson-Schwinger, and phenomenological estimates~\cite{Pospelov:1999rg,Yamanaka:2013zoa,Fuyuto:2013gla,Anselmino:2013vqa,Pitschmann:2014jxa,Goldstein:2014aja,Kang:2015msa,Radici:2015mwa,Radici:2018iag}, which are less accurate, show also good agreement in general with the lattice determinations. The phenomenological determinations calculate the tensor charge as the integral over the longitudinal momentum fraction of the experimentally measured quark transversity distributions, which is expected to improve in the future with the arrival of new data~\cite{Gao:2010av,Goldstein:2014aja,Courtoy:2015haa,Ye:2016prn} and the synergy with the lattice~\cite{Lin:2017stx}.

\begin{figure}[!htb]
\centering
\includegraphics[width=0.95\columnwidth]{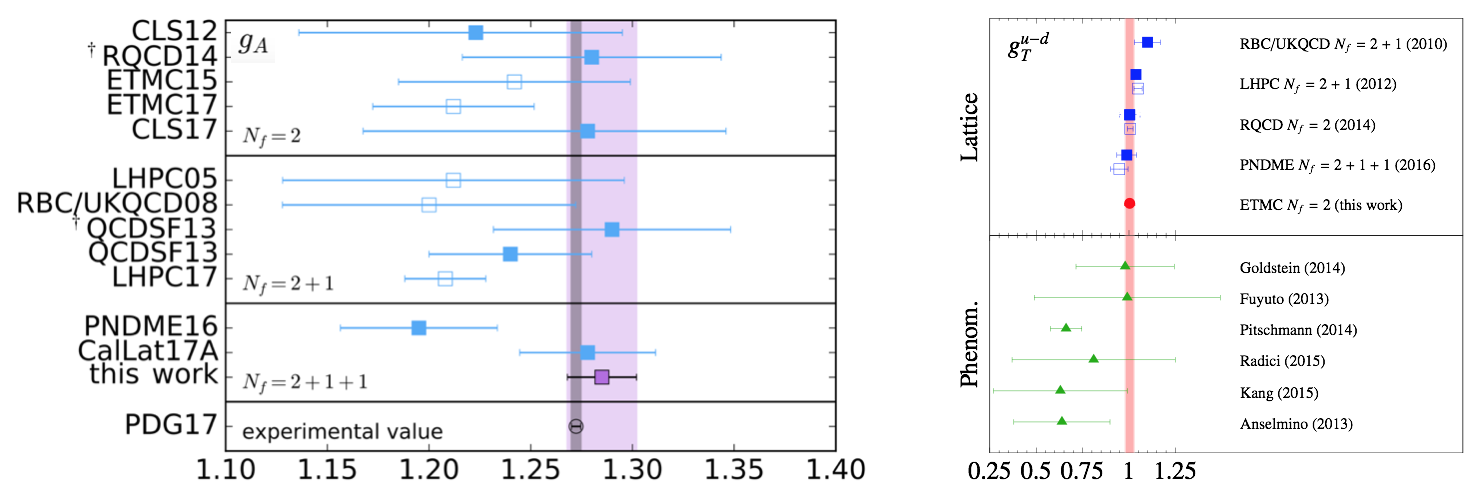}
\caption{
(Left) Summary of $g_A$ lattice results by LHPC05~\cite{Edwards:2005ym}, RBC/UKQCD08~\cite{Yamazaki:2008py}, CLS12~\cite{Capitani:2012gj}, QCDSF13~\cite{Horsley:2013ayv}, RQCD14~\cite{Bali:2014nma}, ETMC15~\cite{Abdel-Rehim:2015owa}, PNDME16~\cite{Bhattacharya:2016zcn}, LHPC17~\cite{Green:2017keo}, CalLat17A~\cite{Berkowitz:2017gql}, ETMC17~\cite{Alexandrou:2017hac}, and CLS17~\cite{Capitani:2017qpc}. 
The results with an open symbol are obtained from only one lattice spacing. 
From Ref.~\cite{Chang:2017oll}. (Right) Comparison of several lattice QCD calculations, phenomenological extractions and model determinations of the tensor charge $g_T$ (in the plot denoted as $g_T^{u-d}$)~\cite{Aoki:2010xg,Green:2012ej,Bali:2014nma,Bhattacharya:2016zcn,Alexandrou:2017qyt,Fuyuto:2013gla,Anselmino:2013vqa,Pitschmann:2014jxa,Goldstein:2014aja,Kang:2015msa,Radici:2015mwa}. Filled squares denote extrapolated values at the physical pion mass, whereas open squares denote lattice results at their lowest pion mass. Solid error bars denote statistical uncertainties whereas dashed error bars show the total uncertainties (statistical and systematic) added in quadrature. From Ref.~\cite{Alexandrou:2017qyt}.}
\label{fig:gAgT}
\end{figure}

Reaching a similar precision in the analogue calculations of the scalar charge was much more complicated, mainly due to much larger statistical uncertainties (see e.g. the results in Refs.~\cite{Green:2012ej,Bhattacharya:2013ehc,Bali:2014nma,Alexandrou:2017qyt}). However, a different method to calculate $g_S$ was followed in Ref.~\cite{Gonzalez-Alonso:2013ura} using the relation
\bea
\label{eq:gSfromCVC}
g_S = g_V \frac{(M_n-M_p)^{QCD}}{m_d-m_u}~,
\eea
which simply follows from CVC, and which involves the nucleon mass splitting in the absence of electromagnetic effects, as indicated by the QCD superindex.
This isospin-symmetry breaking quantity is calculable in the lattice and in fact there has been a significant progress during the last few years~(for a review see e.g. Ref.~\cite{Portelli:2015wna}). Thus, \eref{gSfromCVC} connects two different lattice efforts. Using this relation, the currently most precise determination of the scalar charge, $g_S=1.02(11)$, was obtained \cite{Gonzalez-Alonso:2013ura}.
Although it used an average of available lattice calculations of the nucleon mass splitting $(M_n-M_p)^{QCD}$, the final error is dominated by the determination from the BMW collaboration~\cite{Borsanyi:2013lga}. Moreover, using the {\it ab initio} calculation from the BMW collaboration~\cite{Borsanyi:2014jba} one finds $g_S=1.00(14)$, in perfect agreement with the result of Ref.~\cite{Gonzalez-Alonso:2013ura}. These two calculations~\cite{Borsanyi:2013lga,Borsanyi:2014jba} are the only ones with complete control of all lattice systematics. The direct calculation of the ratio in \eref{gSfromCVC} in the lattice should produce an even more precise determination, as several errors are expected to cancel between numerator and denominator. Somewhat surprisingly, even though some collaborations had extracted both quantities in the past (see e.g. Ref.~\cite{deDivitiis:2013xla}), such calculation has not been performed yet.

More recently, the PNDME collaboration provided a direct calculation of the scalar charge with similar precision, namely $g_S=0.97(12)(6)$~\cite{Bhattacharya:2016zcn}. The consistency with the result obtained with \eref{gSfromCVC} is a highly nontrivial check, as both methods have entirely different systematic errors. Inversely, one can use \eref{gSfromCVC} to produce a competitive and independent result for the nucleon mass splitting in the absence of electromagnetism~\cite{Gonzalez-Alonso:2013ura}, as done in Ref.~\cite{Bhattacharya:2016zcn}.

As a result of these efforts both the scalar and tensor charges are now sufficiently well known so that the subsequent bounds on the quark-level coefficients $\eps_i,\teps_i$ are fully dominated by experimental uncertainties. This represents a non-trivial achievement by the QCD community, which does not happen very often in BSM searches with hadronic probes. Further improvement in the calculation of the hadronic charges will not have any significant impact on the $\beta$-decay bounds, but it will of course strengthen the reliability of current estimates, and the control of all systematics. We include the determination of the charges in the fit in \sref{fit} as gaussianly distributed variables. More involved approaches, such as the R-fit method~\cite{Hocker:2001xe}, which treats theory errors in a more conservative way, would give the same numerical results due to the small size of current errors.

To conclude we consider the pseudo-scalar charge $g_P$. Using PCAC, an analogue relation to \eref{gSfromCVC} is derived, giving $g_P=g_A (M_n+M_p)/(m_d+m_u)=349(9)$~\cite{Gonzalez-Alonso:2013ura}. The large enhancement is due to a charged pion pole in the coupling of a pseudoscalar field to the $du$ vertex in QCD at low energies. Such large factor considerably reduces the suppression of order $q/M\sim 10^3$ of the pseudoscalar bilinear, greatly increasing the sensitivity of $\beta$ decay to pseudoscalar interactions~\cite{Gonzalez-Alonso:2013ura}, as further discussed in~\sref{pseudoscalar}.

\tref{charges} summarizes the values of the hadronic charges used in the numerical analysis (\sref{fit}).

\begin{table}[b]
\centering
\caption{Values of the hadronic charges used in this work. They are given at $\mu=2$ GeV in the $\overline{MS}$ scheme, and errors were added quadratically. See main text for further details.}
\label{tab:charges}
\begin{tabular}{ c  c  c }
 \hline\hline
 Charge	&	Value	&	Ref.	\\ \hline
$g_A$	&	1.278(33)	&	\cite{Berkowitz:2017gql}	\\
$g_T$	&	0.987(55)	&	\cite{Bhattacharya:2016zcn}	\\
$g_S$	&	1.02(11)	&	\cite{Gonzalez-Alonso:2013ura}	\\
$g_P$	&	349(9)	&	\cite{Gonzalez-Alonso:2013ura} \\ \hline\hline
\end{tabular}
\end{table}

\subsection{Nucleus-level EFT ($M_{F,GT}\,C^{(\prime)}_i$)}
\label{sec:nuclearmatrixelements}

In neutron decay, {\it cf.} \eref{nucleonmatching}, one can also perform a general Lorentz decomposition of the nuclear matrix elements and expand in powers of the momentum transfer. The expressions, which can be found in Refs.~\cite{Holstein:1974zf,Cirigliano:2013xha}, are more involved due to the possible spin sequences.
Many of the features discussed for neutron decay also apply to nuclear transitions that are relevant for BSM searches.

At zero-th order in the recoil expansion, the only free parameters in the SM are the nuclear vector and axial-vector charges, which correspond to the Fermi and Gamow-Teller nuclear matrix elements, $M_{F,GT}$. The former is fixed by CVC in the isospin-symmetric limit, whereas the later has to be extracted from data, as nuclear uncertainties make it even more complicated to predict than in the neutron case (see Ref.~\cite{Suhonen:2017krv} for a review).

At higher order in recoil, various induced nuclear form factors appear. The dominant term is usually the nuclear weak magnetism~\cite{Holstein:1974zf,Gell-Mann1958}, related to the hadronic term $\tilde{g}_{T(V)} (q^2)$ in~\eref{nucleonmatchingV}.\footnote{For transitions where the involved nuclei are not members of a common isotopic multiplet, the induced tensor, related to the hadronic term $\tilde{g}_{T(A)} (q^2)$ in~\eref{nucleonmatchingA}, can also give a non-negligible contribution that must be taken into account.} Since its contribution to the observables is now reaching the same order of magnitude as the experimental precision ($\sim 0.1-1.0\%$)
~\cite{Adelberger1999,Gorelov2005, Pitcairn2009, Wauters2009, Wauters2010, Soti2014, Sternberg2015, Fenker2016}, the effect of weak magnetism in nuclei has to be sufficiently well understood \cite{Severijns2017}. This is not a problem for three classes of transitions: 
(i) superallowed pure Fermi $\beta$ decays, where weak magnetism is absent; 
(ii) the neutron and mixed F/GT mirror $\beta$ transitions, all occurring between members of an isospin doublet, where the weak magnetism contribution is given by CVC in terms of the nuclear magnetic moments of the two analog states connected by the $\beta$ transition~\cite{Holstein:1974zf, Calaprice1976}; 
and (iii) for $\beta$ transitions from states that are part of a $T = 1$ multiplet decaying to $T = 0$ states, such as the $^6$He decay, since in this case weak magnetism is related by CVC to the M1 transition strength of the $\gamma$ decay analog to the $\beta$ transition~\cite{Gell-Mann1958,Holstein:1974zf, Calaprice1976}. 
This is why most searches for new physics concentrate on $\beta$ transitions of one of these three types. 
However, when dealing with $\beta$ transitions for which no CVC prediction for weak magnetism is available from experimentally known quantities, one has to rely on systematic studies or on theoretical matrix element calculations \cite{Severijns2018}. To improve the situation, new experimental data, especially for masses $A > 40$ would be valuable as today only limited experimental information is available \cite{Severijns2018}.

Isospin-breaking contributions to $M_F$, which are nuclear-structure dependent, appear at the per-mil level and have to be included for the most precise experimental studies. They are usually encoded as $|M_F|^2=|M_F^0|^2(1-\delta_C)$, where $M_F^0$ is the value given by isospin symmetry. 

Last, non-standard scalar and tensor currents require the calculation of their corresponding nuclear charges, like $g_{S,T}$ in neutron decay~{\it cf.}~\eref{defgS} and \eref{defgT}. In the impulse approximation, where nuclei are treated as collections of free nucleons, the scalar and tensor nuclear charges are equal to the Fermi and Gamow-Teller matrix elements, $M_{F,GT}$, multiplied by the hadronic scalar and tensor charges $g_{S,T}$. Thus, no new nuclear matrix elements have to be calculated when BSM effects are introduced. The meson-exchange, or two-body, corrections to this approximation~\cite{Chemtob:1971pu} are usually neglected, but they might become the dominant theoretical error taking into account the precision achieved in the calculation of the hadronic charges. Lattice QCD studies of nuclei represent an active research field that can be used to study these corrections~\cite{Chang:2017eiq}.

\subsection{Observables}
\label{sec:observables}
Based on the low-energy Effective Lagrangian of~\eref{leff-lowE}, and once the hadronization and nuclearization has been carried out, the phenomenologically relevant $\beta$-decay observables can be calculated. This requires a series of approximations and the precision of the calculation will be both observable and transition dependent. Observables will be linear or quadratic functions of the underlying Wilson coefficients $\epsilon_i$ (or $C^{(\prime)}_i$), and also of certain strong-interaction-dynamics parameters that cannot be calculated with sufficient precision in QCD, such as the axial-vector charge $g_A$. The goal of the $\beta$-decay measurements discussed in this work is to extract with high accuracy and precision the value of these parameters, to learn about the underlying BSM dynamics, as well as about QCD and nuclear physics.

The leading-order expressions for some representative observables is obtained by neglecting recoil and electromagnetic corrections. The distribution in electron and neutrino directions and in electron energy from polarized nuclei is given by~\cite{Jackson:1957zz,Jackson:1957auh}:
\begin{eqnarray}
\label{eq:jtw1}
&&w(\langle \bfvec{J} \rangle \vert E_e, \Omega _e, \Omega_\nu) dE_e d\Omega _e d\Omega _\nu
=  \frac{F(\pm Z, E_e)}{(2\pi )^5} p_e E_e (E_0 - E_e)^2 dE_e d\Omega _e d\Omega _\nu \times  \\
&& \xi \left\{ 1 + a \frac{\bfvec{p}_e \cdot \bfvec{p}_\nu}{E_e E_\nu} +
b \frac{m_e}{E_e}
+ \frac{\langle \bfvec{J} \rangle}{J} \cdot \left[ A \frac{\bfvec{p}_e}{E_e} +
B \frac{\bfvec{p}_\nu}{E_\nu} + D \frac{\bfvec{p}_e \times
\bfvec{p}_\nu}{E_e E_\nu} \right] \right\} ~ ,\nonumber
\end{eqnarray}
\noindent while the distribution in electron energy and angle and in electron polarization from polarized nuclei is given by \cite{Jackson:1957zz, Jackson:1957auh}:
\begin{eqnarray}
\label{eq:jtw2}
&&w(\langle \bfvec{J} \rangle , \bfvec{\sigma} \vert E_e, \Omega _e)dE_e d\Omega _e
=  \frac{F(\pm Z, E_e)}{(2\pi )^4} p_e E_e (E_0 - E_e)^2 dE_e d\Omega _e \times \nonumber \\
&& \xi \left\{ 1 + b \frac{m_e}{E_e} + \frac{\bfvec{p}_e}{E_e} \cdot
\left( A \frac{\langle \bfvec{J} \rangle}{J} + G \bfvec{\sigma } \right) +
 \bfvec{\sigma} \cdot \left[ N \frac{\langle \bfvec{J} \rangle}{J}
 + Q \frac{\bfvec{p}_e}{E_e + m_e} \left( \frac{\langle \bfvec{J}
\rangle}{J} \cdot \frac{\bfvec{p}_e}{E_e} \right) + R \frac{\langle \bfvec{J}
\rangle}{J} \times \frac{\bfvec{p}_e}{E_e} \right] \right\} ~ .
\end{eqnarray}
\noindent In these equations $E_e$, $p_e$, and $\Omega_e$ are the total energy, momentum, and angular
coordinates of the $\beta$ particle and similarly for the neutrino, $E_0$ is the maximum total electron energy, $m_e$ is the electron mass, $\langle \bfvec{J} \rangle$ is the nuclear
polarization of the initial nuclear state with spin $\bfvec{J}$, $\bfvec{\sigma}$ is the spin vector of the $\beta$ particle, and
$F(\pm Z, E_e)$ is the Fermi function, which encodes the dominant Coulomb correction. The upper (lower) sign refers to $\beta^-$ ($\beta^+$) decay and $Z$ is the atomic number of the daughter nucleus.

At leading order the overall factor $\xi$, and the correlation coefficients $a$, $b$, $A$, $B$, etc., depend on the Fermi and pure Gamow-Teller matrix elements, $M_{F,GT}$, on the Lee-Yang coupling constants $C_i^{(\prime)}$, and for some of them on the $\beta$ energy. Their expressions can be found in the seminal articles by Jackson, Treiman and Wyld~\cite{Jackson:1957zz,Jackson:1957auh}, whereas the most general distribution, $w(\langle \bfvec{J} \rangle , \bfvec{\sigma} \vert E_e, \Omega _e,\Omega _\nu)$, which involves 22 coefficients, was obtained a few months later by Ebel and Feldman~\cite{EBEL1957213}. For illustration, the forms of $\xi$, $a$ and $b$ are
\bea
\label{eq:xi}
\xi	&=& |M_F|^2 \left[ |C_V|^2+|C'_V|^2+ |C_S|^2+|C'_S|^2 \right] + |M_{GT}|^2 \left[ |C_A|^2+|C'_A|^2+ |C_T|^2+|C'_T|^2 \right]~, \\
\label{eq:a}
a \times \xi	&=& |M_F|^2 \left[ |C_V|^2+|C'_V|^2 - |C_S|^2-|C'_S|^2 \right] - \frac{1}{3}|M_{GT}|^2 \left[ |C_A|^2+|C'_A|^2- |C_T|^2-|C'_T|^2 \right]~, \\
\label{eq:b}
b \times \xi	&=& \pm 2\,\gamma\,{\rm Re} \left[ |M_F|^2 \left( C_VC_S^* + C'_VC_S^{\prime*}\right) + |M_{GT}|^2 \left( C_AC_T^* + C'_AC_T^{\prime*}\right)  \right]~,
\eea
where $\gamma \equiv \sqrt{1-\alpha^2 Z^2}$ and $\alpha$ is the fine structure constant. For the sake of brevity, we do not include the Coulomb corrections to the $a$ coefficient that can be found in Ref.~\cite{Jackson:1957auh}. 
In these expressions we observe the following generic and important features:
\bi
\item Pure Fermi transitions depend on the Vector and Scalar interaction coupling constants, $C_{V,S}^{(\prime)}$, pure Gamow-Teller transitions depend on the Axial-Vector and Tensor ones, $C_{A,T}^{(\prime)}$, and mixed transitions depend on a combination of all four of them;
\item For pure Fermi or Gamow-Teller transitions the correlation coefficients become independent of the nuclear matrix elements, which only contribute to the normalization $\xi$, thus avoiding the need for a detailed knowledge of the nuclear structure. On the other hand, correlation coefficients in mixed transitions are sensitive to the nuclear matrix elements, but only through the ratio $|M_{GT}|^2 / |M_F|^2$.
\ei
\subsubsection{Linear effects}
\label{sec:linear}
Once again, it is important to identify in the above expressions the effects that are linear (instead of quadratic) in the small non-standard couplings, offering thus the largest NP sensitivity. It is straightforward to see that at linear order in the non-SM couplings, there are only 4 $\mathcal{CP}$-conserving and 3 $\mathcal{CP}$-violating Wilson coefficients left, e.g., %which were chosen to be
\bea
&&~{\rm Re}(C_V+C'_V)~,~{\rm Re}\left(\frac{C_A+C'_A}{C_V+C'_V}\right)~,~{\rm Re}\left(\frac{C_S+C'_S}{C_V+C'_V}\right)~,~{\rm Re}\left(\frac{C_T+C'_T}{C_A+C'_A}\right)~,\\
&&~{\rm Im}\left(\frac{C_A+C'_A}{C_V+C'_V}\right)	,~{\rm Im}\left(\frac{C_S+C'_S}{C_V+C'_V}\right)	,~{\rm Im}\left(\frac{C_T+C'_T}{C_A+C'_A}\right)~.
\eea	
These reduce to two $\mathcal{CP}$-conserving coefficients in the SM limit: ${\rm Re}(C_V+C'_V)$ and ${\rm Re}(C_A+C'_A)$, which correspond to $V_{ud}$ and $g_A$. Neglecting quadratic terms in the NP couplings greatly simplifies the analysis and it represents an excellent approximation in the absence of light right-handed neutrinos, which is a well motivated NP scenario. %the case by construction in the SMEFT.
For illustration, the linearized forms of $\xi$, $a$ and $b$ are
\bea
\label{eq:linearJTW1}
\xi	&\approx& \frac{1}{2}|M_F|^2 |C_V+C'_V|^2 \left( 1+ |\tilde{\rho}|^2 \right)~, \\
\label{eq:linearJTW1a}
a 			&\approx& \frac{ 1 - |\tilde{\rho}|^2/3}{1+ |\tilde{\rho}|^2}~, \\
b 			&\approx& \pm\,2\gamma\,\frac{1}{1+ |\tilde{\rho}^2|} {\rm Re} \left( \frac{C_S+C'_S}{C_V+C'_V} + |\tilde{\rho}|^2 \frac{C_T+C'_T}{C_A+C'_A}  \right)~,
\eea
where
\bea
\label{eq:mixingratio}
\tilde{\rho}\equiv \frac{C_A+C'_A}{C_V+C'_V} \frac{M_{GT}}{M_F}
= - \frac{g_A}{g_V} \frac{M_{GT}}{M_F} \frac{1 + \eL - \eR}{1 + \eL + \eR} ~.
\eea
At this order, $\tilde{\rho}$ can be extracted from e.g. the measurement of $a$. In the SM limit, this quantity is the so-called Fermi/Gamow-Teller mixing ratio $\rho$. The tilde over $\rho$ indicates the additional NP contribution.
Thus, at linear order in NP, the phenomenologically accessible quantity $\tilde{\rho}$ contains (i) the hadronic contribution associated with the axial-vector charge $g_A$; (ii) the nuclear contribution from the matrix elements; (iii) certain NP contributions; and (iv) certain subleading corrections such as the inner radiative corrections, {\it cf.}~\sref{corrections}. It is highly non-trivial that so-many effects affecting the correlation coefficients can be encoded in a single parameter that is phenomenologically accessible for each transition. 

One can explicitly see now the features that were anticipated in~\sref{quark-level-EFT}:
\bi
\item Nonstandard vector interactions, captured by $C_V^{(\prime)}$, only affect the overall normalization, $\xi$. This translates in a contaminated extraction of $V_{ud}$, {\it cf.}~\eref{Vtilde}, which would in turn contribute to the CKM-unitarity test in~\eref{deltaCKMdef}. Thus, superallowed pure Fermi transitions are the best probes for these interactions;
\item Nonstandard axial-vector interactions get fully hidden in the mixing ratio $\tilde{\rho}$, {\it cf.}~\eref{mixingratio}. As a result, a theoretical calculation of this quantity is needed to set a bound on them, which will clearly limit the bound that can be obtained, and which makes neutron decay much better suited than nuclear decays in this case;
\item $\mathcal{CP}$-conserving scalar and tensor interactions modify the $\beta$-decay differential distributions. In the main distribution of~\eref{jtw1}, the only linear effect is the so-called ``Fierz interference term", which has an energy dependence of the form $m_e/E_e$. This is due to the chirality-flipping nature of these operators, together with the sum over the electron polarizations. Thus, the coefficients $a$ and $A$ are given by their SM expressions, the standard Fierz term $b$ is explicitly displayed in~\eref{jtw1}, whereas $B$ takes the form $B_{SM}(\tilde{\rho})+b_B m_e/E_e$. This additional Fierz-like term, $b_B$, is however hard to access experimentally as it involves the neutrino momentum and the nuclear polarization. In contrast, the Fierz term $b$ affects almost any measurement since it does not vanish under angular or energy integration;
\item Last, $\mathcal{CP}$-violating interactions modify $\mathcal{CP}$-odd coefficients, such as $D$ or $R$.
\ei

From the discussion above, it is clear that the Fierz term has a privileged status among the correlation coefficients in the context of NP searches. Somehow strangely, this term was neglected in many BSM searches with $\beta$ decays in the past. Historically, in 1937 Fierz noted that if both $V$ and $S$ (or $A$ and $T$) were present then the $\beta$ spectrum is distorted by an interference term, and it is not anymore a \lq\lq statistical function\rq\rq~\cite{Fierz1937}.\footnote{In this same paper he also showed that there were only 5 (P-conserving!) non-derivative interactions: $A\times A$, $V\times V$, $S\times S$, $P\times P$, and $T\times T$. To prove it he made use of the relations that are nowadays known as Fierz identities. Remarkably, he was only 25 years old when he wrote this article introducing two concepts that nowadays are named after him.} Very soon, this Fierz term was found to be small and the goal became to determine if the main contribution was $V$ or $S$ and $T$ or $A$.\footnote{The situation was actually less clear, since derivative interactions were also considered. In fact, in the period of 1935-1940 the correct interaction was thought to be derivative (Konopinski-Uhlenbeck theory)~\cite{Konopinski:1935zz}, due to several incorrect measurements of the $\beta$ spectrum~\cite{Amaldi:1984as,Pais:1986nu}.}

Once the dominant $V-A$ character was established~\cite{Allen1958}, several experiments in the 60's and 70's verified the $V-A$ predictions through other observables taking often $b=0$ in the analyses. When searches beyond the SM started around the 80's, such practice was maintained for a while, which was not justified when searching for residual non-standard scalar and tensor terms with left-handed neutrinos. The situation has changed around the turn of the century, and the importance of the Fierz term is now fully recognized, which has motivated new experimental and theory efforts concerning the $\beta$ energy spectra, as discussed below.
The story comes therefore full circle: the $\beta$ spectrum, which was the relevant differential distribution at the beginning (leading Pauli to the neutrino proposal), is now again a very important one.

It is to be stressed that setting $b=0$ is justified under certain conditions. For example, in analyses performed within the SM framework, such as the extraction of the axial-vector coupling $g_A$ from measurements of the asymmetry parameter $A$ in neutron decay. Or in specific measurements that are not precise enough to improve current bounds on $b$ and that therefore focus on the sensitivity to interactions entering mainly through quadratic contributions.

\subsubsection{Total decay rates}
\label{sec:totalrates}

When integrating over all kinematic variables the differential distribution in~\eref{jtw1}, one obtains the partial half-life $t$ of the transition, which multiplied by the statistical rate function,
\bea
f\equiv \int F(\pm Z, E_e) p_e E_e (E_0 - E_e)^2 dE_e / m_e^5~,
\label{eq:fdefinition}
\eea
gives the so-called $ft$ value:
\bea
\label{eq:ft}
ft_i = \frac{K}{\xi}\frac{1}{1+ b\,\langle m_e/E_e\rangle}~,
\eea
where $K$ is a combination of fundamental constants, $K/(\hbar c)^6 = 2 \pi^3 \hbar ~ ln 2/(m_e c^2)^5 = 8120.27649(25) \times 10^{-10} \rm{GeV}^{-4}\, s$~\cite{Patrignani:2016xqp}, and $\langle m_e/E_e \rangle$ is the inverse decay energy of each $\beta$ transition averaged over the statistical rate function.\footnote{We follow the definition of $b$ of Ref.~\cite{Jackson:1957auh}, which includes the Coulomb correction factor $\gamma$, {\it cf.}~\eref{b}. This entails a small transition dependence in $b$ beyond the Fermi or Gamow-Teller matrix elements. To avoid this, some works do not include the $\gamma$ factor in $b$, see e.g. Ref.~\cite{Hardy:2014qxa}.} 
 Thus, a non-zero Fierz term induces a $\langle m_e/E_e\rangle$ dependence in the $ft$ values, which are not anymore transition independent as predicted in the SM.
This was first pointed out in 1956 by Gerhart and Sherr~\cite{Gerhart1956}, and used shortly afterwards by Gerhart, who obtained $|b_F|<0.12$~\cite{Gerhart:1958zz}.
During the last 40 years the method has been greatly improved by Hardy and Towner, through the use of more precise measurements and the inclusion of subleading SM corrections (see ~\sref{corrections}). The latest analysis found the impressive result $b_F = -0.0028(26)$~\cite{Hardy:2014qxa}. It is worth noting that the limit was already at the few per-mil level in 1975, namely $b_F = -0.0010(60)$~\cite{Hardy:1975eq}.
Since the factor $\langle m_e/E_e \rangle$ increases almost monotonically with decreasing $Z$ (see~\tref{ftFermi}), the Fierz term can better be probed by comparing the $ft$ values between transitions with low and high endpoint energies, such as $^{10}$C and $^{26m}$Al.

On the other hand, the transition-independent contribution is controlled by the overall normalization $\xi$ which, using Eqs.~(\ref{eq:Vtilde}), (\ref{eq:matchLeeYang}) and~(\ref{eq:linearJTW1}), is given by
\bea
\label{eq:xi-linear}
\xi 	\approx \frac{1}{2}|M_F|^2 |C_V+C'_V|^2 \left( 1+ |\tilde{\rho}|^2 \right)
	\approx |M_F|^2 |G_F \,\tilde{V}_{ud}\, g_V|^2 \left( 1+ |\tilde{\rho}|^2 \right)
	~\xrightarrow{\text{SM}}~ |M_F|^2 |G_F \,V_{ud}\, g_V|^2 \left( 1+ |\rho|^2 \right)~.
\eea
Thus, from the $ft$-values it is possible to extract the contaminated $V_{ud}$ matrix element, $\tilde{V}_{ud}$, as long as $M_F$ and $\tilde{\rho}$ are known. Finally, as explained in~\sref{nucleon-level-EFT}, $g_V=1$ up to quadratic corrections in isospin-symmetry breaking.
Nuclear structure and radiative corrections have to be included in the analysis {\it cf.}~\sref{corrections}, but in this section we neglect them for the sake of simplicity.

We apply now briefly the above results for the $ft$ values to three types of transitions of particular phenomenological relevance, which have been thoroughly studied in the past: (i) superallowed $0^+ \rightarrow 0^+$ pure Fermi transitions, (ii) the nuclear $T = 1/2$ mirror $\beta$ transitions, and (iii) neutron decay.

For superallowed $0^+ \rightarrow 0^+$ pure Fermi $\beta$ transitions, for which $M_{GT} = 0$ and $M_F^2 = 2$ in the isospin-symmetric limit, one has
\bea
\label{eq:Ft-Fermi-TH}
\begin{aligned}
ft_i^{0^+ \rightarrow 0^+}	&=	\frac{K}{\xi^{0^+ \rightarrow 0^+}}\frac{1}{1 + b_F \, \langle m_e/E_e\rangle}\\
\xi^{0^+ \rightarrow 0^+}	&\approx 2\, G_F^2 |\tilde{V}_{ud}|^2	\\
b_F					&\approx -\,2\,\gamma\, {\rm Re} \frac{C_S+C'_S}{C_V+C'_V}\\
\end{aligned}
~~
\begin{aligned}
&\xrightarrow{\text{SM}}~~\frac{K}{\xi^{0^+ \rightarrow 0^+}}~,  \\
&\xrightarrow{\text{SM}}~~2\, G_F^2 |V_{ud}|^2~,\\
&~\xrightarrow{\text{SM}}~~0~.
\end{aligned}
\eea

For superallowed $\beta$ transitions between the isobaric analog states in $T = 1/2$ isospin doublets, so-called mirror $\beta$ transitions, we have $|M_F|^2=1$ and a non-zero Fermi/Gamow-Teller mixing ratio, and thus
\bea
\label{eq:Ft-mirror-TH}
\begin{aligned}
ft_i^{\rm{mirror}}		&=	\frac{K}{\xi_i^{\rm{mirror}}}\frac{1}{1+ b_i \, \langle m_e/E_e\rangle}\\
\xi_i^{\rm{mirror}}		&\approx G_F^2 |\tilde{V}_{ud}|^2 \left( 1+ |\tilde{\rho}_i|^2 \right)\\
b_i					&\approx \pm\frac{2\,\gamma}{1+ |\tilde{\rho}_i^2|} {\rm Re} \left( \frac{C_S+C'_S}{C_V+C'_V} + |\tilde{\rho}_i|^2 \frac{C_T+C'_T}{C_A+C'_A}  \right)\\
\end{aligned}
~~
\begin{aligned}
&\xrightarrow{\text{SM}}~~\frac{K}{\xi_i^{\rm{mirror}}}~,  \\
&\xrightarrow{\text{SM}}~~G_F^2 |V_{ud}|^2 \left( 1+ |\rho_i|^2 \right)~,\\
&~\xrightarrow{\text{SM}}~~0~.
\end{aligned}
\eea
\noindent For these transitions a second measurement is needed to extract the mixing ratio $\tilde{\rho}_i$, such as the measurement of the $\beta$-$\nu$ correlation, $a$, or the $\beta$-asymmetry parameter, $A$. Rearranging~\eref{Ft-mirror-TH} one can define $ft_0^{\rm{mirror}} 	\equiv ft_i \left( 1+  |\tilde{\rho}_i|^2 \right)$ which, similar to the $0^+ \rightarrow 0^+$ case, is transition independent except for the Fierz-term contribution.

These expressions apply in particular to neutron decay, which from a theoretical point of view is the simplest among the $T$ = 1/2 mirror $\beta$ decays, namely
\bea
\label{eq:Ft-n}
\begin{aligned}
f_n \,\tau_n\, \ln{2}		&=	\frac{K }{\xi_n}\frac{1}{1+ b_n \, \langle m_e/E_e\rangle}\\
\xi_n					&\approx G_F^2 |\tilde{V}_{ud}|^2 \left( 1+ 3|\tilde{\lambda}|^2 \right)\\
b_n					&\approx \frac{2\,\gamma}{1+ 3|\tilde{\lambda}|^2} {\rm Re} \left( \frac{C_S+C'_S}{C_V+C'_V} + 3|\tilde{\lambda}|^2 \frac{C_T+C'_T}{C_A+C'_A}  \right)\\
\end{aligned}
~~
\begin{aligned}
&\xrightarrow{\text{SM}}~~\frac{K}{\xi_n}~,  \\
&\xrightarrow{\text{SM}}~~G_F^2 |V_{ud}|^2 \left( 1+ 3|\lambda|^2 \right)~,\\
&~\xrightarrow{\text{SM}}~~0~,
\end{aligned}
\eea
where $\tau_n$ is the neutron lifetime. We wrote here the mixing ratio as $\tilde{\rho}=\sqrt{3}\,\tilde{\lambda}$ to separate the ratio of nuclear matrix elements, which are known in this case ($M_F$ = 1 and $M_{GT} = \sqrt{3}$), from the not-so-well-known hadronic ratio $\lambda$. As in the nuclear transitions, the tilde indicates that NP effects have been absorbed in that ratio.
As mentioned above, a second measurement is needed to extract $\tilde{\lambda}$, typically the $\beta$-asymmetry parameter $A$, and a third one to disentangle the Fierz term. The latter is of course not needed in a SM analysis, since $b = 0$.

\subsubsection{Differential observables}
\label{sec:diff-observables}
%\label{issue-Xtilde}

{\bf $\beta$ spectrum shape.} From the differential distribution in~\eref{jtw1} we see that the leading-order expression for the $\beta$-energy spectrum is given by
\bea
\label{eq:Fierz}
W(E_e) dE_e =  \frac{F(\pm Z, E_e)}{2\pi^3} p_e E_e (E_0 - E_e)^2 dE_e \,\xi \left( 1 + b \frac{m_e}{E_e}  \right) ~ .
\eea
Thus, the Fierz term is the only effect in the shape of the $\beta$-energy spectrum, which can be used to set limits on BSM physics (\sref{spectrum}). Although the effect of the Fierz term in the total integral ($ft$-values) is maximal for very low endpoints, its effect on the shape is very weak in that limit, simply because the factor $m_e/E_e$ becomes almost energy independent, namely $m/E\approx 1$. It was observed recently \cite{Gonzalez-Alonso:2016jzm} that the NP sensitivity of shape measurements is maximal for endpoints energies around 1-2~MeV, and that it decreases very quickly for smaller or larger values.

{\bf $\mathcal{CP}$-conserving correlation coefficients.}
As discussed in~\sref{linear}, $\mathcal{CP}$-conserving correlation coefficients, such as $a$ or $A$, are not linearly sensitive to NP effects, which has important consequences. For definiteness we focus here on $a$, {\it cf.}~\eref{linearJTW1a}. In pure Fermi or Gamow-Teller transitions its numerical value is fixed ($a_F=1$ and $a_{GT}=-1/3$) up to quadratic NP terms, which makes their measurement a unique probe of operators involving right-handed neutrinos ($C_i-C'_i$ terms). In mixed transitions, measurement make possible the determination of the mixing ratio $\tilde{\rho}$, which is a crucial input to use the $ft$ values from mirror decays or the neutron lifetime for BSM searches or to determine $V_{ud}$ in a SM analysis.

Due to the way how experimental asymmetries are measured, it is usually not $X$ itself which is determined but rather
\begin{equation}
\label{eq:Xtilde}
\tilde{X} = \frac{X}{1 + b \langle m_e/E_e \rangle}  ~ ,
\end{equation}
with the average inverse energy depending on the specific isotope and the experimental setup being used. Since most correlation coefficients depend quadratically on the exotic coupling constants while the Fierz interference term has a linear dependence on (left-handed) scalar and tensor couplings, measurements of $\tilde{X}$ are primarily measurements of $b$.

This feature makes possible to easily re-interpret past extractions of various correlation coefficients, usually performed taking $b=0$, as valid constraints in a general BSM setup with a non-zero Fierz term. However, as discussed in detail recently \cite{Gonzalez-Alonso:2016jzm}, Eq.~(\ref{eq:Xtilde}) is not valid for every experimental extraction of a correlation coefficient. In particular it was shown explicitly that it cannot be applied to values of the $\beta-\nu$ angular correlation coefficient $a$ extracted from differential measurements of the recoil momentum distributions \cite{Gonzalez-Alonso:2016jzm}. The prescription has been applied however, in a somewhat undiscriminated way, in several recent global fits \cite{Severijns:2006dr,Konrad2010,Wauters:2013loa,Vos:2015eba} for the re-interpretation of values of $a$ extracted in $^6$He and in neutron decays. The numerical impact of this misinterpretation on the constraints of exotic couplings extracted in global fits is in most cases quite small, simply because the precision achieved so far in measurements of recoil distributions is moderate. The associated constraints are therefore not competitive with determinations of $b$ from other observables that dominate the fits.

{\bf $\mathcal{CP}$-violating correlations coefficients.}
Spin and momentum vectors are odd under the time-reversal operation. Tests of time reversal symmetry in $\beta$ decay require correlations with an odd number of spins and momenta of the particles involved \cite{Jackson:1957zz,Jackson:1957auh}. The $D$ correlation is related to the term $\bf{J}\cdot(\bf{p_e}\times\bf{p_{\nu}})$ in Eq.~(\ref{eq:jtw1}) and is sensitive to an imaginary phase between the $V$ and $A$ couplings. The $R$ correlation is related to the term $\bfvec{\sigma} \cdot (\bf{J} \times \bf{p_e})$ in Eq.~(\ref{eq:jtw2}), requiring the transverse polarization of the $\beta$ particles to be measured, and probes the existence of time reversal violating $S$ and/or $T$ couplings. That is, at linear order in nonstandard couplings we have
\bea
D&\propto& {\rm Im}\left(\frac{C_A+C'_A}{C_V+C'_V}\right)~,\\
R&\propto& {\rm Im}\left(\frac{C_S+C'_S}{C_V+C'_V}\right), ~{\rm Im}\left(\frac{C_T+C'_T}{C_A+C'_A}\right)~.
\eea

Time-reversal violating observables are of special interest for probing the existence of new sources of $\mathcal{CP}$ violation (assuming the $\mathcal{CPT}$ theorem), considered to be intimately related to the observed matter-antimatter asymmetry in the Universe \cite{Sakharov1967}. The $\mathcal{CP}$ violation that is observed in neutral kaons and $B$ mesons, and which is included in the Standard Model via a complex phase in the CKM quark-mixing matrix, leads to effects that are 7 to 10 orders of magnitude smaller than current experimental sensitivities for systems consisting of $u$ and $d$ quarks \cite{Herczeg1997}. This provides a large window to search for other sources of $\mathcal{CP}$ violation, assuming that SM backgrounds, such as e.g.\ final-state interactions (FSI), are well under control.

\subsubsection{Resurrecting the pseudoscalar coupling}
\label{sec:pseudoscalar}
The differential distributions in Eqs.~\eqref{eq:jtw1} and~\eqref{eq:jtw2} do not include the effects from pseudoscalar interactions, parametrized by $C_P^{(\prime)}$. The reason given in most works, including the seminal paper by Jackson {\it et al.}~\cite{Jackson:1957zz}, is the recoil suppression of these contributions. However, it was noted \cite{Gonzalez-Alonso:2013ura} that this effect is compensated to a large extent by the large value of the pseudoscalar charge, $g_P$, {\it cf.}~\sref{hadronic-charges}. This makes $\beta$ decays almost as sensitive to pseudoscalar interactions as they are to scalar and tensor ones, at least {\it a priori}.

However, the sensitivity of the process $\pi\to e\nu$ to such pseudoscalar interaction is orders of magnitudes larger than $\beta$ decays (\sref{leptonicpiondecay}). For this reason, and taking into account the large number of NP couplings present in the standard approach, the pseudoscalar coupling is not included in the numerical analysis in~\sref{fit}.

We discuss here the effect on the neutron lifetime. The linear contribution to the $\beta$ energy spectrum in neutron decay is~\cite{Gonzalez-Alonso:2013ura}
\bea
\label{eq:Pshape}
W(E_e) dE_e =  \frac{F(\pm Z, E_e)}{2\pi^3} p_e E_e (E_0 - E_e)^2 dE_e \,\xi \left( 1 + b_n \frac{m_e}{E_e} + r_P\frac{E_0-E_e}{M}\frac{m_e}{E_e} \right)~,
\eea
where, using both the hadronic- and quark-level couplings,
\bea
\label{eq:Pshape2}
r_P = \frac{\lambda^2}{1+3\,\lambda^2}\frac{C_P+C'_P}{C_A+C'_A} ~=~ - \frac{\lambda}{1+3\,\lambda^2} \,g_P\,\eP ~.
\eea
This generates an extra contribution to the neutron lifetime in~\eref{Ft-n}, namely
\bea
\label{eq:Plifetime}
f_n \,\tau_n\, \ln{2}		&=&	\frac{K }{\xi_n}~\frac{1}{1+ b_n\, \langle m_e/E_e\rangle+\Delta_P}~,\\
\label{eq:Plifetime2}
b_n\, \langle \frac{m_e}{E_e}\rangle + \Delta_P	&\approx& {\rm Re} \left[ 0.34\,\frac{C_S+C'_S}{C_V+C'_V}\langle \frac{m_e}{E_e} \rangle + 1.66\, \frac{C_T+C'_T}{C_A+C'_A}\langle \frac{m_e}{E_e} \rangle + 10^{-4}\, \frac{C_P+C'_P}{C_A+C'_A} \left( -1.5 + 3.8\langle \frac{m_e}{E_e}\,\rangle \right) \right] \\
\label{eq:Plifetime3}
					&\approx& {\rm Re}\left[0.35\,\eS \langle m_e/E_e \rangle- 5.15\,\eT\langle m_e/E_e \rangle+ \, \eP \left( 0.04 -0.10\langle m_e/E_e\,\rangle \right)  \right] ~.
\eea
The small magnitude of the factor multiplying the $\eP$ coefficient is in part due to the recoil suppression, which is not fully compensated by the large $g_P$ value, and in part simply due to a numerical accident.
In the standard approach one neglects the pseudoscalar contribution and sets a bound on the tensor one from the comparison of the neutron lifetime and the ${\cal F}t_{0^+\to 0^+}$ values, which also probe the scalar coupling. It is interesting to proceed the other way around, assuming $\eT=0$ and extracting a bound on the pseudoscalar interaction. From the expression above it is clear that one can obtain such bound by simply rescaling the bound on $\eT$ in the analysis by the factor
\bea
\frac{0.04 -0.10\langle m_e/E_e \rangle}{-5.15\,\langle m_e/E_e \rangle}\approx 0.008~.
\eea
We take advantage of this to set a bound on the pseudoscalar interaction in \sref{fit}.

\subsubsection{Sub-leading corrections}
\label{sec:corrections}
So far, the discussion was based on the leading-order differential distributions in Eqs.~\eqref{eq:jtw1} and~\eqref{eq:jtw2}. Taking into account that the experimental precision is currently at the per-mil level (or even below) for many experiments, these results have to be corrected to include also subleading SM effects. The calculation of these corrections represents a non-trivial task and it has been the subject of many efforts in the past. Excellent reviews can be found in e.g. Refs.~\cite{Holstein:1974zf,Wilkinson:1982hu,Towner:2007np,Sirlin:2012mh,Hayen:2017pwg}. Here we indicate how these corrections fit in the EFT framework, how the observables are modified, and the connection with non-standard effects, which are the main focus of this review.

In addition to the Coulomb corrections contained in the Fermi function, it is useful to separate radiative corrections in long- (``outer") and short-distance (``inner") radiative corrections. The former are electromagnetic corrections to the observables due to the fact that the photon is an active degree of freedom at low energies. They are process- and observable-dependent, and can be further decomposed in (i) a dominant piece that depends trivially on the nucleus (i.e., only on $Z$ and $E_0$), $\delta'_R$, which can be calculated in a model-independent way in QED; and (ii) a small nuclear-structure-dependent piece, $\delta_{NS}$. The literature on their calculation is vast, from the seminal calculation of the corrections to the total rate and the $\beta$ spectrum by Sirlin in 1967~\cite{Sirlin:1967zza}, until more recent developments, such as the corrections to the angular distribution with polarized neutron and electron~\cite{Ivanov:2017mnz}.

On the other hand, short-distance radiative corrections are transition-independent and can be fully encoded in the low-energy coupling constants $C_i,C'_i$. Explicitly, the (axial-)vector matching equations are modified as follows:
\bea
\label{eq:CmatchingAt1loopV}
C^{(\prime)}_V &=& \frac{G_F}{\sqrt{2}} \,  V_{ud} \,g_V \left( 1+ \Delta_R^V \right)^{1/2}  \left(1 + \eL + \eR \pm \teL \pm \teR \right) \left(1-\frac{\delta G_F}{G_F}\right) ~, \\
\label{eq:CmatchingAt1loopA}
C^{(\prime)}_A &=& - \frac{G_F}{\sqrt{2}} \,  V_{ud} \,g_A \left( 1+ \Delta_R^A \right)^{1/2} \left(1 + \eL - \eR \mp \teL  \pm  \teR\right) \left(1-\frac{\delta G_F}{G_F}\right)~,
\eea
where the upper (lower) sign corresponds to the unprimed (primed) coefficients. Factoring out the SM radiative corrections neglects the fact that the four coefficients $\eps_{L,R},\teps_{L,R}$ do not have the same QED running (see~\sref{RGE} for further details). 
$\Delta_R^V$ is particularly important since it is necessary to extract $V_{ud}$. In fact, despite the precise calculation of Ref.~\cite{Marciano:2005ec}, $\Delta_R^V=2.361(38)\%$, it produces currently the largest uncertainty in the extraction of $V_{ud}$~\cite{Hardy:2014qxa}. It is important to note that inner corrections cancel in the SM contribution to correlation coefficients, which therefore only involve the model-independent long-distance electromagnetic corrections, which also cancel to a great extent.

Nuclear corrections, induced by the strong interaction were discussed in~\sref{nuclearmatrixelements}. The isospin-symmetry breaking correction to the Fermi matrix element, $\delta_C$, is particularly important, as it affects the overall strength of the transition and it is thus needed to extract $V_{ud}$.

{\bf Corrected $ft$ values.} For a high-precision determination of $V_{ud}$, nuclear structure and radiative corrections have to be included into~\eref{ft}, leading to a corrected $ft$ value, usually written as $\mathcal{F} t$ and defined as
\bea
\label{eq:Ft}
\mathcal{F} t_i			\equiv ft_i\, (1 + \delta^\prime_R)\, (1 + \delta_{NS} - \delta_C)~,
\eea
where $\delta^\prime_R$ and $\delta_{NS}$ are the integrated radiative corrections discussed above, $\delta_C$ is the isospin-symmetry breaking correction, also introduced above, and the rest of the corrections are included in the statistical rate function $f$. A detailed discussion on the statistical rate function can be found in Appendix A of Ref.~\cite{Hardy2005} (see also Refs.~\cite{Severijns:2008ep,Hayen:2017pwg}).
These quantities are thus transition dependent, but we omitted the subscript $i$ to follow the standard notation. Likewise we also omitted a $V$ superscript in $f$, $\delta_{NS}$ and $\delta_C$, as these quantities are those associated with the vector-mediated transition. The expression of $\xi$ in terms of the underlying fundamental parameters, ~\eref{xi-linear}, has to be modified as follows~\cite{Severijns:2008ep}
\bea
%\label{eq:xi}
|\tilde{\rho}|^2	&\to&	\frac{f_A}{f_V}\frac{1+\delta_{NS}^A}{1+\delta_{NS}^V}|\tilde{\rho}|^2 ~,%\approx \frac{f_A}{f_V}|\tilde{\rho}|^2 ~,
\eea
where $f_V$ and $f_A$ are the statistical rate functions for the vector and axial-vector parts respectively. Finally, the expression of $\tilde{\rho}$ itself in~\eref{mixingratio} is also modified to include the following corrections~\cite{Behrens1982,Severijns:2008ep,NaviliatCuncic:2008xt}:
\bea
%\label{eq:xi}
\tilde{\rho}\equiv \frac{M_{GT}}{M_F}\frac{C_A+C'_A}{C_V+C'_V}
= - \frac{M^0_{GT}}{M^0_F}\frac{g_A}{g_V} \left[ \frac{(1- \delta_C^A) (1 + \Delta_R^A)}{(1- \delta_C^V)(1 + \Delta_R^V)}\right]^{1/2}  \frac{1 + \eL - \eR}{1 + \eL + \eR} ~,
\eea
which do not have any practical consequence, as these corrections are much smaller than the errors affecting the theoretical calculations of axial nuclear and hadron charges. For instance, the  difference between $\Delta_R^A$ and $\Delta_R^V$ is of order 0.1\%~\cite{Fukugita:2004pq}, which is an order of magnitude below the precision of current lattice calculations of $g_A$. Thus, and for sake of simplicity, we omit these corrections hereafter.
The precise extraction of $V_{ud}$ has motivated the calculation of the remaining corrections, $\delta^\prime_R$, $\delta_{NS}$, and $\delta_C$, as well as the statistical rate functions, $f_{V,A}$, for the relevant transitions~\cite{Towner:2007np, Severijns:2008ep, Towner2010, Towner2010b, Hardy:2014qxa, Towner2015a}.

In summary, the leading-order expressions for the $ft$ values and neutron lifetime given in~\sref{totalrates} are replaced by:
\bea
\label{eq:FtpureF}
\mathcal{F} t_i^{0^+ \rightarrow 0^+}		&=& \frac{K}{\xi^{0^+ \rightarrow 0^+}}\frac{1}{1+ b_F \, \langle m_e/E_e\rangle} \approx \frac{K}{2\, G_F^2 |\tilde{V}_{ud}|^2\,(1+\Delta_R^V)}\frac{1}{1+ b_F\, \langle m_e/E_e\rangle}~,
								\\
\label{eq:Ftmirror}
{\cal F}t_i^{\rm{mirror}}				&=& \frac{K}{\xi_i}\frac{1}{1+ b_i \, \langle m_e/E_e\rangle}
								\approx \frac{K}{G_F^2 |\tilde{V}_{ud}|^2 \,(1+\Delta_R^V)\left[ 1+ (f_A/f_V)|\tilde{\rho}_i|^2 \right]}\frac{1}{1+ b_i \, \langle m_e/E_e\rangle}
								~,\\
\label{eq:nlifetime}
f_n\, \tau_n\, \ln{2} (1+\delta^\prime_R)	&=& \frac{K }{\xi_n}\frac{1}{1+ b_n \, \langle m_e/E_e\rangle} ~
								\approx \frac{K}{G_F^2 |\tilde{V}_{ud}|^2 \,(1+\Delta_R^V)\left( 1+  3|\tilde{\lambda}|^2 \right)}\frac{1}{1+ b_n \, \langle m_e/E_e\rangle} ~.
\eea
In mirror decays we neglected the correction $(1+\delta_{NS}^A)/(1+\delta_{NS}^V)$, which multiplies the mixing ratio, as done in current analyses~\cite{Severijns:2008ep,NaviliatCuncic:2008xt}. In neutron decay,~\eref{nlifetime}, we neglected the $f_A/f_V$ ratio, which represents a negligible correction~\cite{Wilkinson:1982hu}. Numerically the SM expression for the neutron lifetime can be written as
\bea
\tau_n 	=	\frac{4908.6\pm 1.8}{|V_{ud}|^2 \left( 1+  3\lambda^2 \right)} ~,
\label{eq:Ft-n-num}
\eea
where we used $f_n = 1.6887(1)$ for the neutron statistical rate function~\cite{Towner2010b,Czarnecki:2018okw}, and $\delta^{\prime(n)}_R = 1.4902(2)\%$ for the long-distance electromagnetic corrections~\cite{Towner2010b}. A precise SM relation can be obtained combining~\eref{FtpureF} and~\eref{nlifetime}, namely~\cite{Czarnecki:2018okw}
\bea
\tau_n\,\left( 1+  3\lambda^2 \right)	=	\frac{2\,\mathcal{F} t^{0^+ \rightarrow 0^+}}{f_n\,\log{2}\,(1+\delta^{\prime(n)}_R)} ~,
\label{eq:Ft-n-num2}
\eea
where the not-so-well-known correction $\Delta_R^V$ does not enter.

{\bf SM corrections to the $\beta$ spectrum shape.} As shown in~\eref{Fierz}, measurements of the $\beta$ spectrum shape give us another handle to search for exotic scalar and tensor currents encoded in the Fierz term. Since various subleading corrections modify also the $\beta$ spectrum shape, with some of them generating terms of the form $1/E_e$, it is crucial to have them under control. This has motivated a new compilation for the theoretical description of the $\beta$ spectrum with high accuracy even at very low kinetic energies \cite{Hayen:2017pwg}. The accuracy can reach a few parts in $10^{-4}$ provided nuclear aspects related to higher order corrections are sufficiently well known.
Within the SM, the $\beta$-energy spectrum can be written as~\cite{Holstein:1974zf,Hayen:2017pwg}
\bea
\label{eq:beta-spectrum}
W(E_e) dE_e = p E_e (E_0 - E_e)^2 F(\pm Z, E_e) B(Z, E_e) f_1(E_e) dE_e  ~ ,
\eea
where $f_1(E_e)$ is a spectral function that includes hadronic corrections \cite{Holstein:1974zf}
and $B(Z,E_e)$ accounts for all other corrections not explicitly shown in Eq.~(\ref{eq:beta-spectrum})
\cite{Hayen:2017pwg}.
The spectral function can be written as~\cite{Holstein:1974zf}
\bea
\label{eq:shapefactor}
f_1(E_e) = C_0 + C_1\,E_e + C_{-1}/E_e + C_2\,E_e^2~,
\eea
where the $C_i$ are small calculable coefficients that depend on the nuclear form factors.
A prominent role is played by weak magnetism, which is the dominant induced form factor for most
transitions and which contributes to $C_0$, $C_1$ and $C_{-1}$ in the spectral
function~\cite{Holstein:1974zf,Calaprice1976,Severijns2018}.
The (small) quadratic term in~\eref{shapefactor} arises mainly from the $q^2$ term in
the Gamow-Teller form factor \cite{Holstein:1974zf,Calaprice1976}.
The extraction of weak magnetism from $\beta$-spectrum shape measurements (instead of its inclusion as initial input) and its agreement with the QCD prediction would represent a clear and pedagogical cross-check that all experimental systematics are well under control, as proposed in Ref.~\cite{Huyan:2016zfp}. In other words, the small SM contribution to the Fierz term could be extracted from the data. In neutron decay it has the value $(C_{-1})_n^{SM}\approx -1.35(1)\times 10^{-3}\,m_e$~\cite{Ando:2004rk,Bhattacharya:2011qm}. This includes not only the weak magnetism contribution but also kinematical corrections.

Finally, it is interesting to note that the radiative corrections to the $\beta$ spectrum are qualitatively different from those needed in the total rates. For example, the isospin-symmetry breaking corrections $\delta_C$, which require nuclear-structure calculations, are absorbed in the overall normalization of the spectrum. Thus, determinations of the Fierz term are sensitive to different theory errors in each case.

\subsubsection{CKM unitarity test}
\label{sec:CKMunitarity}

Once all subleading corrections have been taken into account, the measurement of the $ft$ values and the neutron lifetime make possible to determine the CKM element $V_{ud}$. 
If NP effects are present they contaminate this extraction, {\it cf.}~\eref{Vtilde}, which can only be probed through a test of the unitarity of the CKM matrix. Focusing on the first row, where the highest precision is achieved, we can define the following quantity:
\bea
\label{eq:deltaCKMdef}
\Delta_{CKM} \equiv  |\tilde{V}_{ud}|^2+|\tilde{V}_{us}|^2+|\tilde{V}_{ub}|^2 - 1~,
\eea
which is zero in the absence of nonstandard effects. Here the tilde over the CKM elements denotes that they contain NP effects, as in \eref{Vtilde}. This probe, which is also known as a quark-lepton (or Cabibbo) universality test, allows one to set strong bounds on both nonstandard left- and right-handed weak currents, parametrized by $\eL$, $\delta G_F$, and $\eR$, and it is quadratically sensitive also to other interactions, {\it cf.}~\eref{xi}.

Using the value $|\tilde{V}_{ud}| = 0.97417(21)$ from the latest study of pure Fermi superallowed $\beta$ decays~\cite{Hardy:2014qxa}, with the PDG recommended values for the $us$ and $ub$ CKM matrix elements, $|\tilde{V}_{us}| = 0.2248(6)$ and $|\tilde{V}_{ub}| =$~0.00409(39)~\cite{Patrignani:2016xqp}, one finds at 90\% CL
\bea
\label{eq:deltaCKMpdg}
\Delta_{CKM} = -0.44(81)\times 10^{-3}~,
\eea
where the $V_{ub}$ contribution has almost no impact. The error decomposition is given by $\Delta_{CKM}= -0.44(68)_{V_{ud}}(45)_{V_{us}}\times 10^{-3}$. 
These values are however extracted assuming no other NP contributions are present, apart from those absorbed in the CKM matrix elements. For example, the $V_{ud}$ value given above is obtained assuming $b=0$. In a general framework where such effects are present, $\Delta_{CKM} = -0.12(84)\times 10^{-3}~$ \cite{Gonzalez-Alonso:2016etj}, which is slightly less stringent.

The $V_{us}$ matrix element is obtained from kaon decays, which extract either the product of $|V_{us}|$ and the semileptonic-decay form factor at zero momentum transfer, $f_{+}(0)$ (in $K_{\ell3}$ decays), or the ratio of CKM matrix elements $|V_{us}|/|V_{ud}|$ multiplied by the ratio of decay constants $f_K / f_\pi$ (in $K_{\ell2}$ decays). Important progress was made both in experiments (see e.g. Refs.~\cite{Patrignani:2016xqp,Antonelli:2010yf, Moulson:2014cra}), as well as in the determination of the form factors, the latter due to a rapid expansion in large-scale numerical simulations in lattice QCD~\cite{Aoki:2016frl, Bazavov2014a, Bazavov2014b}. As a result the precision of $V_{us}$ has improved by more than a factor of 3 over the last decade. According to Ref.~\cite{Hardy2013}, the uncertainty on $|V_{us}|$ could be improved by a factor 2
if the current uncertainty of 0.3\% on the lattice calculations could be decreased to 0.1\%.

The error of $|V_{ud}|$, which has the largest impact on the uncertainty of the unitarity sum, is dominated by the transition-independent part of the radiative correction, $\Delta_R^V$, followed (with an about twice smaller fractional uncertainty) by the nuclear structure dependent terms, $\delta_{NS}$~-~$\delta_C$. A reduction of the $\Delta_R^V$ uncertainty by a factor 2 would reduce the $|V_{ud}|^2$ uncertainty by 30\% \cite{Hardy:2014qxa}. However, a further improvement in $\Delta_R^V$ seems to be a huge challenge for theory and as far as we know no new calculations of this correction are currently underway. A small improvement in $|V_{ud}|$ can also be obtained from a reduction in the uncertainty associated with the nuclear-structure-dependent corrections, $\delta_{NS} - \delta_C$.
\subsection{Connection with high-energy physics}
\label{sec:HEP}
If the new fields introduced by the theory that supersedes the SM are not only heavier than the hadronic scales but also heavier than the electroweak scale, then an effective Lagrangian can also be used at such energies. If the electroweak symmetry breaking is linearly realized (like in the SM), the resulting theory goes by the name of Standard Model EFT (SMEFT), and its Lagrangian has the following structure~\cite{Buchmuller:1985jz}
\bea
\label{eq:BW}
{\cal L}_{\rm SMEFT} = {\cal L}_{SM} + \sum_i \frac{w_i}{v^2}  {\cal O}_i^{(6)} + \ldots ~,
\eea
where $v=(\sqrt{2}G_F)^{-1/2}\simeq 246$ GeV is the vacuum expectation value of the Higgs field, $\cO_i^{(6)}$ are $SU(3)\times SU(2)_L\times U(1)_Y$-invariant dimension-six operators built with SM fields and generated by the exchange of the new fields that were integrated out, and $w_i$ are the associated Wilson coefficients.
There is also a (lepton-number violating) dimension-5 operator, which generates Majorana neutrino masses, but it has no impact for the observables we discuss in this work. 
In this normalization the dependence on the NP scale $\Lambda$ has been absorbed in the $w_i$ coefficients, namely $w_i \propto v^2 / \Lambda^2$. The dots in \eref{BW} refer to higher-dimensional operators that are suppressed by additional powers of $v/\Lambda$.

There are infinite operator bases $\{\cO_i^{(6)}\}$ one could use but the physics is of course independent of this choice. We work here in the so-called Warsaw basis \cite{Grzadkowski:2010es}\footnote{One difference is that for operators with the $SU(2)_L$ singlet contraction of fermionic currents we omit the superscript~${}^{(1)}$.}, minimally extended to include the operators with right-handed neutrinos that generate the $\teps_i$ coefficients at low-energies~\cite{Cirigliano:2012ab,Liao:2016qyd}. In this basis there are twelve effective operators,
which are listed in \tref{warsaw},
that generate a tree-level contribution to nuclear and neutron $\beta$ decay, either through a modification of the $W$ boson vertex to fermions or introducing a new four-fermion interaction. The operators affecting the extraction of $G_F$ from muon decay are also listed, as they are relevant for the CKM-unitarity test.
\begin{table}[t]
\centering
\caption{SMEFT operators relevant for this work. The fields $\ell,q,e,u,d$ denote the fermionic SM gauge multiplets, $\nu$ is the right-handed neutrino, and $\varphi$ is the $SU(2)$ Higgs doublet. Moreover, $\tilde{\varphi}^a\equiv \epsilon_{ab}(\varphi^b)^*$, with $\epsilon_{12}=+1$. For further details about the conventions see Ref.~\cite{Grzadkowski:2010es}.\vspace{0.3cm}
%$\epsilon_{jk}$ is totally antisymmetric with $\epsilon_{12}=+1$
\label{tab:warsaw}}
\begin{tabular}{|c|c|}
\hline
${\cal O}_{\varphi \ell}^{(3)}$	& $({\varphi^\dag i\,\raisebox{2mm}{${}^\leftrightarrow$} \hspace{-4mm} D_\mu^{\,I}\,\varphi})(\bar \ell \tau^I \gamma^\mu \ell)$\\
${\cal O}_{\varphi q}^{(3)}$	& $({\varphi^\dag i\,\raisebox{2mm}{${}^\leftrightarrow$} \hspace{-4mm} D_\mu^{\,I}\,\varphi})(\bar q \tau^I \gamma^\mu q)$\\
${\cal O}_{\varphi u d}$		& $i(\widetilde{\varphi}^\dag D_\mu \varphi)(\bar u \gamma^\mu d)$\\
${\cal O}_{\ell\ell}$			& $(\bar \ell \gamma_\mu \ell)(\bar \ell \gamma^\mu \ell)$\\
\hline
\end{tabular}
\hspace{0.6cm}
\begin{tabular}{|c|c|}
\hline
${\cal O}_{\ell q}^{(3)}$		& $(\bar \ell \gamma_\mu \tau^I \ell)(\bar q \gamma^\mu \tau^I q)$\\
${\cal O}_{\ell edq}$			& $(\bar \ell^a e)(\bar d q^a)$\\
${\cal O}_{\ell equ}$			& $(\bar \ell^a e) \eps_{ab} (\bar q^b u)$\\
${\cal O}_{\ell equ}^{(3)}$		& $(\bar \ell^a \sigma_{\mu\nu} e) \eps_{ab} (\bar q^b \sigma^{\mu\nu} u)$\\
\hline
\end{tabular}
\hspace{0.6cm}
\begin{tabular}{|c|c|}
\hline
${\cal O}_{e\nu ud}$			& $ (\overline{e} \gamma^\mu \nu) (\overline{u} \gamma_\mu  d)$\\
${\cal O}_{\ell\nu uq}$		& $(\overline{\ell} \nu) (\overline{u} q)$\\
${\cal O}_{\ell\nu qd}$		& $(\bar{\ell}^a \nu)\epsilon_{ab}(\bar{q}^b d)$\\
${\cal O}_{\ell\nu qd}^{(3)}$	& $(\bar{\ell}^a\sigma^{\mu\nu} \nu)\epsilon_{ab}(\bar{q}^b\sigma_{\mu\nu}d)$\\
${\cal O}_{\varphi \nu e}$		& $i(\widetilde{\varphi}^\dag  D_\mu \varphi) (\overline{\nu}\gamma^\mu e)$\\
\hline
\end{tabular}
\end{table}

At order $\Lambda^{-2}$ and at tree level the Wilson coefficients of both theories are related as follows (matching equations)~\cite{Cirigliano:2009wk,Cirigliano:2012ab,Gonzalez-Alonso:2017iyc}\footnote{When comparing SMEFT results from different works it is important to note the conventions adopted, {\it e.g.}, the normalization of the Wilson Coefficients (and the definition of $v$), whether the up- or down-quark Yukawa was taken to be diagonal in the basis where SMEFT coefficients are defined, or the use of $\tilde{\varphi}$ or $\varphi^T$ in the definition of the ${\cal O}_{\varphi u d}$ operator.}
\begin{subequations}
\bea
\eL &=&  \eLv  +   \eLc
~~~~~~~~~~~~~~~~~~~~~~~~~~~~~~~~~~~~~~~~
\teL ~=~  + \frac{1}{2}[w^\dagger_{\varphi \nu e}]_{11}~,
 \\
\eLv &=& [w_{\varphi \ell}^{(3)}]_{11} + \frac{ [V \, w_{\varphi q}^{(3)} ]_{11}}{V_{ud} }~,
 \\
\eLc  &=&  -  \frac{ [V \, w_{\ell q}^{(3)} ]_{1111}}{V_{ud} }~,
\\
\eR  &=&  + \frac{1}{2} \frac{ [w_{\varphi u d}]_{11}}{V_{ud} }
~~~~~~~~~~~~~~~~~~~~~~~~~~~~~~~~~~~~~
\teR ~=~ -  \frac{1}{2} \frac{ \left[w_{e \nu u d}\right]_{1111}}{V_{ud}} ~,
\\
\eSP  &=&  -\frac{1}{2} \frac{ [w_{\ell equ}^\dagger \pm V \, w_{\ell edq}^\dagger ]_{1111}}{V_{ud} }
~~~~~~~~~~~~~~~~~~~
\teSP  ~=~   - \frac{1}{2}\frac{ [ w_{\ell\nu uq} \mp V \, w_{\ell\nu qd}]_{1111}}{V_{ud} }~,
\\
\eT  &=&  -   \frac{1}{2} \frac{ [w_{\ell equ}^{(3) \ \dagger}]_{1111}}{V_{ud} }
~~~~~~~~~~~~~~~~~~~~~~~~~~~~~~~~~~~
\teT  ~=~     \frac{1}{2} \frac{ [V \, w_{\ell\nu qd}^{(3)} ]_{1111}}{V_{ud} }~.
\eea
\label{eq:matchingeqs}%
\end{subequations}
where $V$ denotes the CKM matrix, and both $\eps_i$ and $w_i$ coefficients are evaluated at the electroweak scale. Their renormalization-scale dependence is discussed in \sref{RGE}. Operators containing flavor indices are defined in the flavor basis where the down-quark Yukawa matrices are diagonal, and matricial notation only affect quark indexes. For later convenience, we split the $\epsilon_L$ coefficient into an SMEFT vertex correction, $\eLv$, and an SMEFT contact interaction, $\eLc$. Finally, the linear NP modification to the muon lifetime is given by
\bea
\label{eq:deltaGF}
\frac{\delta G_F}{G_F}
&=& [w_{\varphi \ell}^{(3)}]_{11} + [w_{\varphi \ell}^{(3)}]_{22} - [w_{\ell\ell}]_{1221}~,
\eea
where we neglected a tiny contribution suppressed by $m_e/m_\mu$ associated to the %1221 flavor component of the
operator ${\cal O}_{\ell e} = (\bar \ell \gamma_\mu \ell)(\bar e \gamma^\mu e)$, which is related to the muon decay parameter $\eta$~\cite{Gonzalez-Alonso:2010vnm}. Although the inclusion of the $\eta$ term in the analysis of muon decay data produces an important increase of the Fermi constant error~\cite{Danneberg:2005xv}, its impact is still negligible for $\beta$-decay studies. Finally, we sum over all flavor indices in~\eref{BW} and eliminate redundancies making the necessary coefficients equal, e.g. $[w_{\ell\ell}]_{2112}=[w_{\ell\ell}]_{1221}$.

The use of the SMEFT framework has several advantages:
\bi
\item It allows us to understand the implications of collider searches for low-energy experiments and vice versa. It also connects $\beta$-decay searches with neutral-current low-energy experiments, such as atomic parity violation, due to the $SU(2)$ gauge invariance;
\item It makes the connection with NP models easier, as the NP bounds can be given at the high-energy scale where the new particles are active. There is actually an important effort to automatize the one-loop matching between the SMEFT and NP models~\cite{Henning:2014wua,delAguila:2016zcb,Ellis:2016enq,Henning:2016lyp,Fuentes-Martin:2016uol,Zhang:2016pja};
\item Despite its few assumptions the SMEFT comes with predictions, which are a consequence of the symmetries and its field content. In particular, the traditional SMEFT predicts $\teps_i=0$, because it does not contain right-handed neutrinos. Also, it predicts a lepton-independent right-handed coupling, $\eR$, up to $\cO(v/\Lambda)^4$ corrections \cite{Cirigliano:2009wk}.
\ei
These advantageous features come however at a cost, which is just a loss of generality. In fact the SMEFT entails a few additional assumptions with respect to the low-energy EFT in \eref{leff-lowE-linear}, which might not be correct. These include the absence of non-standard states below the weak scale, the linear realization of the electroweak symmetry breaking or the smallness of higher-dimensional SMEFT operators. Furthermore the analysis of LHC data using the SMEFT assumes that the new particles are sufficiently heavy to be integrated out even at such high scales. For models where that is not the case, a model-dependent analysis becomes necessary to evaluate the complementarity with $\beta$-decay searches.

{\bf The flavor-universal limit.} 
As pointed out in Ref.~\cite{Cirigliano:2009wk}, the theoretical description is significantly simplified if the flavor symmetry $U(3)^5$ of the SM gauge Lagrangian is approximately respected in the SMEFT. This symmetry corresponds to the freedom to make $U(3)$ rotations in family space for each of the fermionic gauge multiplets ($\ell,e,\nu,q,u,d$). 
In this limit %, some times called flavor-blind limit,
only flavor-universal operators with left-handed fermions contribute to $\beta$ and muon decays, which implies that only $\eL$ and $\delta G_F$ are present in their low-energy Lagrangians. Their effect can be entirely absorbed inside $V_{ud}$ and $G_F$, so that the SM analysis of the data holds. The only NP probe left is then the CKM unitarity test of \eref{deltaCKMdef}, which probes the following combination of operators
\bea
\Delta_{\rm CKM} = 2~\left( - w_{\varphi \ell}^{(3)} + w_{\varphi q}^{(3)}- w_{\ell q}^{(3)} + 2\,w_{\ell\ell}^{(3)}        \right)~.
\label{eq:DeltaCKMsmeft}
\eea
The first three coefficients are simply defined through $[w_I]_{ij(kl)}=w_I \delta_{ij}(\delta_{kl})$, whereas for the purely leptonic operator there are two $U(3)^5$-invariant contractions: $[w_{\ell\ell}]_{ijkl}=w_{\ell\ell}^{(1)} \delta_{ij}\delta_{kl} + w_{\ell\ell}^{(3)} \left( -\delta_{ij}\delta_{kl} + 2\,\delta_{il}\delta_{jk}\right)$.
\subsubsection{Renormalization-scale dependence}
\label{sec:RGE}
The inner radiative correction, $\Delta_R^V$, contains the (log-enhanced) QED running, from the electroweak to the hadronic scale, of the Wilson coefficient associated to the $(V-A)\times(V-A)$ operator in the quark-level Lagrangian in~\eref{leff-lowE}. Once NP terms are introduced, there are several major differences with respect to the situation in the SM. First, in contrast to the vector operator, (pseudo)scalar and tensor operators do run with QCD as their anomalous dimension is not zero.
Secondly, the running of the coefficients has to be performed to scales much higher than the electroweak scale, both to match to specific models and to compare with searches carried out at the LHC. And last, (pseudo)scalar and tensor operators mix in the QED running up to the electroweak scale, as well as in the electroweak running to higher scales~\cite{Voloshin:1992sn,Campbell:2003ir,Alonso:2013hga,Aebischer:2017gaw,Gonzalez-Alonso:2017iyc}.\footnote{In Ref.~\cite{Gonzalez-Alonso:2017iyc} some disagreements were noted in the diagonal elements of the electroweak anomalous dimension matrix between Ref.~\cite{Campbell:2003ir} and Ref.~\cite{Alonso:2013hga}, confirming the results of the latter. This does not have any phenomenological relevance as the running of the diagonal elements is fully dominated by QCD.}

Working at three loops in QCD~\cite{Gracey:2000am} and one loop in QED~\cite{Aebischer:2017gaw,Gonzalez-Alonso:2017iyc}, and taking into account finite threshold corrections~\cite{Chetyrkin:1997un,Misiak:2010sk}, the following running below the weak scale is found \cite{Gonzalez-Alonso:2017iyc}
\bea
\left(
\begin{array}{c}
\eL\\
\eR\\
\eS\\
\eP \\
\eT \\
\end{array}
\right)_{\!\!\!\mbox{($\mu=2$~GeV)}}
\!\!\!\!\!=
\left(
\begin{array}{ccccc}
 1 & 0 &0&0&0\\
 0 & 1.0046&0&0&0\\
 0 & 0 & 1.72 & 2.46\times 10^{-6} & -0.0242 \\
 0 & 0 & 2.46\times 10^{-6} & 1.72 & -0.0242 \\
 0 & 0 & -2.17 \times 10^{-4} & -2.17 \times 10^{-4}& 0.825 \\
 \end{array}
\right)
\!\!\left(
\begin{array}{c}
\eL\\
\eR\\
\eS\\
\eP \\
\eT \\
\end{array}
\right)_{\!\!\!\mbox{($\mu=m_Z$)}}~,
\label{eq:RGEepsilon}
\eea
where the $\epsilon_L$ coefficient is not scale-dependent because it is defined with respect to the SM contribution, {\it cf.}~\eref{leff-lowE-linear}. The low-energy scale is set to $\mu=2$ GeV which is usually chosen to extract the associated lattice form factors.
The running above the weak scale of the SMEFT coefficients associated to the (pseudo)scalar and tensor operators is~\cite{Gonzalez-Alonso:2017iyc}
\bea
\left(
\begin{array}{c}
w_{ledq} \\
w_{\ell equ}\\
w^{(3)}_{\ell equ} \\
\end{array}
\right)_{\!\!\!\mbox{($\mu=m_Z$)}}
=
\left(
\begin{array}{ccc}
1.19 & 0. & 0. \\
 0. & 1.20 & -0.185 \\
 0. & -0.00381 & 0.959 \\
\end{array}
\right)
\left(
\begin{array}{c}
w_{ledq}\\
w_{\ell equ} \\
w^{(3)}_{\ell equ} \\
\end{array}
\right)_{\!\!\!\mbox{($\mu=1$ TeV)}}~,
\label{eq:RGEsmeft}
\eea
where $\mu=1$ TeV is used as a representative NP and LHC scale. An accidentally large $(2,3)$ element is found, which makes low-energy probes of tree-level (pseudo)scalar interactions also sensitive to tensor ones.

The importance of taking into account the QCD running is clear from the diagonal elements above, which include a sizable two-loop correction. The size of the three-loop piece is much smaller, indicating a small error from the neglected terms. As expected from its one-loop QED/electroweak nature, the operator mixing represents a much smaller effect, with the above-mentioned exception. It cannot be neglected however given the very precise bounds on $\epsilon_i$ currently obtained. This effect is particularly relevant for $\eP$, which is strongly constrained by leptonic pion decays, which are thus also sensitive to scalar/tensor terms through radiative corrections (\sref{leptonicpiondecay}).

Other finite corrections are expected in the matching of the underlying NP model to the SMEFT, in the matching between the various EFTs below that scale, and in the long-distance radiative corrections. These additional pieces are not logarithmically enhanced and thus should represent a subleading correction, especially for very high NP scales.

\subsubsection{Model-dependent studies}
\label{sec:specificmodels}
Specific NP models generate only a few of the Wilson coefficients, which moreover are typically correlated by the underlying dynamics. In this section we briefly review some examples that are relevant for $\beta$-decay studies and that were considered in the past. We refer to Refs.~\cite{Herczeg:2001vk,Marciano:2011zz} for further details and examples.

{\bf Low-energy (V-A)$\times$(V-A) interactions.} Such contributions are easily generated in many well-known NP models. This is ultimately due to the fact that they are not in conflict with the approximate $U(3)^5$ flavor symmetry. In the low-energy quark-level EFT of~\eref{leff-lowE} they are parametrized by $\eL$, which generates $C_{V,A}^{(\prime)}$ at the nucleon level. In most models one typically generates a similar contribution to muon decay, parametrized by $\delta G_F/G_F$. It is convenient to work with the difference of both contributions because (i) it is the quantity that is phenomenologically accessible:
\bea
\Delta_{CKM} \supset 2\left( \mbox{Re}\, \eL - \frac{\delta G_F}{G_F} \right)~,
\eea
and (ii) several NP contributions cancel in the difference, such as the universal corrections to the $W$ propagator. Thus, this is the quantity calculated in most works. Some of the models that have been constrained through this test are:
\bi
\item R-parity conserving Minimal Supersymmetric SM (MSSM), where the effect is generated by one-loop (box and vertex-corrections) diagrams with the superpartners in the internal lines~\cite{Barbieri:1985ff,Barbieri:1985kv,Hagiwara:1995fx,Kurylov:2001zx,Bauman:2012fx}. The main effect comes from a possible mass difference between squark and sleptons;
\item Models with $Z'$ bosons, contributing at one loop to muon and $\beta$ decay, and giving~\cite{Marciano:1987ja}:
	\bea
	\Delta_{CKM} %= 0.01\,\lambda\,\frac{\log{\frac{m_{Z'}^2}{m_W^2}}}{\frac{m_{Z'}^2}{m_W^2}-1}
	\approx -0.01\,\lambda\,\frac{m_W^2}{m_{Z'}^2}\log{\frac{m_W^2}{m_{Z'}^2}}~,
	\eea
	where $\lambda$ is a model-dependent parameter. As an interesting and recent example, this parameter happens to be very large ($\lambda\approx16$) in a model that has been proposed to accommodate some flavor anomalies~\cite{Gauld:2013qja}, leading to a $2.7$ TeV limit on the $Z'$ mass;%, which is a crucial constraint for the model;
\item In the R-parity violating MSSM, tree-level contributions to muon and $\beta$ decay are possible, namely~\cite{Barger:1989rk,RamseyMusolf:2000qn}:
\bea
\Delta_{CKM} \approx  |\lambda'_{11k}|^2 \frac{v^2}{2\,M_{\tilde{d}_R^k}^2} - |\lambda_{12k}|^2 \frac{v^2}{2\,M_{\tilde{e}_R^k}^2}~,
\eea
where $\lambda^{(')}$ and $M_{\tilde{f}}$ denote the R-parity violating couplings and sfermion masses, respectively. It has been shown that the $\Delta_{CKM}$ constraint allows one to probe new regions of the parameter space of the model~\cite{Barger:1989rk,RamseyMusolf:2000qn} (see Ref.~\cite{Barbier:2004ez} for a review). This framework can be seen as a particular leptoquark realization. The implications of the CKM-unitarity constraint in general leptoquark frameworks were studied in Refs.~\cite{Barger:2000gv,Herczeg:2001vk}.
\ei

{\bf Scalar and tensor interactions.} Due to their chirality-flipping nature, in many NP models these interactions follow a Yukawa-like structure, which implies extremely suppressed BSM effects in processes involving light fermions, such as $\beta$ or muon decays. This is for example realized in models with an extended Higgs sector~\cite{Haber:1978jt,Herczeg:2001vk}. Moreover, even in NP models generating sizable scalar and tensor interactions with light fermions, they typically generate at the same time pseudoscalar interactions of similar magnitude. This is the case in the MSSM with left-right mixing among first and second generation sleptons and squarks~\cite{Profumo:2006yu}, or the R-parity violating MSSM~\cite{Herczeg:2001vk,Yamanaka:2009hi}, where the effect is generated at loop and tree level respectively. For such models the strong bounds obtained from leptonic pion decays on pseudoscalar interactions (\sref{leptonicpiondecay}) imply similarly strong bounds on scalar and tensor ones. This would make them unobservable in $\beta$ decay except maybe in small regions of the parameter space, where (apparently) unnatural cancellations occur.

{\bf Right-handed currents.}
A well-known class of NP theories that generates right-handed currents are the so-called left-right (LR) symmetric models. Forty years after their introduction, they remain a viable and actively studied SM extension. In the simplest framework the gauge group is $SU(2)_L\times SU(2)_R\times U(1)$ and an extended scalar sector is also introduced to spontaneously break the gauge symmetry down to $U(1)_{em}$. In the low-energy quark-level EFT in~\eref{leff-lowE} right-handed currents are parametrized by $\eR$, which contributes to $C_{V,A}^{(\prime)}$ at the nucleon level.
In the simplest framework (``manifest LR symmetry") the couplings of $W_{L,R}$ to fermions are equal, non-standard ${\mathcal CP}$-violating effects are absent, and the only new parameters relevant for $\beta$-decay studies are the $W_2$ mass and the mixing angle $\zeta$, defined by
	\bea
	W_L &=& c_\zeta W_1 + s_\zeta W_2~,\\
	W_R &=& - s_\zeta W_1 + c_\zeta W_2~,
	\eea
where $W_{1,2}$ are the mass eigenstates. This framework generates a right-handed current at tree-level through this mixing, namely $\eR\approx -\zeta$~\cite{Beg:1977ti,Holstein:1977qn,Herczeg:2001vk}. As a result, they contribute to the phenomenologically extracted $V_{ud}$ value~\cite{Towner:1995za}, {\it cf.}~\eref{Vtilde}, which can be probed through the CKM unitarity test, namely:
\bea
\label{eq:CKMinLR}
\tilde{V}_{ud} %\approx V_{ud} \left[ 1 + \mbox{Re} \left(  \eL+\eR  \right) \right] 
\approx V_{ud} \left( 1 -\zeta \right)~\to~\Delta_{CKM}=-2\,\zeta~.
\eea
If light enough right-handed neutrinos are present, additional interactions are also produced: $\teL\approx -\zeta,~\teR\approx m^2_{W_1}/m^2_{W_2}$~\cite{Beg:1977ti,Holstein:1977qn,Herczeg:2001vk}. They provide access to the NP scale, although with small sensitivity due to the quadratic dependence.
The simple setup above was studied in the original papers~\cite{Beg:1977ti,Holstein:1977qn}, whereas the implications of $\beta$-decay data for more complex realizations were analyzed afterwards~\cite{Langacker:1989xa,Herczeg:2001vk}.

\vspace{0.2cm}
The above discussion makes clear the qualitative difference between the CKM-unitarity test and the rest of $\beta$-decay NP searches. The former constrains vector interactions that are known to be generated at an observable level in well-known scenarios. The latter access more exotic NP interactions, which are allowed in a general EFT setup, but which are difficult to build in concrete NP models.

The various models discussed in this section are approximately described by the SMEFT approach if the new particles are sufficiently heavy. In that case, the interplay of the bounds obtained from $\beta$ decay with other electroweak precision tests or collider searches can be derived from the SMEFT analysis in~\sref{alternativeprobes}. Otherwise, model-dependent studies are needed.

%%%%%%%%%%%%%%%%%%%%%%%%%%%%%%%%%%%%%%%%%%%%%%%%%%%%%%%%%%%%%%%%%%%%%%%%%%%%%%%%
\section{Experimental Activity}
\label{sec:exp}

This section reviews recently published as well as ongoing and planned measurements and is organized
according to the different observables, i.e.\ $ft$ values, $\beta$-spectrum shape, and correlation
coefficients. For each observable the relevant physics information that is obtained or can be addressed
is pointed out, and prospects for further improvements in precision are discussed as well.
An overview of the experiments that are discussed here is given in Table~\ref{tab:expSummary}
at the end of this section.

\subsection{$\mathcal{F}t$ values \label{sec:Ft-values}}

The statistical rate function, $f$, which enters the $ft$ value of a $\beta$ transition %, ~\eref{ft}),
depends on its total transition energy $Q_{EC}$. On the other hand, the partial half-life of the decaying nucleus is obtained from the half-life, $t_{1/2}$, of the parent nucleus and the branching ratio, $BR$, of the observed transition, corrected for the electron capture probability, $\varepsilon / \beta^+$:
\begin{equation}
t = \frac{t_{1/2}} {BR} \left(1 + \frac{\varepsilon}{\beta^+} \right)  ~  .
\label{eq:t}
\end{equation}
The $\varepsilon / \beta^+$ values are typically of the order 0.1-0.2\% and can be calculated to a few percent \cite{Bambynek1977, Hardy2005} so that they do not contribute perceptibly to the overall uncertainties.
The determination of the $ft$ value thus requires the measurement of three observables: the half-life, the branching ratio, and the $Q_{EC}$ value.

The corrected $ft$ value, $\mathcal{F} t$, is obtained by including also the various theoretical corrections that induce a small transition dependence of the $ft$ values in the SM. The corrections not included in $f$ are shown in~\eref{Ft}. As discussed in \sref{totalrates}, the ${\cal F}t$ values of the superallowed $0^+ \rightarrow 0^+$ pure Fermi transitions, of neutron decay, and recently also of the $T = 1/2$ mirror $\beta$ transitions, have been used to extract $V_{ud}$ and/or to set bounds on the corresponding Fierz terms.

\subsubsection{Superallowed Fermi $\beta$ transitions \label{sec:zero-plus}}

As shown in~\eref{FtpureF}, the $\mathcal{F}$t value for superallowed $0^+ \rightarrow 0^+$ pure Fermi transitions is given in the SM by
\be
\label{eq:Ft-Fermi}
2\, \mathcal{F} t^{0^+ \rightarrow 0^+}
= \frac{K}{ G^2_F\, V^2_{ud}\, (1 + \Delta_R^V)}~.
\ee

Extensive reviews of the experimental results can be found in Refs.~\cite{Towner2010, Hardy2014}.
The most recent evaluation of half-life, decay-energy, and branching-ratio measurements related to
14   superallowed $0^+ \rightarrow 0^+$ decays found compatible values, with
$ \overline{\mathcal{F} t}^{0^+ \rightarrow 0^+} =$ 3072.27(72) s \cite{Hardy:2014qxa}, thereby verifying the CVC prediction at the level
of $1.2 \times 10^{-4}$. The 14 individual values are included in the fits discussed in~\sref{fit} and are listed in~\tref{ftFermi}.

Despite a very modest improvement in the statistical uncertainty, the present $\overline{\mathcal{F} t}^{0^+ \rightarrow 0^+}$ value \cite{Hardy:2014qxa} is almost a factor of two more precise than about a decade ago \cite{Hardy2005}.
This is caused by a drastic decrease of the systematic error associated with the nuclear corrections $\delta_C$, which passed from dominant to negligible. The origin is the enlargement of the dataset and the improvement of the $\mathcal{F} t$ value in many transitions that do not provide the most precise value, but that make possible to test the various nuclear-structure approaches. Such experimental progress was possible mainly due to the new and very precise values for
the $Q_{EC}$ transition energies extracted from Penning trap-based mass measurements \cite{Blaum2006}, and new
and very precise half-life measurements such as from Refs.~\cite{Grinyer2008, Blank2010, Finlay2011, Park2012, Ball2014}.
Five Penning trap mass measurement facilities contribute to the study of superallowed Fermi and mixed Fermi/Gamow-Teller transitions:
CPT (Argonne National Laboratory) \cite{Savard1997, Clark2003}, ISOLTRAP (ISOLDE-CERN) \cite{Mukherjee2008},
JYFLTRAP (University of Jyv\"askyl\"a) \cite{Eronen2012}, LEBIT (Michigan State University)
\cite{Ringle2009} and TITAN (TRIUMF-ISAC) \cite{Brodeur2012}. An extreme precision of 7~$\times$~10$^{-6}$ has been
achieved with the JYFLTRAP by alternating between the mother and daughter nuclei \cite{Eronen2009}.
Nevertheless, many previously performed reaction measurements like \ $(p,n)$, $(p,\gamma)$, $(n,\gamma)$,
and $(^{3}{\rm He},t)$, still make significant contributions to the world averages of the $Q_{EC}$
values \cite{Hardy2014}. Precise half-life measurements are challenged by impurity activities,
and rate-dependent dead-time, gain variation, and pile-up effects and require either $\beta$ or $\gamma$ detection
methods. New fast real-time data digital acquisition techniques, with an effective dead-time smaller than 1~$\mu$s, such as
used in Refs.~\cite{Flechard2010,Hughes2018}, will certainly open up new opportunities for high-precision half-life
measurements.

The above-given average $\mathcal{F} t$ value, when combined with the muon lifetime \cite{Tishchenko2013}, yields \cite{Hardy:2014qxa}
\be
\label{eq:Vud-Fermi}
|V_{ud}| = 0.97417(21) \ {\rm (pure\ Fermi\ transitions)} ~.
\ee
\noindent Together with precise determinations of $|V_{us}|$ and $|V_{ub}|$, provides a $10^{-4}$-level test of unitarity from the first row of the CKM matrix, see~\eref{deltaCKMpdg}. As discussed in~\sref{HEP} this translates in a powerful NP bound in many BSM scenarios. It is worth reminding that if the Fierz term is not assumed to be zero (as usually done), then the precision of the CKM unitarity test is relaxed (\sref{CKMunitarity}).

Measurements of $Q_{EC}$, $t_{1/2}$, and $BR$ with reduced uncertainties for the already well-measured superallowed Fermi transitions are regularly being reported while, in addition, more transitions are being characterized as well thereby extending the experimental dataset. As discussed in~\sref{CKMunitarity}, considering anticipated improvements to both $|V_{ud}|$ and $|V_{us}|$, the uncertainty on the unitarity sum will likely be reduced to $\pm$0.00040 in the foreseeable future~\cite{Hardy2013}.

The $\mathcal{F} t^{0^+ \rightarrow 0^+}$ values also allow one to search for exotic scalar currents, which would induce a dependence with the average inverse decay energy of each $\beta$ transition, $\langle m_e/E_e \rangle$, %averaged over the statistical weight $p_e E_e (E_0 - E_e)^2$
as discussed in \sref{totalrates}, $cf.$ \eref{FtpureF}. 
The sensitivity is largest for $^{10}$C and $^{14}$O since $\langle m_e/E_e \rangle$ decreases with increasing $Z$ and mass number $A$. Especially for $^{10}$C the $\mathcal{F} t$ value, and so the sensitivity to scalar currents, can still be improved substantially as its uncertainty is dominated by the precision on the branching ratio.
Any measurement of the branching ratio for this transition has to take into account the pile-up of the 511 keV
annihilation quanta with respect to the 1021.7 keV $\gamma$ ray from $^{10}$C decay. A new measurement \cite{Blank2016}
was performed at ISOLDE using a precisely calibrated Ge detector \cite{Blank2015}. The pile-up was determined
using the $^{19}$Ne decay, which has a half-life and $\beta$-decay $Q_{EC}$ value close to $^{10}$C. The data are still
being analyzed and correspond to an absolute statistical precision of 0.004\% on a branching ratio of about 1.5\% \cite{Blank2018}.
Even though being about twice less precise this will provide an important cross-check on the currently adopted value.

\subsubsection{Superallowed mirror $\beta$ transitions \label{sec:mirror}}

It was recently pointed out that the superallowed $\beta$ transitions between the isobaric analog states in $T = 1/2$ isospin doublets can also contribute to the determination of $|V_{ud}|$ \cite{NaviliatCuncic:2008xt}. From~\eref{Ftmirror} one can define, in the SM,
\begin{equation}
{\cal F}t_0^{\rm mirror} \equiv ft (1+\delta^\prime_R)(1+\delta_{\rm NS}-\delta_C)
\left[ 1 + (f_A/f_V)\rho^2 \right] =
\frac{K}{G^2_F V^2_{ud} (1+\Delta_R^V)} ~ ,
\label{eq:Ft-mirror}
\end{equation}
where, similar to~\eref{Ft-Fermi}, the left-hand side includes all transition-dependent terms. As the nuclear structure of mirror nuclei differs from the pure Fermi emitters, a precise test of CVC in the superallowed mirror decays would provide a strong consistency test of
the nuclear-structure-dependent corrections $\delta_{\rm NS}$ and $\delta_C$, and would contribute to an improved determination of $|V_{ud}|$.
However, an additional measurement is now required in order to determine the mixing ratio $\rho$. This could be a measurement of the $\beta$-$\nu$ correlation, $a$, or the $\beta$-asymmetry parameter, $A$. The precision will ultimately be limited also here by the uncertainty of the nucleus-independent radiative correction $\Delta_R^V$.

Using available experimental data for $\rho$ in the mirror decays of $^{19}$Ne, $^{21}$Na, $^{29}$P, $^{35}$Ar and $^{37}$K, the average value was found to be $\overline{{\cal F}t}_0^{\rm mirror} = 6173(22) \ {\rm s}$ \cite{NaviliatCuncic:2008xt}. This provided a new independent test of CVC at the $4\times 10^{-3}$ level and enabled extracting the value $|V_{ud}| = 0.9719(17)$.
The precision obtained depends primarily on the uncertainties of the mixing ratios, $\rho$, for the different transitions. An overview of the precision for $|V_{ud}|$ that can be obtained for each of the mirror $\beta$ transitions up to $^{45}$V, by measuring the $\beta$-$\nu$ correlation, $a$, or the $\beta$-asymmetry parameter, $A$, with a relative precision of 0.5\% and combining this with the transitions' ${\cal F}t$ value, is presented in Ref.~\cite{Severijns2013}.

In the past few years, new and precise half-lives, $Q_{EC}$ values and branching ratios have been determined for many of the mirror $\beta$ transitions \cite{Severijns2018}. Measurements of the $\beta$-$\nu$ correlation coefficient in the decays of $^{19}$Ne and $^{35}$Ar have been performed with the LPCTrap at GANIL \cite{Couratin2013, Lienard2015} and are currently being analyzed. A high-precision result for the $\beta$-asymmetry parameter, $A = -0.5707(19)$, has recently been reported in the decay of $^{37}$K, corresponding to ${\cal F}t_0^{\rm mirror} = 6137(28) \ {\rm s}$ \cite{Fenker2016, Fenker2018} (\sref{asymmetry}).
With this new result and the other mirror $\beta$-decay data in Ref.~\cite{NaviliatCuncic:2008xt} the
new weighted average for all mirror transitions is $\overline{{\cal F}t}_0^{\rm mirror} = 6159(17) \ {\rm s}$, corresponding to
\begin{equation}
|V_{ud}| = 0.9730(14) \ {\rm (mirror\ transitions)} ~ .
\label{eq:Vud-mirrors}
\end{equation}
\noindent This result, which is dominated by the $\beta$-asymmetry parameter
for $^{19}$Ne and $^{37}$K, is slightly more precise than the result quoted in Ref.~\cite{NaviliatCuncic:2008xt}
and a factor of about 7 less precise than the value from the superallowed $0^+ \rightarrow 0^+$
pure Fermi transitions in \eref{Vud-Fermi}.

\subsubsection{Neutron lifetime}
\label{sec:neutron lifetime}

Neutron decay is the simplest among the $T$ = 1/2 mirror decays with the advantage
that no nuclear structure effects are present. Nonetheless, the precision on $|V_{ud}|$ is ultimately limited also by the uncertainty of the radiative correction $\Delta_R^V$.
As shown in \eref{Ft-n-num}, within the SM the $V_{ud}$ matrix element can be extracted from
the neutron lifetime, $\tau_n$, and the factor $\lambda$ which can be obtained from correlation measurements in neutron decay (\sref{correlations}).
However, obtaining accurate results for $\tau_n$ and
$\lambda$ turned out to be difficult and very challenging experimentally
\cite{Abele2008, Paul2009, Konrad2010, Wietfeldt2011, Dubbers2011, Baessler2014, Young2014}.
Using the values adopted by the PDG \cite{Patrignani:2016xqp}, i.e.\ $\tau_n = (880.2 \pm 1.0)$ s and
$\lambda = 1.2723(23)$, results in $|V_{ud}| = 0.9758(15)$, in agreement with, but about
7 times less precise, than the value from the superallowed Fermi decays in Eq.~(\ref{eq:Vud-Fermi}). Using the
most precise results for both observables, i.e.\ $\tau_n = 877.7(7)_{\rm stat}(^{+4}_{-2})_{\rm syst}$ s \cite{Pattie2017} and
$\lambda = 1.2761(^{+14}_{-17})$ \cite{Mund2013}, leads to $|V_{ud}| = 0.9748(11)$. 
To reach a precision similar to the current extraction from nuclear decays one needs determinations of $\tau_n$ and $\lambda$ with approximately $0.2$ s and $0.01\%$ precision respectively.\footnote{The latter translates into an approximate precision of $0.05\%$ in the $A_n$ or $a_n$ parameters.}

The parameter $\lambda$ has traditionally been determined primarily from measurements of the $\beta$-asymmetry
parameter, $A_n$. However, as will be discussed in~\sref{neutron asymmetry}, a shift in the resulting
values has occurred, simultaneously with a reduction of the total size of the corrections by more than an order
of magnitude. The $\beta$-$\nu$ correlation coefficient, $a_n$, provides an independent means
of determining $\lambda$, and moreover with different systematics. New experiments to determine $a_n$ were recently
initiated as well (\sref{beta-nu neutron}). 

Neutron lifetime experiments have recently been reviewed in Refs.~\cite{Wietfeldt2011,Dubbers2011,Baessler2014, Young2014}.
The neutron lifetime, $\tau_n$, is measured with two very different methods. In the ``beam method"
it is obtained from the ratio $\tau_n = N / r$ between the number, $N$, of neutrons present
at any time in a given volume (obtained from a cold neutron beam passing through this volume), and
the rate, $r$, of neutron decays observed in this same volume. The difficulty of this type of measurements
resides in combining the absolute measurement of the number of decay protons detected,
with an absolute measurement of the number of neutrons present in the same volume \cite{Byrne1996, Yue2013}.
The protons are stored in a Penning trap volume and sent at regular times onto a detector to be counted.
Neglecting small corrections, all neutrons passing
through the volume interact with a well-characterized $^6$Li absorber. Detectors surrounding this absorber
count the rate of $\alpha$ particles and tritons produced in the $^6$Li(n,$\alpha$)$^3$H reaction.
The uncertainty on the cross section for this reaction is the main issue in such experiments \cite{Yue2013}.

In the ``storage method" a number of ultracold neutrons, $N_0$, are stored during some time interval,
$\Delta t$, in either a material or a magnetic volume (``bottle"). When opening the bottle
the number of surviving neutrons is
\begin{equation}
N(\Delta t) = N_0 e^{-\Delta t / \tau_n} ~ ,
\end{equation}
\noindent provided the only source of neutron losses in the bottle is $\beta$ decay.
However, experiments in material bottles typically suffer from a number of losses other than $\beta$ decay.
Thus, one does not directly measure the neutron lifetime, but a neutron storage time, $\tau_s$:
\begin{equation}
\tau_s^{-1} = \tau_n^{-1} + \tau_w^{-1} + \tau_0^{-1}  ~ .
\end{equation}
\noindent Here, $\tau_w^{-1} $ is the total loss rate due to interactions with the bottle wall by capture on the
walls, losses due to imperfections of the wall coating, or mechanical imperfections such as gaps, welds, etc.,
and $\tau_0^{-1}$ the loss due to all other mechanisms like scattering from residual gas in the bottle
\cite{Young2014}.
The difficulty with this method is then to understand all loss mechanisms through processes other
than $\beta$ decay which is, however, a challenging task
\cite{Pokotilovski1999, Lamoreaux2002, Daum2011, Nesvizhevsky2013}.

Recently, bottle experiments started making use of magnetic trapping to avoid material wall effects.
Magnetic trapping experiments suffer mainly from three other sources of systematic effects, which seem to
be smaller in size than the above mentioned wall-related effects \cite{Young2014}:
(i) depolarization, because only one spin state is trapped, with the other spin state being accelerated onto
the surfaces of the magnetic field-generating structures; (ii) so-called quasi-bound neutrons, with energies
above the trappable energy; and (ii) ``quasi-stable" trajectories in which a small fraction of the neutrons may
be trapped in orbits that do not drain to a detector when the trap is emptied.

\begin{figure}[!htb]
\begin{center}
\includegraphics[width=\columnwidth]{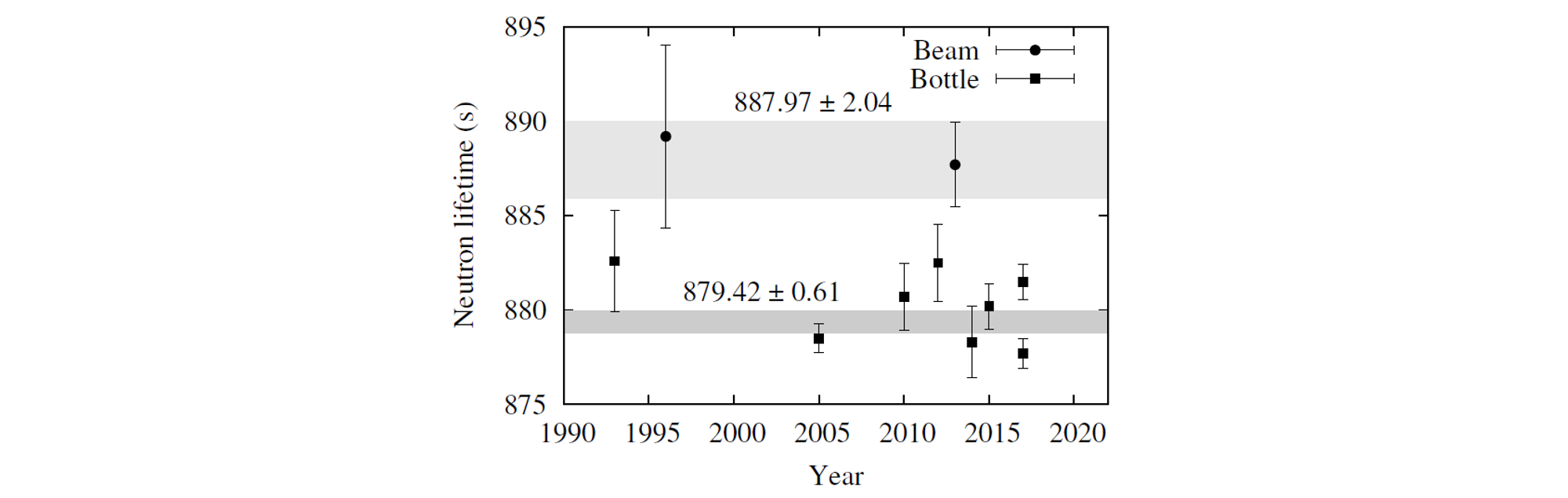}
\caption{Overview of neutron lifetime results, separated into ``beam" and ``bottle" experiments (see also~\tref{neut_averages}). The ``bottle" experiments are performed with ultracold neutrons (UCN) stored in either a material bottle, a gravitational trap, or recently also a magneto-gravitational trap. Note the about four standard deviations tension between the weighted average values of both types of experiments. The uncertainty of the average of the ``trap" measurements was scaled by a factor $\sqrt{\chi^2/\nu}\approx 1.52$ following the PDG prescription (\sref{exp_data}).
}
\label{fig:neutron-tau}
\end{center}
\end{figure}

Before 2005, the beam and bottle methods yielded comparable results. The measurement of Serebrov {\rm et al.} \cite{Serebrov2005}, subsequent re-analyses of a number of previous results, as well as several new measurements, have changed the picture completely. This is shown in Fig.~\ref{fig:neutron-tau}, which shows the most recent measurements~\cite{Byrne1996, Yue2013,Mampe1993, Serebrov2005, Pichlmaier2010, Steyerl2012, Ezhov2014, Arzumanov2015, Pattie2017, Ser17} (the individual values are listed in~\tref{neut_averages}).
The average value from bottle measurements is now systematically lower than for the beam method (Fig.~\ref{fig:neutron-tau}), with $\tau_{\rm beam} -\tau_{\rm bottle} = (8.6 \pm 2.1)$~s.\footnote{Throughout this work we add statistical and systematic errors in quadrature. Adding them linearly would reduce the tension between the various determinations~\cite{Ser17}.}
This calls for a next generation of neutron lifetime experiments with smaller and better understood systematic corrections and possibly also higher precision.\footnote{It was recently pointed out that the tension between both classes of measurements, as well as the the reactor neutrino flux deficiency~\cite{Serebrov:2018mva}, could be explained by a dark decay channel for the neutron~\cite{Fornal:2018idm}. The simplest version of the model seems to be however incompatible with various laboratory and astrophysical observations~\cite{Tang:2018eln,McKeen:2018xwc,Baym:2018ljz,Motta:2018rxp,Pfutzner:2018ieu}.}
We discuss below currently ongoing or planned experiments with both methods, although most are bottle experiments.

{\bf Beam experiments.} At the National Institute of Standards and Technology (NIST) in Gaithersburg, USA, an upgraded version of
the previously performed beam experiment is being prepared \cite{Wietfeldt2014}. In a staged approach the collaboration expects
to gradually increase the attainable precision while simultaneously obtaining information on different sources of systematics
that could help resolve the current discrepancy between the beam and bottle experiments.
In a first phase, aiming at a precision of about 1.0~s, the existing apparatus will be upgraded so as to improve on (i) the understanding and
control of neutron absorption and scattering in the $^6$LiF deposit and its substrate, (ii) detection of recoil
protons from the faint neutron beam halo, and (iii) nonlinearities of the Penning trap. The subsequent construction of a new and larger
version of the apparatus, based on similar principles but with an optimized design, is expected to reduce the major systematic uncertainties
to below 0.1~s \cite{Wietfeldt2014}.

At the J-PARC fundamental neutron physics beamline, a beam experiment (LiNA) using a Time Projection Chamber (TPC)
is being prepared \cite{Arimoto2015}. The TPC consists of a drift cage with a multi-wire proportional chamber (MWPC)
inside the vacuum vessel. In a previous neutron lifetime experiment with a TPC at the Institute Laue-Langevin (ILL)
in Grenoble, a precision of 3.1~s was reached \cite{Grivot1988, Kossakowski1989}.
The new design that will be used at J-PARC focuses on reducing the background and improving signal
efficiency. Short neutron bunches with a length of approximately half the TPC will be used
and decay events will be analyzed only during the periods when the neutron pulse is completely confined inside the TPC.
The neutron intensity will be obtained by detecting protons produced in the ($n, p$) reaction on $^3$He diluted
in the gas filling of the TPC \cite{Arimoto2015}, thereby assuring that the same fiducial volume can be used for both the decay measurement
and the determination of the neutron intensity. The experiment envisages an uncertainty of about 1.0~s and later $<$~0.3~s \cite{Otono2017}.

{\bf Bottle experiments.} Several laboratories have developed new and improved bottle experiments. As mentioned above, the
Gravitrap experiment, located at the ILL, produced one of the most precise neutron lifetime results to date, i.e.\
$\tau_n = 878.5 \pm 0.7_{\rm stat} \pm 0.3_{\rm syst}$~s \cite{Serebrov2005}. This value was about 6 standard deviations
lower than the world average of all previous experiments. However, losses in the Gravitrap experiment, and as a consequence
also systematic corrections and uncertainties, were about an order of magnitude smaller than previous
material trap experiments. This was due to the use of an improved coating for neutron storage with Fomblin grease, which has a low neutron
capture cross-section, at cryogenic temperatures of about 120~K.
Since then, an improved version of the Gravitrap experiment was built that uses a much larger storage volume
and with the Fomblin wall coating being cooled to about 80 K to further reduce wall losses. In addition,
lower operating pressures (typically about $2 \times 10^{-6}$ mbar) are used to reduce or even eliminate the
correction due to scattering losses on residual gas. The experiment is aiming at a precision 
of 0.2 s \cite{Serebrov2013}. First results yielded a value of $\tau_n = (881.5 \pm 0.7_{\rm stat} \pm 0.6_{\rm syst})$~s
with storage times as long as about 865~s \cite{Ser17}.
The apparatus is now being modified to reach even lower trap temperatures of around 10 K, which
should further reduce the loss factor and allow obtaining storage times even closer to the neutron lifetime.

Gravitrap is presently the only storage experiment with a material vessel. All other new bottle experiments
use magnetic storage of ultracold neutrons. This offers the attractive possibility to minimize losses due
to interactions with the walls by confining the ultracold neutrons in an inhomogeneous magnetic field.

In 2000, a magnetic trapping experiment at NIST used a quadrupole trap defined by a set of superconducting
coils \cite{Huffman2000, Brome2001, Oshaughnessy2009}. It measured neutron decays $\it{in ~ situ}$ by
detecting the decay electrons in superfluid $^4$He at a temperature near 300 mK.
The helium served also to produce ultracold neutrons from a mono-energetic cold neutron beam.
Scintillation light from the decay electrons was counted in photomultiplier tubes outside the helium bath.
The measurement was hampered by systematic uncertainties, the most important ones being ultracold neutron absorption by $^3$He in the superfluid $^4$He bath,
imperfect background subtraction, and quasi-bound neutrons \cite{Young2014}.

The first experiment to successfully use magnetic storage with permanent magnets and reach a competitive
precision  on the neutron lifetime was the Ezhov experiment at the ILL using a magneto-gravitational trap \cite{Ezhov2005, Ezhov2009}.
The first trap was composed of small NdFeB permanent magnets with Fe-Co poles between them, all arranged radially in a 20-pole
geometry forming a cylindrical trapping volume, open at the top, and with a height of 57 cm and a fiducial
storage volume of about 10 liters. The number of trapped neutrons after a given storage time was measured.
Depolarization, which is the dominant loss mechanism from the trap, was monitored by measuring the number of
depolarized neutrons that leaked through the magnetic shutter after having flipped their spin.
The first result, $\tau_n = (878.3 \pm 1.9)$~s, was obtained with a record neutron storage time of $(874.6 \pm 1.7)$~s
and with corrections due to leakage being at the 0.5\% level \cite{Ezhov2014, Bazarov2016}.
A new trap made of permanent magnets but with an approximately 9 times larger volume is under construction,
aiming at a precision of about 0.3~s \cite{Ezhov2014}.

A significantly larger magneto-gravitational trap is being developed by the Munich group for the PENeLOPE experiment
\cite{Materne2009}. The toroidal geometry of the trap is realized by a number of superconducting coils
producing a storage field of up to 1.8 T, in a volume of about 700 liters
and over a meter high. A vertical guiding field makes decay protons emitted upward to spiral
along the field lines and be detected by large area avalanche photodiodes at the top of
the storage vessel. Protons emitted downward are almost all reflected via the magnetic
mirror effect and redirected upward to the detectors. Instead of neutrons, protons are detected to
avoid the problem of wall losses and depolarization. The surviving ultracold neutrons are monitored
by ramping down the storage field and counting the drained neutrons. The
experiment aims at an uncertainty on the neutron lifetime of about 0.1 s \cite{Materne2009}.

The UCN$\tau$ experiment at the Los Alamos Neutron Science Center uses a very large-volume (600 liters)
magneto-gravitational trap \cite{Walstrom2009, Young2014, Morris2017}. It confines ultracold neutrons
from above by gravity, and from below and laterally by a bowl-shaped, so-called Halbach, array of permanent magnets.
A set of ten electromagnet coils provides a longitudinal field perpendicular to the
direction of the Halbach array field. The neutron lifetime is determined by filling the trap
from the bottom, then cleaning the stored neutron spectrum with an absorber of $^{10}$B on a ZnS substrate,
and finally counting the surviving neutrons with an ``internal" detector that is lowered into the trap
after a defined holding time. The detector consists of commercial ZnS(Ag) screens with about 3.25 mg/cm$^2$
of phosphor coated with about 20 nm of boron enriched to 95\% $^{10}$B \cite{Morris2017}.
In the reaction $^{10}\rm{B}$($n, \alpha$)$^{7}\rm{Li}$ the $\alpha$ particle and $^7$Li are produced
more or less back-to-back so that that at least one of them stops in the ZnS(Ag) screen producing scintillation
light. A first result of $\tau_n = (878.1 \pm 2.8_{\rm stat} \pm 0.6_{\rm syst})$~s was reported \cite{Morris2017}, whereas a more precise value of $\tau_n = (877.7 \pm 0.7_{\rm stat} ~ ^{+0.4}_{-0.2 ~ \rm syst})$~s has been recently published \cite{Pattie2017}. This is the first measurement of the neutron lifetime that does not require corrections that are larger than the quoted uncertainties.
A final precision of 0.1~s or even smaller is being aimed at \cite{Walstrom2009}.

The HOPE experiment \cite{Leung2009, Leung2016, Young2014}
at the ILL contains twelve octupole trapping potential modules formed by NdFeB magnets,
inside a set of three solenoidal magnets in a cryostat. The central solenoid provides a uniform guiding
field in the storage volume, while the bottom one serves as a magnetic shutter. The
top solenoid can be used to focus electrons and protons from neutron decays within the storage
volume onto a particle detector. Decays can also be monitored by draining
surviving neutrons from the 2 liter storage vessel onto a detector. The design and the small physical
volume of the trap allows for monitoring losses and characterizing systematic effects related
to depolarization and spectral effects. A statistical uncertainty of about 0.5 s in 50 days of running is
estimated \cite{Leung2009}. The proof of principle was successfully demonstrated \cite{Leung2016}.

Finally, the $\tau$SPECT experiment is also based on magnetic storage \cite{Karch2017}. It is installed at
the ultracold neutron source at the TRIGA reactor in Mainz and uses the magnet system of the $a$SPECT
experiment \cite{Zimmer2000, Gluck2005}. In $\tau$SPECT the magnet is mounted in a horizontal geometry
with high field solenoids on either end of a roughly uniform field region. Ultracold neutrons surviving
in the 8 liters storage volume are counted with an external detector. The experiment is currently taking data.
In the ongoing first phase a sensitivity of about 1 s, including systematics, is pursued. The goal in a second phase,
with additional detectors for monitoring losses from the trap, is to reach a 0.2 s sensitivity \cite{Young2014}.

\subsection{$\beta$-spectrum shape \label{sec:spectrum}}

Many measurements of the $\beta$-spectrum shape have been performed in the 1950's and 1960's. A compilation
can be found in Ref.~\cite{Daniel1968}. After the 1970's very few measurements have been performed.

As discussed in~\sref{corrections}, the spectral function $f_1(E_e)$ contains hadronic corrections that
have to be included in the $\beta$-spectrum shape~\cite{Holstein:1974zf}. In a recent effort the description
of several of these was extended and new contributions were added, while providing also a fully analytical description \cite{Hayen:2017pwg}. The accuracy to which the $\beta$ spectrum can be calculated depends on a number of nuclear aspects (e.g.\ nuclear deformation), the contribution of which can typically be up to about $10^{-3}$ \cite{Hayen:2017pwg}.
A recent review about the reliability of assumptions made in the calculations of $\beta$-energy
spectra can be found in Ref.~\cite{Mougeot2015}.
The slope of the spectral function, $df_1/dE_e$, is typically of the order of $\pm 0.5\% \, $MeV$^{-1}$, where the upper(lower) sign refers to $\beta^-$($\beta^+$) decay. This is mainly determined by the linear term that is dominated by the weak magnetism form factor, $b_{\rm WM}$. For Gamow-Teller transitions, to first order in recoil,
neglecting the induced tensor form factor, and assuming no second class currents,
the coefficient $C_1$ in~\eref{shapefactor} can be written as~\cite{Holstein:1974zf}
\be
C_1 = \frac{2}{3M} \left( 5 + \frac{2b_{\rm WM}}{c} \right) ~ ,
\label{eq:C1}
\ee
\noindent with $M$ the mass of the recoiling nucleus and $c = g_A \, M_{GT}$ the Gamow-Teller form factor.

Especially noteworthy are a series of $\beta$-spectrum shape measurements aimed at testing the CVC theory~\cite{Feynman1958, Wu1964, Grenacs1985}, which is the basis for the electroweak unification~\cite{Glashow1959, Salam1959, Weinberg1967}. As mentioned in~\sref{nuclearmatrixelements}, CVC relates the weak magnetism term of the $\beta$-decay branches from an isotopic triplet to the isovector M1 width of the analog $\gamma$ transition. The classical test for this, originally proposed by Gell-Mann \cite{Gell-Mann1958}, is a measurement of the spectral function of the mirror $\beta$ decays of $^{12}$B and $^{12}$N. The measurements of Lee, Mo and Wu~\cite{Lee1963} confirmed the CVC theory, even after a more encompassing reanalysis of their data~\cite{Wu1977}. The initial results had been challenged~\cite{Calaprice1976} based on an error in the Fermi functions used in their analysis. As a cross-check a measurement of the $\beta$-spectrum shape of $^{20}$F was suggested~\cite{Calaprice1976}, which was predicted to have a large slope in the spectral function of about 1\% MeV$^{-1}$. Subsequent measurements with $^{20}$F \cite{Genz1976, Calaprice1978, VanElmbt1987, Hetherington1989} confirmed the CVC theory. Additional confirmation came later from $\beta$ angular correlations for spin-aligned $^8$Li and $^8$B~\cite{Sumikama2011}, and a correlation with a delayed $\alpha$ or $\gamma$ ray \cite{Tribble1975, McKeown1980, DeBraeckeleer1995}.

{\bf Radioactive source activity standardization}.
At the CEA in Saclay a project was set up to measure the shape of $\beta$ spectra with
metallic magnetic calorimeters~\cite{Enss2000, Fleischmann2005}. These cryogenic detectors consist
of an absorber and a thermometer that are usually operated at temperatures between 10 and 50 mK
\cite{Loidl2010}. The absorber consists of two 12~$\mu$m thick gold foils of
$0.76 \times 0.66$~mm$^2$ with the activity of typically about 10~Bq deposited between these foils.
This provides both a 4$\pi$ detection geometry
and a 100\% detection efficiency. The cuboid ($50 \times 50 \times 15$~$\mu$m$^3$)
Au:Er thermometer is fixed to this absorber by ultrasonic cold welding so as to assure a
metallic contact and efficient heat transfer~\cite{Loidl2010}. The main aim of this project is
to provide information assisting dose calculations for medical purposes \cite{Mougeot2014a},
for decay heat calculations in the nuclear power industry \cite{Loidl2010, Kossert2011}, and to reduce
uncertainties in activity measurements with liquid scintillators in ionizing radiation metrology
\cite{Kossert2015}. Spectrum shape measurements performed on $^{241}$Pu and $^{63}$Ni
\cite{Loidl2014, Mougeot2014} with $\beta$-endpoint energies of 20.8~keV and 67.2 keV, respectively,
have allowed investigating the theoretical corrections at such low energies.
The results indicated clearly the importance of the atomic
exchange effect in the lowest energy part of the spectrum \cite{Mougeot2012, Mougeot2014}. In
this atomic exchange effect, a $\beta$ particle is emitted directly into a bound state of the daughter
atom, thereby expelling an initially bound electron into the continuum. Experimentally this process
cannot be distinguished from regular $\beta$ decay with a final state containing
one electron in the continuum, requiring a correction to the spectrum shape.
Such technical developments are however of limited interest for the topic of the present
review since the end-point energies of the measurements performed so far are too low for
sensitive searches of exotic couplings through the Fierz term \cite{Gonzalez-Alonso:2016jzm}
and the low activity of the source makes the collection of $10^8$ events \cite{Gonzalez-Alonso:2016jzm}
very difficult.

{\bf Fundamental properties of the weak interaction.}
Spectrum shape measurements for studying fundamental properties of the weak interaction relate mainly to
(i) searches for exotic scalar and tensor currents beyond the standard model (Fierz term), and
(ii) investigating the weak magnetism form factor \cite{Calaprice1976, Severijns2018}.
The contribution from the Fierz term, $b$, to the $\beta$ spectrum is proportional to $m_e / E_e$,
while the dominant contribution of weak magnetism is proportional to $\pm E_e/M$.
However, this functional difference does not enable to reliably separate the
two effects from a fit of an experimental spectrum because the two terms become strongly correlated.
The kinematic sensitivity of $\beta$-decay differential spectra in different transitions was discussed in
Ref.~\cite{Gonzalez-Alonso:2016jzm}, finding a maximum NP sensitivity for endpoints energies around 1-2~MeV,
such as the neutron or $^6$He.

{\it Nuclear decays.}
Taking advantage of recent developments in both detection and data-acquisition techniques, several new $\beta$ spectrometers for fundamental physics studies have been constructed in the past few years.
 
Also at CEA Saclay, a $\beta$ spectrometer using a silicon semi-conductor detector has recently been installed
\cite{Bisch2014}. Its main component consists of a 500~$\mu$m thick silicon Passivated
Implanted Planar Silicon (PIPS) detector with an active surface area of 450~mm$^2$,
a very thin entrance window of 50 nm Si equivalent and low leakage current. It operates at liquid
nitrogen temperatures and under ultra-high vacuum. With about 1.2~keV energy loss of electrons
in the dead layer, and a 5 keV energy resolution (FWHM at 77~K for
85~keV conversion electrons from $^{109}$Cd), the cut-off in energy of this detector is about
10~keV. Initial measurements with a $^{36}$Cl source showed encouraging results~\cite{Bisch2014}.
The main goal of this spectrometer is the study of the spectrum shape of forbidden $\beta$ transitions.
The $\beta$ spectra from $^{14}$C, $^{60}$Co, $^{99}$Tc, and $^{151}$Sm have been measured
\cite{Bisch2014a}. At present, the setup is being modified into a 4$\pi$ version so as to
reduce backscattering effects. This will include two 1~mm thick silicon PIPS detectors with a thin
freeze-dried radioactive source deposit on a 500 nm thick film \cite{Mougeot2017}.
 
At Madison a superconducting $\beta$ spectrometer has been developed \cite{Knutson2014}. Electrons
or positrons from a source are focused on a 1~cm diameter, 5~mm thick Si(Li) detector operated at liquid
nitrogen temperature and with an energy resolution of 10 keV FWHM.  The spectrometer has a momentum resolution
of about 2\% and a peak solid angle of 0.5 sr. Stray laboratory magnetic fields and the Earth magnetic field are canceled with a loosely wound solenoid surrounding the spectrometer. The inner mechanics of the spectrometer has been machined such as to minimize electron scattering. To minimize source scatter the activity is typically contained in a 3~mm diameter spot at the center of a 13~$\mu$m thick aluminum foil that is suspended across a 15~mm diameter hole in the copper source holder that is cooled to typically 140~K. A lead shield, about 9~cm thick and located between the source and the detector, prevents $\gamma$ rays from the source to reach the detector. First measurements have been performed for the allowed $0^+ \rightarrow 1^+$ pure Gamow-Teller $\beta$ decay of $^{14}$O. The average slope of the shape function over the energy range from 1.9 to 4.0~MeV was found to be $-0.0290(8)_{\rm stat}(6)_{\rm syst}$. This is consistent with the value of $-0.0285$ predicted by CVC which uses for the weak magnetism matrix element the value obtained from the experimental M1 decay strength of the 2313 keV $\gamma$ ray from the analog (first excited) state in $^{14}$N \cite{George2014, Towner2005}.
For the $0^+ \rightarrow 0^+$ pure Fermi decay of $^{66}$Ga, the spectrum was found
to be consistent with that of an allowed spectrum, with an average slope
of $-0.0026(10)_{\rm stat}(30)_{\rm syst}$ per MeV \cite{Severin2014},
in agreement with the expected absence of weak magnetism in pure Fermi transitions.
 
A new approach to measure the $\beta$-spectrum shape consists to implant the radioactive isotope sufficiently deep into
a large active detector so that all radiation from the source, except for part of the bremsstrahlung radiation,
is contained inside the detector \cite{Huyan:2016zfp, Huyan2018}. Most importantly, this avoids issues with
(back)scattering in the source and on the detector, a major problem in, and often even limiting, $\beta$-spectrum
shape measurements \cite{Hetherington1989, Wauters2009}.
In the setup at Michigan State University, radioactive nuclei are produced by a fragmentation
reaction, with the fragments being mass separated to obtain an isotopically pure beam~\cite{Huyan:2016zfp}.
One of the attractive aspects of this approach is that the candidates for the measurements of
the spectrum shape are selected such as to have the smallest uncertainties related to hadronic
corrections while maintaining, at the same time, an acceptable kinematic sensitivity~\cite{Naviliat-Cuncic2016}.
For $^{6}$He ions, the beam
is implanted at a depth of 12~mm into either a 10 $\times$ 5 $\times$ 5~cm$^3$ CsI(Na) or a
$\varnothing$~7.6~$\times$~7.6~cm$^2$ NaI(Tl) scintillation detector.
Monte-Carlo simulations are performed to quantify
the fraction of the $\beta$-particle energy that escapes the detector through photon radiation \cite{Huyan2018}.
The simulated absorbed energy spectrum is then convoluted with the
resolution function of the detector, determined with calibration $\gamma$ sources. First measurements have been
performed with $^{6}$He, which decays via a $0^+ \rightarrow 1^+$ pure Gamow-Teller transition. The first
goal is to observe the effect of the weak magnetism term in the $\beta$-spectrum shape. As the ground state
of $^6$He forms an isospin triplet with the (unbound) $0^+$ ground state of $^6$Be and the $0^+$ excited state
at 3563 keV in $^6$Li, CVC relates the  weak magnetism term for the $\beta$ decay of $^6$He to the M1 decay
strength of the 3563 keV $\gamma$ ray from the analog state in $^6$Li. This gives $C_1^{\rm{CVC}} = 0.650(7) \%$ MeV$^{-1}$,
which corresponds to a total deviation of 2.3 \% over the full spectrum, {\it cf.} Eqs.~(\ref{eq:shapefactor}) and
(\ref{eq:C1})~\cite{Huyan:2016zfp}. Using half of the total available data, a relative uncertainty
of 6\% has been reached on the weak magnetism form factor. This includes the correlation with the
extraction of the detector gain, which is an instrumental parameter obtained in an auto-calibration
procedure from the same fits \cite{Huyan2018b}.
In a second phase, if the experimental techniques passes the CVC test, the weak magnetism will be included in the theoretical description of the spectrum, to search for tensor couplings through the Fierz term. From the same half of the total data, an absolute statistical error of 0.002 has been reached on the Fierz term. The analysis of systematic effects is ongoing.
In the meantime, measurements with this method have also been performed
with $^{20}$F \cite{Naviliat-Cuncic2016} and a new value of the half-life has
been published \cite{Hughes2018}.

Another approach to $\beta$-spectrum shape measurements is the recent development by the Leuven and
Krakow teams of a multi-wire drift chamber (MWDC), called miniBETA \cite{Lojek2009, Severijns2014, Lojek2015}.
This project is based on experience gained with two larger-sized MWPCs~\cite{Bodek2001, Ban2006} to determine the $R$ and $N$ correlations in free neutron decay \cite{Kozela2009, Kozela2012}
and a small prototype described in Ref.~\cite{Lojek2009}. The miniBETA spectrometer is operated with a He/isobutane gas mixture,
with the concentration ratio ranging from 70\%/30\% to 90\%/10\%, at a pressure down to 300 mbar \cite{Lojek2015}. The layout (\fref{miniBETA}) consists of
10 planes with each 8 anode (signal) wires separated with 24 cathode (field) wire planes, leading to a hexagonal
cell geometry with each cell consisting of a very thin anode wire (NiCr alloy, 25 $\mu$m diameter) surrounded
by six cathode field wires. The signal planes are each separated by 15 mm, with the wires within a plane
being separated by 17.32~mm \cite{Lojek2015}.
\begin{figure}[!htb]
\begin{center}
\includegraphics[width=\columnwidth]{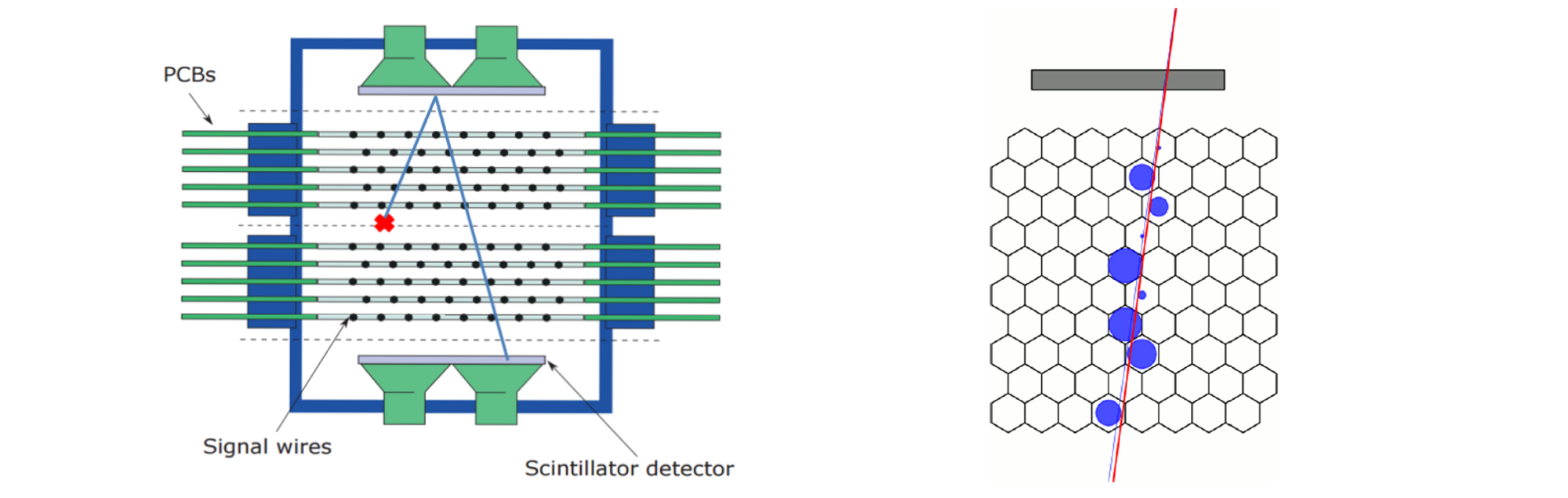}
\caption{Left panel: Layout of the miniBETA spectrometer. Only the signal wires, mounted on ten PCB boards and with the source (red cross) installed at the center, are shown here. Right panel: a muon track (red line) going through all ten planes combined as a single detector. The track is obtained as the straight line that best fits to the blue disks, the surface of which represents the amount of charge collected on the anode wire in the center of a particular hexagonal cell. The gray rectangle at the top is the scintillator that triggers the read-out of the MWDC. See text for details.}
\label{fig:miniBETA}
\end{center}
\end{figure}
In the standard layout the source is installed between two MWDCs
with five planes of anode wires and a fiducial volume of 20 $\times$ 20 $\times$ 10 cm$^3$ each, that are
themselves placed between two plastic scintillators. The source is normally installed at the center but can
be moved in the central plane between the detectors for calibration purposes. In an alternative layout
all 10 anode wire planes are combined into a single MWDC, with the source at one side of it, between the scintillators.
The latter provide the time reference signal for the drift time measurement. A typical track obtained with a muon
passing the detector is shown in the right panel of Fig.~\ref{fig:miniBETA}. The energy of the $\beta$ particles is
obtained either directly from the energy deposited in the scintillators, or from the curvature of their trajectories
when a vertical magnetic field is applied. Simulations have shown that in the latter case energy resolutions
between 10 and 20 keV for the energy interval from about 100 keV up to several MeV can be obtained.
The use of the drift chamber technology assures scattering effects to be inherently small,
due to the 5 $\mu$m thick source foil, the helium and isobutane gas filling, and the 25 $\mu$m diameter
wires. The use of a hexagonal cell structure minimizes
the number of wires needed for proper operation of the MWDC. The miniBETA geometry provides a solid angle
several orders of magnitude larger than for ``classical" magnetic spectrometers \cite{VanElmbt1987, Knutson2014}.
Further, the observation of $\beta$ particle tracks provides an additional direct means to recognize particles
that are backscattered on one of the scintillation detectors (left panel of Fig.~\ref{fig:miniBETA}).
This allows, in addition, for detailed studies of the backscattering probability as a function of energy and incidence
angle. The detector is currently in its final commissioning phase \cite{Perkowski2018}. A position resolution of
about 0.5 mm is obtained in the plane perpendicular to the wires, and of about 6 mm along the wires (by
charge-division). For most cells a total cell efficiency of 95\% was observed, and up to 98\% for radii
$r <$ 7 mm . A precision between 0.1\% and 1\% on the spectral function is being aimed at \cite{Perkowski2018}.

The UCNA-Nab-Leuven collaboration is currently pursuing measurements of the $\beta$-spectrum shape
of $^{45}$Ca \cite{Wexler2016} using the UCNA spectrometer \cite{Plaster2012} (Fig.~\ref{fig:UCNA})
at Los Alamos National Laboratory.
With a magnetic field of up to 1 T between two segmented arrays of hexagonal silicon detectors (replacing the MWPCs)
and the source installed in the mid plane of the UCNA apparatus, the setup provides a 4$\pi$ magnetic spectrometer.
The experiment uses a data acquisition system that registers the pulse shape for each event and synchronizes
the timing from the signals coming from both detector arrays.
Adding the energy signals from both detectors enables to reconstruct the full energy of the $\beta$
particles, free of any loss due to scattering. Highly-segmented Si wafer detectors \cite{Brodeur2016}
with a thickness of 1.5 mm and 2.0 mm and about 3 keV energy resolution are used for this. Good performance of
pure Si detectors in magnetic fields was previously demonstrated in fields up to 13 T \cite{Kraev2005, Wauters2009a}.
The $^{45}$Ca source was dried onto a very thin 500 nm foil in a 0.5 inch plastic holder to minimize scattering.
The isotope $^{45}$Ca is decaying with a half-life of 162.7 d via a $7/2^- \rightarrow 7/2^-$ almost pure (i.e.\
$>$ 98.5\%, taking into account possible isospin impurities) Gamow-Teller $\beta$ transition, with an endpoint energy
of 256.7 keV. Data taking has been finished and analysis is currently ongoing. A per-mil level sensitivity to the Fierz term is aimed at \cite{Hayen2018}.

Probably the largest effort ever to measure the shape of a $\beta$ spectrum, albeit focused only on a very narrow window
near the spectrum endpoint, is the development of the KATRIN spectrometer \cite{Katrin2005, Otten2008}. The goal is the
determination of the neutrino mass from tritium decay with a sensitivity of about 200 meV
\cite{Bonn1999, Kraus2005, Otten2008, Aseev2011}.
Despite the large statistics, such a shape measurement is insensitive to exotic currents because these induce a negligible energy variation in the small window studied. There are proposals to modify the KATRIN setup so that the entire spectrum can be measured, which could be used to search for sterile neutrinos with keV masses~\cite{Mertens:2014nha,Mertens:2014osa}. Because of the small endpoint energy of tritium decay, the sensitivity to exotic currents would be however an order of magnitude smaller than that of the transitions discussed above, as shown in Fig. 2 of Ref.~\cite{Gonzalez-Alonso:2016jzm}. This limits severely the bounds that can be obtained from such a modified KATRIN setup, even if the small sensitivity could be partly compensated by very large statistics and the reasonably clean theoretical description of tritium decay. 
This conclusion is in contradiction with those from Ref.~\cite{Ludl:2016ane}, which claims that the bounds on $\eL$ and $\epsilon_{R,S,T}$ could be improved by about six and five orders of magnitude, respectively. There are several reasons behind this disagreement. First of all, the bounds in Ref.~\cite{Ludl:2016ane} are extracted from the total effect to the spectrum, not from the shape distortion, and without taking into account the uncertainty of the SM normalization. Once this is properly done there is no improvement whatsoever on $\epsilon_{L,R}$ because (i) their only effect is a shift in the normalization, i.e. in $\tilde{V}_{ud}$, {\it cf.}~\eref{Vtilde}\footnote{As shown in~\eref{mixingratio}, there is an additional effect of these couplings that gets completely hidden inside the mixing ratio.}; and (ii) as a result they can only be probed through a CKM unitarity test, which is already limited by the uncertainty of the inner radiative correction $\Delta_R^V$ (see~\sref{CKMunitarity}). 
Second, the analysis in Ref.~\cite{Ludl:2016ane} does not take into account either that the mixing ratio $\rho$ has to be determined accurately from a correlation coefficient measurement, where non-standard effects must be also considered. 
On the other hand the predicted $\epsilon_{S,T}$ bounds in Ref.~\cite{Ludl:2016ane} are weakened by about an order of magnitude once these issues are properly taken into account, but they would still be very stringent. Finally, such strong bounds come however from a totally unrealistic sensitivity of $10^{-7}$, considered in a different context to smooth spectral distortions\footnote{Although Ref.~\cite{Mertens:2014osa} considers the possibility of looking for a localized kink in the spectrum with sensitivities down to $10^{-6}$ at 90\% CL, this approach is largely insensitive to smooth spectral modifications, which are the signature of exotic interactions.}, along with neglecting uncertainties in the theoretical description of the energy spectrum, which are at the level of $10^{-4}-10^{-3}$~\cite{Mertens:2014nha}. 

% It would be interesting to study the $\epsilon_{S,T}$ bounds that can be realistically expected from such modified KATRIN setup once these issues have been properly addressed.}

{\it Neutron decay.}
No dedicated measurement of the spectrum shape in the decay of unpolarized neutrons, trying to establish either
the effect of weak magnetism or search for a non-zero Fierz term, has been performed so far. The contribution
of weak magnetism in neutron decay has neither been singled out in measurements of correlation
coefficients, to which it is expected to cause typically about 1\% electron-energy dependence~\cite{Wilkinson:1982hu,Abele:2008zz,Gardner2013}.
The only reported effort, which included weak magnetism as a free parameter in the analysis of PERKEO data on
the asymmetry parameter in polarized free neutron decay, resulted in an error of about
seven times the expected effect \cite{Dubbers2011}.
Initial preparations for a similar determination of the weak magnetism form factor from a dedicated measurement of the energy-dependence of the $A$ parameter with the PERKEO III spectrometer were described in Ref.~\cite{Kaplan2007}.

Concerning the Fierz term, Ref.~\cite{Hickerson2017} used recently the 2010 UCNA data set \cite{Mendenhall2013} that provided a precision measurement of the $\beta$ asymmetry parameter, $\tilde{A}_n$, to carry out a first direct spectral extraction of the Fierz interference term  in neutron decay. The result
\be
b_n = 0.067(5)_{\rm stat}(^{+90}_{-61})_{\rm syst} ~ ,
\ee
\noindent is consistent with the SM with the uncertainty being dominated by the absolute energy
reconstruction and the linearity of the beta spectrometer energy response \cite{Hickerson2017}. Whereas this
dataset has yielded the $\beta$ asymmetry parameter, $\tilde{A}_n$, with a fractional error $<$ 1\% \cite{Mendenhall2013}
leading to a precise determination of $\lambda$,
this result for $b_n$ has no significant impact on existing limits for tensor couplings,
indicating the importance to control the detector response for an accurate extraction of the Fierz term.
Concerning future efforts, the Nab experiment is aiming at an accuracy of $\Delta b_n \simeq 3 \times 10^{-3}$ \cite{Pocanic2009}.
There are also plans to extract the Fierz term from electron spectra measured with e.g.\ a retardation
spectrometer of the $a$SPECT type \cite{Zimmer2000, Gluck2005} or a $\bfvec R \times \bfvec B$ spectrometer
\cite{Konrad2015} installed at the planned PERC facility that is being set up at the Technische Universit\"at
M\"unchen \cite{Dubbers2008, Konrad2012}.

Limits on $b_n$ can also be obtained from a fit to the $\beta$-asymmetry spectrum, $\tilde{A}_n(E_e)$, which should be less sensitive to instrumental nonlinearities than the unpolarized spectrum, as
only ratios of count rates are involved \cite{Dubbers2011}.
A fit to some older $\tilde{A}_n$ data of PERKEO,
with $\lambda$ and the Fierz term as free parameters, gave $|b_n| < 0.19$ (95\% C.L.), with however a
strong correlation between $\lambda$ and $b_n$ \cite{Dubbers2011}.
An analysis of the existing UCNA and PERKEO III $\beta$ asymmetry data set could yield $b_n$ with an absolute uncertainty 
below $0.05$~\cite{Hickerson:2013,Plaster:private} and around $0.03$~\cite{Markisch2017} respectively.
See also Ref.~\cite{Gardner:2013aya} for a similar study of the sensitivity to $b_n$ of future precise $\tilde{A}_n(E_e)$ and $\tilde{a}_n(E_e)$ data.

Much stronger limits on $b_n$ can be obtained in a global fit to neutron and/or nuclear decay data that has been integrated over the entire energy range~\cite{Severijns:2006dr, Dubbers2011, Faber2009, Konrad2010,Pattie:2013gka,Wauters:2013loa,Vos:2015eba}. Such limits rely on the effect of the Fierz term in the total rates and the $\tilde{X}$ effect in the correlation coefficients, {\it cf.}~\eref{Xtilde}. The value obtained depends on how many parameters ($C_i$ and $C'_i$) are left free in the analysis and the specific dataset used as input.

\subsection{Correlation measurements \label{sec:correlations}}

Correlation observables depend on specific and different combinations of the coupling constants for
the possible weak interaction types \cite{Jackson:1957zz,Jackson:1957auh}. They provide complementary information
on the structure of the weak interaction and allow performing a wide range of symmetry tests. Observables and
isotopes can be selected to maximize the sensitivity to a specific type of interaction or symmetry,
with minimal dependence on nuclear structure related aspects. Using pure Fermi or Gamow-Teller
transitions renders the correlation coefficients independent of the corresponding nuclear matrix elements that are
then, to first order, independent of nuclear structure effects. In the following a brief overview will be given
of recent and ongoing or planned experiments (see also Refs.~\cite{Severijns:2006dr, Dubbers2011, Holstein2014, Vos:2015eba}).

\subsubsection{$\beta-\nu$ correlation \label{sec:beta-nu}}

The $\beta$-$\nu$ correlation, related to the parameter $a$ in
Eq.~(\ref{eq:jtw1}), does not require the nucleus or neutron to be polarized. Its measurement is complicated, however,
by the fact that the neutrino momentum must be determined. This can be done by directly observing the recoiling
daughter nucleus \cite{Johnson1963, Behr2009, Flechard2011} or by observing its decay products,
i.e.\ $\gamma$ rays \cite{Egorov1997}, protons \cite{Schardt1993, Adelberger1999}, or
$\alpha$ particles \cite{Clifford1989, Sternberg2015}.

The $\beta$-$\nu$ correlation has been crucial to determine the nature of the weak interaction \cite{Allen1958}.
Nowadays it remains an important observable to determine the mixing ratio $\rho$ in mixed transitions (i.e. $\lambda$ in the neutron decay), and to search for new physics.
Namely, its measurement allows searching for possible scalar or tensor
current contributions via the quadratic dependence of $a$ on the coupling constants. This gives information that is
independent of the handedness of such currents. A linear dependence appears if $\tilde{a}$ is measured.

{\bf Nuclear decays.}
In the past decade several new results for the $\beta$-$\nu$ correlation have become available while new
projects have been initiated as well. Many of these use either atom traps \cite{Behr2014} or ion traps
\cite{Paul1990, Blaum2006} that provide almost ideal source conditions for weak interaction experiments. They offer
well-localized and cooled samples of particles in vacuum, that can often even be purified $\it in ~ situ$. They also
permit almost undisturbed observation of the recoil ions, and reduce significantly the effects of scattering
for $\beta$ particles, which is usually limiting experiments with radioactive sources embedded in a material
\cite{Wauters2009, Wauters2010}.
The $\beta$-$\nu$ correlation measurements in nuclear decays that we include in the fits described
in~\sref{fit} are listed in~\tref{nuclear}.

The experiment with $^{21}$Na in a Magneto-Optical Trap (MOT) at Berkeley, which had at first yielded a value
deviating by about three standard deviations from the SM \cite{Scielzo2004}, was scrutinized
and repeated \cite{Vetter2008}. This showed that the original result had been affected by the formation of
Na dimers in the MOT when too many $^{21}$Na atoms were present at a given time, with the breakup
of a dimer upon the $\beta$ decay of one of the constituent atoms affecting the value of the
$\beta$-$\nu$ correlation coefficient. The new and more precise value obtained, i.e.\
$a = 0.5502(60)$, is consistent with the SM prediction of $a = 0.553(2)$ \cite{Vetter2008}.
The latter is calculated using the ${\cal F}t$ values of the $^{21}$Na and $0^+\to 0^+$ transitions,
which allows extracting the mixing ratio. However, in contrast to what is reported in Ref.~\cite{Vetter2008},
the direct comparison between the measured value and the quoted
SM prediction cannot be used to extract constraints on exotic couplings because these also contribute to the
${\cal F}t$ value.

A new measurement of the $\beta$-$\nu$ correlation for $^{38}$K with the TRIUMF Neutral Atom Trap (TRINAT) setup \cite{Gorelov2005}
is planned aiming at a 0.1\% uncertainty \cite{Behr2005, Behr2014}.
The key feature will be a new electric field configuration allowing the collection of all ions
onto the ion detector, thereby reducing the errors from the ion detection efficiency as a function of angle and position
on the detector, as well as from the knowledge of the electric field. The electric field will also completely
separate the charge states +1 and +2 in time-of-flight. This requires a much larger ion detector,
a larger diameter for the field electrodes, a larger vacuum chamber, and also a larger $\beta$ detector. The recoil momentum
spectrum will also be measured by coincidences between the recoil ions and shake-off electrons \cite{Behr2005, Behr2014}.

At the Hebrew University of Jerusalem a specially-designed MOT for neon isotopes is being developed \cite{Ron2014}.
It will be installed at the SARAF facility at the Soreq Nuclear Research Center \cite{Mardor2018}. Measurements of the
$\beta$-$\nu$ correlation, and later possibly also of other correlation coefficients, are planned with $^{18}$Ne, $^{19}$Ne,
and $^{23}$Ne \cite{Mardor2018}.

At the University of Washington, Seattle, a high-intensity source of $^6$He
atoms with $\sim 10^9$ decays/s, was developed \cite{Knecht2011, Mardor2018}. It was first used for a high-precision
measurement of the $^6$He half-life and to extract the value of $g_A$ in this nucleus \cite{Knecht2012}.
A $^6$He MOT was developed for determining the $\beta$-$\nu$ correlation from coincidences between the $\beta$
particles and recoil ions aiming at a 0.1\% uncertainty \cite{Knecht2010}.

At the Weizmann Institute of Science, an electrostatic ion beam trap is being developed to store $^6$He ions produced
via the $^{9} {\rm Be}(n,\alpha){\rm ^{6}He}$ reaction and subsequently observe the $\beta$-$\nu$ correlation via
$\beta$ particle-recoil ion coincidences \cite{Zajfman1997, Hass2011, Aviv2012}. The system is actually being commissioned
\cite{Mardor2018}.
Storage times of $\sim$1.2 s for O$^+$ and CO$^+$ ions and of $\sim$0.6 s for $^4$He$^+$ ions have been obtained
and a first $\beta$ spectrum has been observed as well \cite{Mukul2017}.

Several experiments have been performed with the LPCTrap setup at GANIL, Caen
\cite{Rodriguez2006, Ban2013}. Analysis of a first measurement with $^6$He \cite{Flechard2008,
Flechard2011}, with a total of about 10$^5$ $\beta$-recoil coincidence events recorded yielded
$a = - 0.3335(73)_{\rm stat}(75)_{\rm syst}$, in agreement with the SM value $a_{GT} = -1/3$.
A subsequent measurement with more than an order of magnitude higher statistics is still being analyzed
\cite{Velten2011, Lienard2015, Fabian2015}. In the meantime measurements with $^{19}$Ne and $^{35}$Ar have been
performed as well \cite{Ban2013, Lienard2015}.

A very interesting byproduct of these measurements with LPCTrap is related to the
charge state distributions extracted for the recoiling $^6$Li, $^{19}$F and $^{35}$Cl
daughter nuclei \cite{Couratin2012, Couratin2013, Lienard2015, Fabian2018}. For $^6$Li the experimental shake-off
probability was found to be in perfect agreement with simple quantum mechanical calculations \cite{Couratin2012}.
For $^{35}$Cl the experimental charge state distributions compared very well to theoretical calculations
based on the sudden approximation and accounting for subsequent Auger processes, allowing to identify the
ionization reaction routes that lead to the formation of all charge states \cite{Fabian2018}. For $^{19}$F
such excellent consistency was not found, which was attributed to theoretical shortcomings of the Independent
Particle Model used, which apparently does not provide accurate enough ionization probabilities for systems with low nuclear
charge. Similar information was also obtained in the $\beta$-$\nu$ correlation measurements with $^{38m}$K at
TRIUMF \cite{Gorelov2000}, $^{21}$Na at Berkeley \cite{Scielzo2003} and $^{6}$He at Seattle \cite{Hong2017}.

At Argonne National Laboratory the $\beta$-$\nu$ correlation in the predominantly Gamow-Teller $\beta$ decay
of the $^8$Li ground state to the first excited state in $^8$Be was determined by inferring the momentum of
the neutrino from the kinematic shifts of breakup $\alpha$ particles \cite{Li2013} in the Beta-Decay Paul Trap
(\fref{Argonne-BPT}) \cite{Scielzo2012}. The result of a first measurement was equivalent to
$a = -0.3307(60)_{\rm stat}(67)_{\rm syst}$~\cite{Li2013}.
A subsequent experiment with an upgraded detector system and improved statistics yielded
$a = -0.3342(26)_{\rm stat}(29)_{\rm syst}$~\cite{Sternberg2015}, consistent with the SM.
\begin{figure}[!htb]
\begin{center}
\includegraphics[width=\columnwidth]{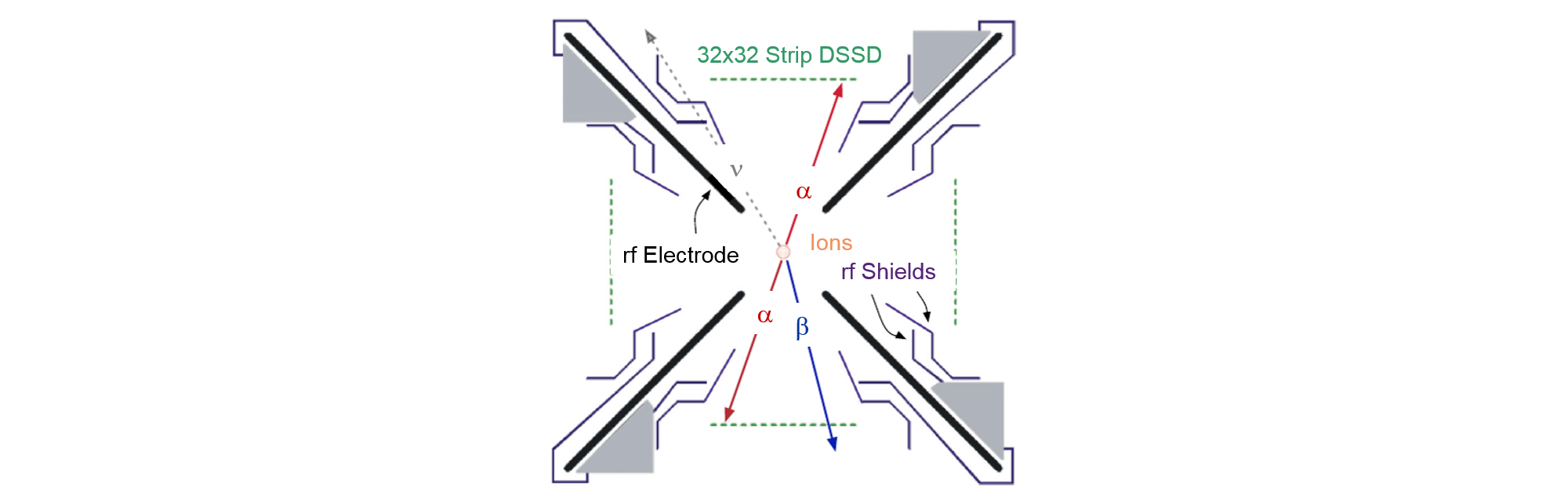}
\caption{Layout of the Beta-Decay Paul Trap. The rf-electrodes are shown, as well as the rf-shields to shield the 1-mm-thick double-sided silicon detectors (DSSD) with 32 strips on each side from rf interference. Both $\alpha$ and $\beta$ particles could
be identified by their energy deposition within a single detector pixel. The energy resolution was better than 20 keV for most strips. From Ref.~\cite{Sternberg2015}.}
\label{fig:Argonne-BPT}
\end{center}
\end{figure}

Penning ion traps are also used for correlation measurements. They provide a strong radial confinement of
charged particles making possible a near-$4\pi$ acceptance. This enables collecting a large
amount of statistics required for reaching competitive precision. The WITCH project \cite{Beck2003, Kozlov2006} at
ISOLDE was set up to determine the $\beta$-$\nu$ correlation coefficient from the measured energy spectrum of
the recoiling daughter ions from $^{35}$Ar $\beta$ decay. Initial results were obtained \cite{Kozlov2008,
Beck2011, VanGorp2014, Finlay2016} together with important information about the behavior of ion clouds in
a Penning trap \cite{Coeck2006, VanGorp2011, Porobic2015}. However, radiation-induced background turned out
to be a limiting factor. Attempts to control this effect created serious doubts about reaching a competitive precision
\cite{Finlay2016}.

The TAMUTRAP project \cite{Mehlman2013, Mehlman2015} was set up to measure the $\beta$-$\nu$ correlation
coefficient for  $T$~=~2, $0^+ \rightarrow 0^+$ superallowed $\beta$-delayed proton emitters in the mass $A =$20 to
50 region, searching for scalar currents. The experiment uses a cylindrical Penning trap providing up to
4$\pi$ acceptance for $\beta$ particles and protons. The MeV-energy protons and the higher-energetic $\beta$ particles
can escape the shallow axial potential well, and are guided by the strong magnetic radial confinement toward
the end of the trap where position sensitive silicon detectors are installed. The experiment will
observe the emitted positrons and protons in coincidence and determine the kinematic shifts of protons emitted
in two opposite directions with respect to the $\beta$ particles as a function of the energy of the latter.
For a pure vector interaction a large kinematic shift is expected, typically an order of magnitude larger than
for a pure scalar interaction.
\begin{figure}[!hbt]
\begin{center}
\includegraphics[width=\columnwidth]{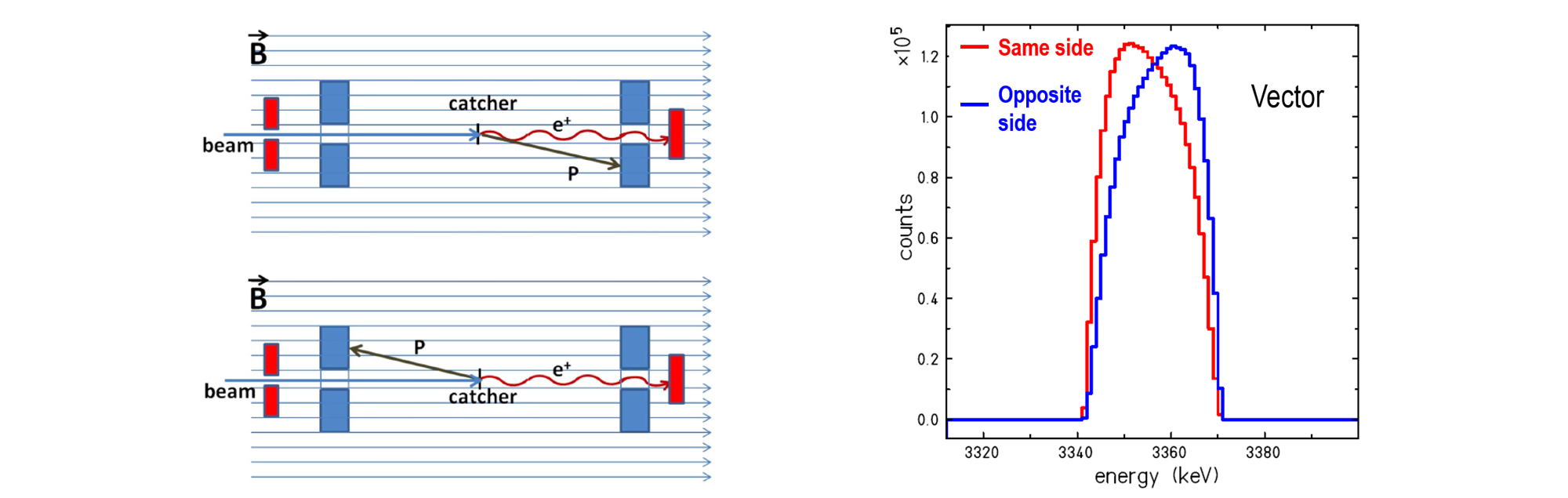}
\caption{Left panel: arrangement of the proton and $\beta$ detectors around the beam catcher foil. Right panel: expected kinematic shift for protons emitted parallel and anti-parallel with respect to the positron emission direction in the decay of $^{32}$Ar, for a pure Vector interaction. From Ref.~\cite{Severijns2017}.}
\label{fig:WISARD}
\end{center}
\end{figure}
This provides increased sensitivity to the $\beta$-$\nu$ correlation compared to previous measurements
\cite{Schardt1993, Adelberger1999}, which did not observe the $\beta$ particles and determined
the kinematic broadening instead of the shift of the $\beta$-delayed proton peak. A precision of
about 0.1\% is being aimed at \cite{Melconian2014}.

A project with the same aim, called WISARD, is being prepared at ISOLDE \cite{Blank2016, Severijns2017}.
There the superconducting magnet system of the former WITCH project \cite{Beck2003} will be used to provide
a large solid angle to detect the $\beta$ particles and protons from the $^{32}$Ar mother isotopes. These will
be implanted into a thin foil (\fref{WISARD}). Rings of $\beta$ and proton detectors are envisaged to record the decay products.
A precision of 0.1\% on the $\beta$-$\nu$ correlation coefficient is aimed at \cite{Severijns2017}.

At the TwinSol facility \cite{Becchetti2003} of the University of Notre Dame, $\beta$-$\nu$ correlation
measurements on the light mirror $\beta$ transitions of $^{11}$C, $^{13}$N, $^{15}$O, and $^{17}$F are
planned \cite{Brodeur2016, OMalley2016}. Such measurements would further extend the series of measurements
that contribute to the determination of the $|V_{ud}|$ quark-mixing matrix element in mirror $\beta$ transitions
(\sref{mirror}). All four decays have a similar sensitivity to $|V_{ud}|$ for
a given uncertainty on the $\beta$-$\nu$ correlation \cite{Severijns2013}.
Information on possible scalar and tensor currents could be obtained as well. The radioactive ion beams will be produced
in transfer reactions with beams from the tandem accelerator, will then be separated by TwinSol and finally be stopped
in a large-volume gas catcher \cite{Savard2003}. A radio-frequency quadrupole \cite{Darius2004} will serve to
cool and bunch the radioactive ions. These bunches will be sent to a multi-reflection time-of-flight spectrometer
to separate contaminant ions of different masses. The purified bunches will be sent to and captured in a Paul trap,
called NSLtrap, similar to the LPCTrap \cite{Rodriguez2006, Ban2013}. The $\beta$-$\nu$ correlation will be obtained
from the ToF spectrum of the recoil ions measured in coincidence with the $\beta$ particles. Based on the available
production rates, a 0.5\% relative uncertainty on the $\beta$-$\nu$ correlation coefficient seems feasible
\cite{Brodeur2016}. A major challenge will be the required long trapping times, in the range of minutes, due to
the rather long half-lives (i.e.\ 65 s to 20 min) for these four mirror nuclei \cite{Brodeur2016}.

{\bf Neutron decay.}
\label{sec:beta-nu neutron}
As mentioned above, the measurement of the $\beta$-$\nu$ correlation enables extracting $\lambda = g_A/g_V$, and provides a means
to search for possible scalar or tensor contributions. For the first purpose, the sensitivity of $a_n$ to $\lambda$
is close to that of the $\beta$-asymmetry parameter, $A_n$, i.e.\ $|da_n/d\lambda| = 0.30$ versus $|dA_n/d\lambda| = 0.37$.
Several experiments are currently ongoing, each with their own specificity. The results that we include in the different fits described in~\sref{fit},
are listed in Table~\ref{tab:neut_averages}.

\begin{figure}[!htb]
\begin{center}
\includegraphics[width=\columnwidth]{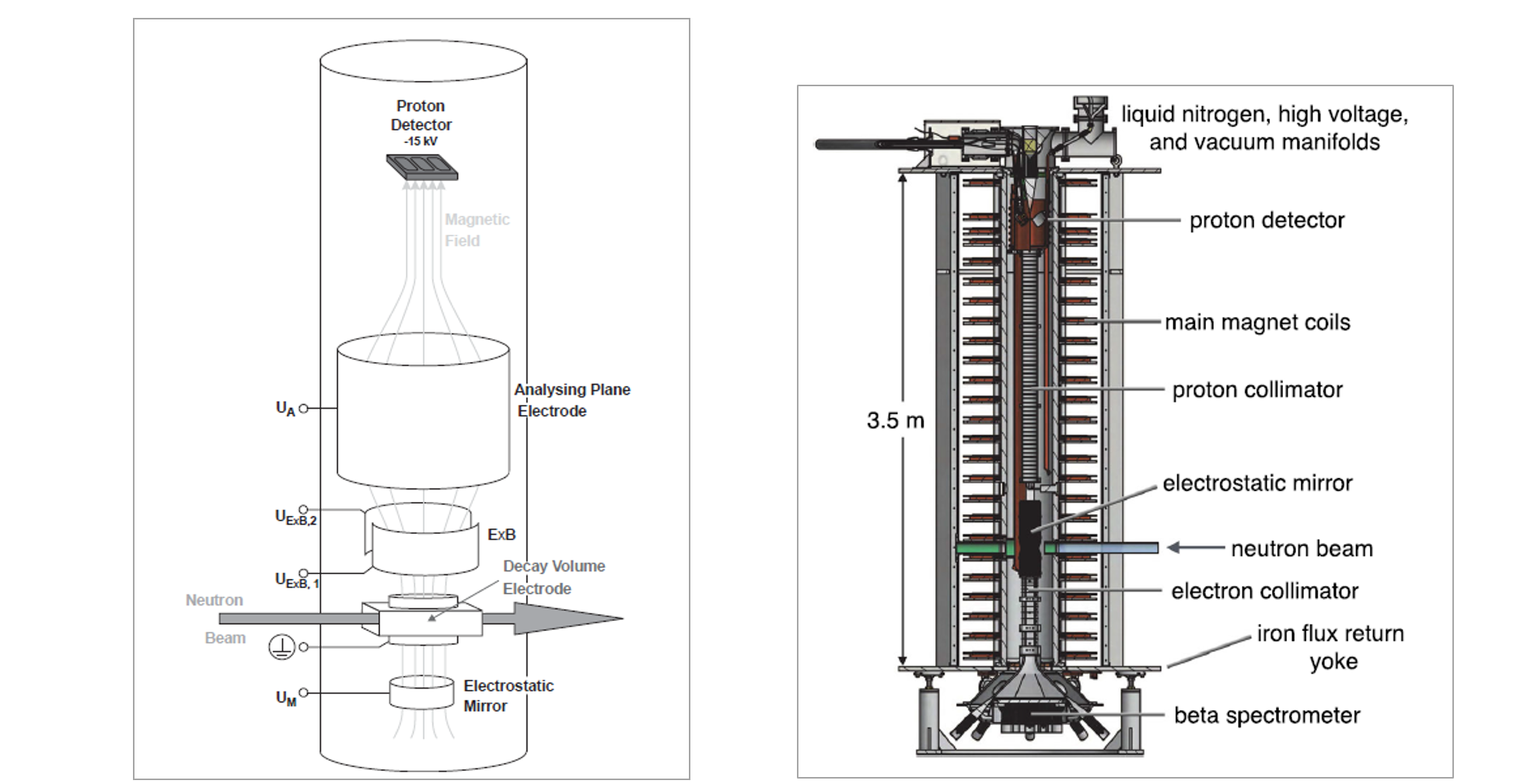}
\caption{Layout of the $a$SPECT (left panel) and of the $a$CORN spectrometers (right panel). See text for details. From Refs.~\cite{Simson2009} and \cite{Darius2017}.}
\label{fig:aSPECT-aCORN}
\end{center}
\end{figure}

The $a$SPECT experiment \cite{Zimmer2000, Gluck2005}, located at ILL, measures the shape of the proton spectrum from neutron decay
(Fig.~\ref{fig:aSPECT-aCORN}, left panel).
Decay protons from a beam of neutrons spiral around the magnetic field lines and are guided towards the analyzing plane of
a spectrometer where a variable retardation voltage is applied \cite{Baessler2008}. Protons with sufficient kinetic energy can
pass the analyzing plane and are subsequently accelerated to 15 keV and magnetically focused onto a silicon drift detector.
Protons initially emitted away from the spectrometer are reflected by an electrostatic mirror.
Observing the proton count rate for different retardation voltages allows determining the proton
spectrum shape from which the correlation coefficient $a_n$ can be obtained.
The background can in principle be determined by performing
measurements with a retardation voltage above the proton endpoint energy of about 750 eV. Background generated
from trapped electrons or ions in the Penning-like traps in the center of the spectrometer turned out to be a
major issue requiring modifications to the spectrometer \cite{Baessler2008, Simson2009}. Data analysis and studies of
systematic effects are being finalized and will lead to a final relative uncertainty of 1\% to 1.5\% \cite{Heil2017}.

The $a$CORN spectrometer at the new high-flux neutron beam line at NIST
\cite{Wietfeldt2005, Wietfeldt2009, Collett2017} uses a very different principle and relies on the three-body
decay kinematics of the free neutron. In this spectrometer (\fref{aSPECT-aCORN}, right panel) neutrons decay in a weak magnetic field with the field lines guiding
the decay electrons and protons towards the detectors. The proton detector is located at the top of the setup and the electron
detector at the bottom. A system of proton and electron collimators assures that only electrons and protons with their
momenta almost parallel to the magnetic field arrive on the detectors. The electron detector situated at the bottom of the setup observes
only electrons with their momentum pointing downward. Neutron decays with
the neutrino momentum being parallel to the electron momentum result in upward moving recoil protons with a large momentum thus having
a relatively short time-of-flight. Decays with the neutrino
momentum anti-parallel to the electron momentum result in recoil protons with low momentum, emitted mostly towards
the bottom, i.e. away from the proton detector. These are reflected upwards on an electrostatic mirror that leads, together with
their low momentum, to a longer time-of-flight. As a consequence, for each electron energy two groups of protons, fast and
slow ones, arrive at the proton detector. It can be shown that for each electron energy the $\beta$-$\nu$ correlation
coefficient is proportional to the asymmetry in the count rates of the events in the fast and slow proton group
\cite{Wietfeldt2005, Collett2017}. Since the correlation coefficient is deduced here from an asymmetry,
it is actually $\tilde{a}_n$ that is extracted. The first result led 
to $\tilde{a}_n = -0.1090(30)_{\rm stat}(28)_{\rm syst}$
\cite{Darius2017}, slightly improving on previous measurements
\cite{Str78, Byr02}. Data taking is continuing and it is expected that an ultimate uncertainty of about 1\% on $\tilde{a}$
will be achieved \cite{Darius2017}.

The Nab spectrometer at the Los Alamos National Laboratory also measures the electron energy and proton
time-of-flight in neutron decay to determine the $\beta$-$\nu$ correlation coefficient. The original concept involved a
symmetric two-arm design of the spectrometer Refs.~\cite{Bowman2005, Pocanic2009}. Later, a new strategy
was identified leading to an asymmetric variant \cite{Baessler2013}. Here, a small fraction of the cold neutrons passes through
the decay volume situated in an about 1.7~T magnetic field. Decay protons have to pass through a
field pinch, at 4~T, acting as a filter region, above the fiducial volume. They can then arrive on the detector at the top
of the setup, after having been accelerated to an energy of 30~keV.
Decay electrons are guided by the magnetic field lines to either the upper or the lower detectors. The slope of the inverse squared
proton time-of-flight spectrum for a given electron kinetic energy is, within the range allowed by the neutron decay kinematics,
proportional to the $\beta$-$\nu$ correlation coefficient \cite{Baessler2013, Young2014}. In order to observe both the up to
750~keV electrons as well as the 30~keV protons, two Si detectors with an area of about 100~cm$^2$ have been developed.
They are highly-segmented, with thin front dead layer, energy thresholds below 10~keV, and energy resolution of $<$~3~keV FWHM
\cite{Broussard2017}. The goal of the Nab experiment is to determine $a$ with a relative statistical uncertainty of $0.1\%$ %$10^{-3}$
while reducing the total systematic uncertainty to the same level \cite{Baessler2013, Young2014}.

\subsubsection{$\beta$-asymmetry parameter \label{sec:asymmetry}}

The $\beta$-asymmetry parameter, $A$, requires the initial state to be polarized.
The asymmetry parameter can be used to perform precise tests of parity violation, thereby
searching for e.g.\ right-handed $V+A$ currents.
Alternatively, assuming maximal violation of parity, it can provide information on possible scalar
or tensor currents as the actual observable is in fact $\tilde{A}$. In  mixed transitions
it provides a means to determine the Gamow-Teller/Fermi mixing ratio, $\rho$.
The type of physics to be addressed and the sensitivity that can be obtained, depend 
on the specific $\beta$ transition that is selected \cite{Naviliat-Cuncic1991, Govaerts1995, Severijns2013}.

{\bf Nuclear decays.} %\label{sec:beta asymmetry nuclear}}
In the past decade four measurements of the $\beta$-asymmetry parameter in nuclear decays have been
reported, with the nuclei being polarized either by low temperature nuclear orientation
\cite{Postma1986, vanWalle1986, Wouters1990} or by optical pumping on atoms trapped in a MOT
\cite{Behr2009, Behr2014, Behr2014a}.

The low temperature nuclear orientation measurements were built on experience gained in earlier experiments
studying isospin-mixing \cite{Schuurmans2000, Severijns2005, Golovko2005}. Geant4-based simulations were used to take
into account effects of the polarizing magnetic field as well as scattering of the $\beta$ particles in the source foil and
on the detectors~\cite{Kraev2005, Wauters2009a}. The pure Gamow-Teller
$\beta^-$ transitions of $^{114}$In \cite{Wauters2009}, $^{60}$Co \cite{Wauters2010} and $^{67}$Cu
\cite{Soti2014} have recently been used to search for tensor currents.
The nuclear properties and $\beta$ transitions of these isotopes as well as the experimental conditions
like sample foil thickness, $\beta$-particle detectors, magnetic field strengths,
were intentionally chosen to be very different in order to have minimal overlap of systematic effects and
corrections to be applied \cite{Severijns2014}. All three results (listed in
Table~\ref{tab:nuclear} in~\sref{fit}) are in agreement with the corresponding SM values \cite{Severijns2014},
and were included in the fits discussed in~\sref{fit}.

The sensitivity attainable with the low-temperature nuclear orientation method used in these experiments
appears to have reached a limit. The main limitations are the precision with which
the nuclear polarization can be determined and the effects of scattering in the sample foil.
In order to reach improved precisions significant technical modifications to the experimental setups
would be required.

Recently, first results for the asymmetry parameter using laser optical pumping in a
MOT, obtained with TRINAT, have been reported
\cite{Fenker2016, Fenker2018}.
Atoms of the mirror nucleus $^{37}$K were confined in an alternating-current MOT and were spin-polarized to 99.13(9)\%
via optical pumping \cite{Fenker2016b}. The nuclear polarization was measured $\it in ~ situ$ by photoionizing
a small fraction of the polarized atoms and then using the optical Bloch equations to model the
population distribution (\fref{TRINAT}) \cite{Fenker2016b}. Following previous developments \cite{Melconian2007, Fang2011}
this was the first time the nuclear polarization in an atom trap could be determined with a precision smaller
than 0.1\%. This was an essential achievement in order to determine angular correlation parameters to this
level of precision.
\begin{figure}[!htb]
\begin{center}
\includegraphics[width=\columnwidth]{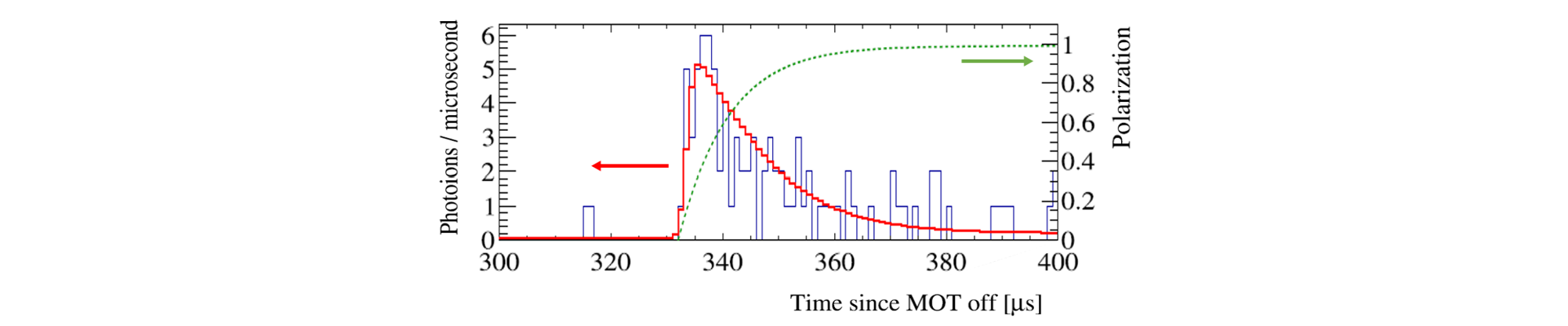}
\caption{Increase in nuclear polarization (dotted green curve) during optical pumping of $^{37}$K atoms in the TRINAT MOT trap. When the atoms are fully polarized they are all driven into one particular (so-called stretched) hyperfine state of the $^{37}$K atomic ground state that cannot be accessed by the pump laser anymore, leaving in the end no atoms in the excited state. The increase of polarization shown in the figure is thus obtained from the observed decrease of the number of photoions from the excited state (full red line) when probing this state with a different laser frequency, while the atoms are driven from the excited state to the stretched hyperfine state of the ground state with the optical pumping laser, by using the Bloch equations. Adapted from Ref.~\cite{Fenker2016b}.}
\label{fig:TRINAT}
\end{center}
\end{figure}
The result for the asymmetry parameter in the $3/2^+ \rightarrow 3/2^+$ mirror $\beta$ decay of $^{37}$K,
$\tilde{A}= -0.5707(13)_{\rm stat}(13)_{\rm syst}(5)_{\rm pol}$ \cite{Fenker2018}, constitutes the most
precise measurement of the $\beta$ asymmetry parameter from a nucleus to date. Statistics and
systematics contribute similarly to the uncertainty, with a small contribution from the uncertainty in
determining the nuclear polarization. The systematics is dominated by the uncertainty on the background
correction, the  position and velocity of the trapped atom cloud and the choice of the Geant4
physics list package for the simulations \cite{Fenker2018}.
The above result has a slightly higher sensitivity to right-handed currents compared to previous
measurements of the $\beta$-particle longitudinal polarization in the decays of $^{12}$N and $^{107}$In
\cite{Severijns1993, Severijns1998, Allet1996, Thomas2001}. It also allowed the extraction
of a new independent value for the $V_{ud}$ element of the CKM matrix, $|V_{ud}| = 0.9748(22)$, which,
together with the $\beta$-asymmetry result for $^{19}$Ne \cite{Calaprice1975} now dominates the data set
from the mirror $\beta$ transitions leading to $V_{ud}$, Eq.~(\ref{eq:Ft-mirror}). The authors of
Ref.~\cite{Fenker2016b} estimate that by further increasing the
light polarization and decreasing the transverse magnetic field, which will both increase the average nuclear
polarization and decrease its uncertainty, a polarization uncertainty of $\sim$0.04\% could be achieved
in planned new measurements, allowing for an uncertainty of $\sim$0.1\% on this and other polarized
correlation parameters.

{\bf Neutron decay.} \label{sec:neutron asymmetry}
The $\beta$-asymmetry parameter in neutron decay is the prime observable to determine the ratio $\lambda = g_A/g_V$
between the axial-vector and vector coupling constants. For many years, measurements of this parameter have
been limited by large corrections, that were moreover difficult to determine precisely.
The three oldest results that are still considered
by the PDG~\cite{Bopp1986, Liaud1997, Yerozolimsky1997} %),
are subject to corrections in the range
of 15 to 30\%, have a final uncertainty in the 1-2\% range, and are in addition systematically lower than the more
recent ones \cite{Abele1997, Abele2002, Mund2013,Liu2010, Liu2010a, Plaster2012, Mendenhall2013, Bro17}
by as much as 3 to 5\%. The required corrections are related mainly to the angular acceptance of the
electron detector, beam-related background, and the neutron beam polarization.
A previous detector response/acceptance problem was largely overcome by using two electron detectors
placed in a strong magnetic field, such that each detector essentially covers a solid angle of
2$\pi$, with the fiducial decay volume in between \cite{Abele1997, Plaster2012}.
Beam-related background was efficiently reduced in the most recent experiments by using pulsed-beam
methods \cite{Markisch2009}. The uncertainty related to the neutron-beam polarization was reduced
with the use of supermirror polarizers \cite{Schaerpf1989}, and
crossed supermirrors \cite{Kreuz2005}, or by transporting the low-energy ultracold neutrons through
a magnetic field spin-state selector \cite{Holley2012}
producing close to 100\% neutron-beam polarization. Opaque spin filters using polarized Helium-3
\cite{Zimmer1999} now allow the characterization of the neutron-beam polarization with 0.1\% precision
\cite{Mund2013, Mendenhall2013}.
The PERKEO II \cite{Abele1997} and UCNA \cite{Plaster2008} (\fref{UCNA}) spectrometers, located at ILL and LANSCE respectively, implemented these improved geometries and methods, and were able to reach $\sim0.5\%$ relative uncertainties on $A_n$ \cite{Abele1997, Abele2002, Mund2013,Liu2010, Liu2010a, Plaster2012, Mendenhall2013, Bro17}. The results are shown in~\tref{neut_averages}, along with the previous extractions that we also use in the fits in~\sref{fit}. The UCNA experiment plans to take more data to eventually reach a precision of 0.14\% \cite{Plaster2018}.

\begin{figure}[!htb]
\begin{center}
\includegraphics[width=\columnwidth]{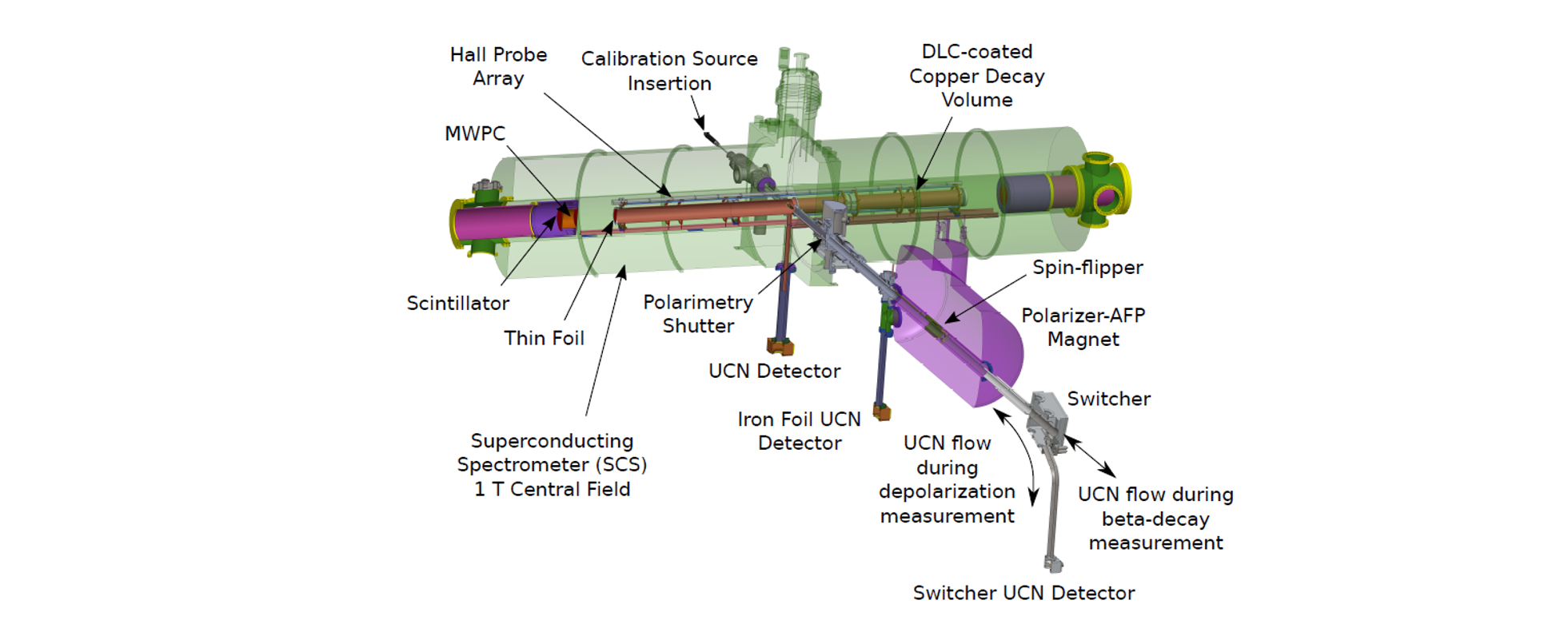}
\caption{Layout of the UCNA experimental apparatus. From Ref.~\cite{Bro17}.}
\label{fig:UCNA}
\end{center}
\end{figure}

The PERKEO III spectrometer \cite{Markisch2009} uses more advanced concepts and differs in two main respects from the PERKEO II and UCNA spectrometers. First, it uses a strong magnetic field aligned along the neutron beam so as to make the fiducial decay volume much larger. At the exit of the spectrometer, the magnetic field lines turn outward to large plastic scintillators for detecting the decay electrons. Second, the neutron beam is pulsed, selecting a small wavelength slice of the neutron beam what reduces the velocity spread. Data for the analysis are then only taken while a neutron beam pulse is contained in the fiducial volume, thus suppressing the amount of beam-related background. Analysis of the PERKEO III data is almost finished and close to unblinding. A combined relative uncertainty of $18 \times 10^{-4}$ was reached with similar statistical and systematic errors, and with systematics being dominated by the neutron-beam polarization, the magnetic mirror effect and detector non-linearity and stability \cite{Markisch2017}.

Recently, the construction of a next generation neutron spectrometer, the PERC instrument \cite{Dubbers2008, Konrad2012} has started at the FRM-II research reactor in Munich. In PERC, the fiducial decay volume is essentially an 8 m long section of neutron guide with a strong magnetic field along the guide. The neutrons at the end of it are directed into a beam stop while electrons and protons from neutron decays in the 8 m long neutron decay channel are being guided by the magnetic field lines towards a detector. Beam-related background is again suppressed by not recording data when the neutron beam bunch hits the beam stop. Depending on the neutron decay observable to be measured, a suitable electron or proton detector has to be installed at the end of the PERC decay channel. Alternatively, a $R \times B$ spectrometer \cite{Wang2013, Konrad2015} could be used as a multipurpose detection setup. PERC aims at improving the sensitivity of neutron decay studies by about one order of magnitude \cite{Konrad2012}. Statistical and systematic relative uncertainties below $3 \times 10^{-4}$ are aimed at for $A_n$, but also for $a_n$, $B_n$, and the proton asymmetry parameter, $C_n$ \cite{Dubbers2008}.

\subsubsection{$\nu$-asymmetry parameter \label{sec:recoil}}

The neutrino-asymmetry parameter, $B$, requires not only the initial state to be polarized but also the observation of the $\beta$ particle and recoil momenta in order to infer the neutrino kinematics. It is mainly used to search for right-handed currents, and in neutron decay also to extract $\lambda$ although it is not very sensitive in this case, i.e.\ $|dB_n/d\lambda| = 0.076$. Note that the $B$ parameter contains a Fierz-like term~\cite{Jackson:1957auh,Jackson:1957zz}, so that it can be written as $B(E_e)= B_{SM} + b_B m_e/E_e$. As a result the $\tilde{B}$ parameter takes the form:
\be
\tilde{B} = \frac{B_{SM} + b_B \langle m_e/E_e \rangle }{1 + b \langle m_e/E_e \rangle}~.
\ee

A measurement of $B$ in the superallowed mirror $\beta$ decay of $^{37}$K was performed at TRINAT
\cite{Melconian2007}. Like for the measurement of the $\beta$-asymmetry
parameter on the same isotope (\sref{asymmetry}), the atoms were
cooled and confined in the MOT, with optical pumping and photoionization techniques
being used to produce and measure, $\it in ~ situ$, a 96.5(8)\% spin-polarized sample. The resulting value,
$\tilde{B} =0.755(20)_{\rm stat}(13)_{\rm syst}$ \cite{Melconian2007}, is consistent
with the SM prediction, $\tilde{B} = -0.7692(15)$ \cite{Melconian2007} and constitutes the first
measurement of the neutrino-asymmetry parameter in a nuclear decay.
The sensitivity of this result to right-handed currents is similar to that from other experiments
(\sref{asymmetry} and~\sref{polarization}).
However, the authors of Ref. \cite{Melconian2007} consider that this result
could be improved by about an order of magnitude in view of the much higher precision to
which the nuclear polarization can now be determined \cite{Fenker2016b}, and by further using a different
MCP-detector setup, the position calibration of which constituted the larger part of the systematic
error \cite{Melconian2007}.

A recent measurement of $B$ in neutron decay with the PERKEO II spectrometer resulted in $\tilde{B}_n = 0.9802(50)$ \cite{Schumann2007}, confirming two earlier results from the
Gatchina group \cite{Kuznetsov1995, Serebrov1998}, as well as a less  precise result from PERKEO II \cite{Kreuz2005a},
but with significantly smaller corrections. The weighted average of all results, i.e.\ $\tilde{B}_n = 0.9805(30)$
(see~\tref{neut_averages}), is used in the fits of~\sref{fit}. These various values are internally consistent, but they are in $\sim2\sigma$ tension with the $A_n$ determinations, as can be clearly seen in Fig. 21 of Ref.~\cite{Dubbers2011}. This has relevant consequences for BSM fits with tensor currents with right-handed neutrinos (see~\sref{fit-RH}), as this determination plays a major role in probing them.

New measurements of $B_n$ are planned at different laboratories. At Los Alamos the UCNB experiment is expected to yield a combined statistical and systematic precision approaching $1 \times 10^{-4}$ \cite{Wilburn2009}. With PERC (\sref{neutron asymmetry}) a statistical as well as systematic relative uncertainty of $3 \times 10^{-4}$ is envisaged \cite{Dubbers2008}. Finally, with the BRAND apparatus (\sref{R}) that would be installed at the ILL-Grenoble and later at the European Spallation Source (ESS) \cite{ESS2018} a statistical precision below $1 \times 10^{-3}$ and ultimately of $2 \times 10^{-4}$ is aimed at \cite{Bodek2011, Bodek2016}.

\subsubsection{Recoil-Asymmetry parameter}

The TRINAT MOT \cite{Behr2014} was also used to measure the angular distribution of recoiling daughter
nuclei, extracting the recoil-asymmetry parameter, $A + B$, from the $1^+ \rightarrow 0^+$ pure Gamow-Teller
$\beta$ decay of spin-polarized $^{80}$Rb \cite{Pitcairn2009}. The result for this observable, $\tilde{A} + \tilde{B} = 0.015(29)_{\rm stat}(19)_{\rm syst}$,
which is zero in the SM \cite{Jackson:1957zz, Treiman1958}, directly probes the presence of possible tensor type interactions. In $^{80}$Rb the values of the
weak magnetism and induced tensor form factors cannot be deduced from CVC. The measured momentum
dependence of the asymmetry was used to constrain these higher-order SM corrections, with the
remaining uncertainty on these being the dominant systematic error on the result for $A + B$.

The angular distribution of recoiling protons from the decay of polarized free neutrons, i.e.\ the quantity
$C_n = -x_C(A_n + B_n)$, where $x_C = 0.27484$ is a kinematical
factor \cite{Treiman1958, Gluck1996}, was measured
with the PERKEO II spectrometer \cite{Abele1997}, resulting in $\tilde{C}_n = 0.2377(26)$ \cite{Schumann2008}.
This enabled to deduce $\lambda = 1.275(16)$, which is consistent with the world average value derived from
measurements of the $\beta$-asymmetry parameter in neutron decay, i.e.\ $\lambda = 1.2723(23)$ \cite{Patrignani:2016xqp},
but is almost one order of magnitude less precise. The overall uncertainty of about 1.2\% on $\lambda$ is
dominated by detector calibration and statistics. Higher precision could probably be reached with the new
PERC spectrometer that is currently being prepared \cite{Dubbers2008, Konrad2012}.

\subsubsection{Polarization-Asymmetry correlation \label{sec:polarization}}

The longitudinal polarization $P$ of the $\beta$ particle emitted by unpolarized nuclei is given by~\cite{Jackson:1957zz}
\bea
%P=\frac{G\,\frac{p_e}{E_e}}{1+b \frac{m_e}{E_e}}~,
P=\frac{G}{1+b\,m_e/E_e}\cdot \frac{p_e}{E_e}~,
\eea
where $G$ is the coefficient introduced in~\eref{jtw2}.
In the past, $P$ has been compared in two pairs of pure Fermi and pure Gamow-Teller decays, i.e. in the positron decays of
$^{14}$O and $^{10}$C \cite{Carnoy1990, Car91} and the electron decays of $^{26}$Al and $^{30}$P
\cite{Wic87}. The values obtained for the quantity $P_F/P_{GT}$ provide strong constraints on
right-handed charged weak currents \cite{Car91}.

Relative measurements of the longitudinal polarization of $\beta$ particles emitted by polarized nuclei, the so-called polarization-asymmetry correlation \cite{Quin1989} were performed in $^{107}$In~\cite{Severijns1993}, $^{12}$N~\cite{Allet1996,Thomas2001} and $^{21}$Na~\cite{Schewe1997,Schewe1997a} . In such experiments, the longitudinal polarization of $\beta$ particles were measured for particles emitted anti-parallel or parallel to the nuclear polarization direction. The results from these experiments were used as a probe of right-handed interactions. The measurements performed with $^{12}$N~\cite{Thomas2001} and $^{107}$In~\cite{Severijns1993} provide stringent tests of maximal parity violation in $\beta$ decay. A review can be found in Ref.~\cite{Severijns2014}.

It was argued that measurements of the positron
polarization ratio for either polarized and unpolarized nuclei, or for nuclei polarized in two
opposite directions, for selected $\beta$ transitions (e.g.\ $^{12}$N and $^{107}$In, or
the mirror $\beta$ decays of $^{17}$F, $^{21}$Na, $^{25}$Al and $^{41}$Sc)
could provide sensitivities to the $W_R$ mass ranging from about 500 GeV/c$^2$ up to 1.7 TeV/c$^2$
(at zero mixing), for degrees of nuclear polarization ranging from 80\% to 100\%
and when measuring at the maximum of the $\beta$-particle energy spectrum~\cite{Govaerts1995}. Such polarizations are rather difficult to produce with nuclei implanted in suitable materials.
Alternatively, new levels of sensitivity could be achieved by exploiting the fact that the $\beta$ particle polarization is energy dependent and, for polarized nuclei, it can even cancel at a
non-zero energy which depends on the nuclear polarization. An experiment along these lines,
with $^{36}$K and $^{37}$K polarized by optical
pumping is currently being prepared at NSCL, East Lansing \cite{Naviliat-Cuncic2017}.

\subsubsection{Correlations linearly sensitive to imaginary terms \label{sec:imaginary}}
Spin and momentum vectors are both odd under the time-reversal operation. Tests of the time reversal symmetry in $\beta$ decay thus require correlations with an odd number of spins and momenta of the particles involved \cite{Jackson:1957zz,Jackson:1957auh}. Two of these have been investigated experimentally: the $D$ correlation that is sensitive to an imaginary phase between the $V$ and $A$ couplings, and the $R$ correlation that requires the transverse polarization of the $\beta$ particles to be measured and probes the existence of time reversal violating $S$ and/or $T$ couplings.

{\bf D-correlation.}
%\label{sec:D}
The $D$ coefficient can only be probed in mixed Fermi/Gamow-Teller transitions and has been measured so far for only two systems, $^{19}$Ne and the neutron. For $^{19}$Ne, the measurements were performed by the Princeton team in the 80's leading to the combined result $D = 0.0001(6)$ \cite{Calaprice1985}. The FSI contribution to $D$ for $^{19}$Ne was calculated to be $D_{\rm FSI} = 2.6 \times 10^{-4} \, p_e/p_{\rm max}$ \cite{Callan1967}, which is of the same order as the experimental uncertainty.

In neutron decay a series of precise measurements were performed more recently. The TRINE experiment at the ILL yielded $D_n  = -0.00028(64)_{\rm stat}(30)_{\rm syst}$ \cite{Soldner2004}, whereas the emiT experiment at NIST \cite{Lising2000} yielded a final result of $D_n = -0.000094(189)_{\rm stat}(97)_{\rm syst}$ \cite{Mumm2011, Chupp2012}. The world average value, $D_n  = -0.00012(20)$ \cite{Patrignani:2016xqp}, has a three times smaller precision than $^{19}$Ne. The FSI contribution to $D_n$ is estimated at $D_{\rm FSI} = 1.2 \times 10^{-5}$ \cite{Chupp2012}, and can be calculated to $1\%$ or better~\cite{Ando:2009jk},
leaving ample room for reaching a higher experimental sensitivity.
The world average is used in the fits (Table~\ref{tab:neut_averages}).

The available results for both $^{19}$Ne and the neutron are consistent with the absence of time-reversal-violating $V, A$ currents and constrain time-reversal-violating scalar and tensor interactions that arise in certain extensions to the SM such as leptoquarks \cite{Chupp2012}.
A new project at the JYFL-Jyv\"askyl\"a and GANIL-SPIRAL2 laboratories (called MORA) aims at measuring the $D$ correlation with $^{23}$Mg with a sensitivity at the few $10^{-4}$ level \cite{Delahaye2017}.

{\bf R-correlation and transverse electron polarization.}
\label{sec:R}
The $R$ coefficient has been investigated in the decays of $^{19}$Ne, $^{8}$Li and the neutron. The measurement on $^{19}$Ne at Princeton University in 1983 yielded a result with moderate precision, $R = -0.079(53)$ \cite{Schneider1983}. A series of measurements with $^8$Li at the Paul Scherrer Institute between 1996 and 2003 provided a much more precise value $R = 0.0009(22)$ \cite{Sromicki1996, Sromicki1999, Huber2003}.

Recently, the first measurement of the $R$ coefficient in neutron decay was performed as well \cite{Kozela2009, Kozela2012}. The experiment was carried out at the FUNSPIN cold neutron beam line at the Paul Scherrer Institute \cite{
Schebetov2003, Zejma2005}. A specially designed Mott polarimeter \cite{Ban2006} allowed measuring simultaneously both transverse components of the electron polarization, thus yielding the $R_n$ and
the ($\mathcal{CP}$-conserving) $N_n$ correlation coefficients. The SM value of this last coefficient is non-zero and served as a test of the sensitivity and as a control of systematic effects for the experimental apparatus. The Mott polarimeter consisted of two identical modules located  at both sides of the neutron beam. Each module contained a MWPC for electron tracking, a removable thin Pb Mott scatterer foil, and a scintillator hodoscope for electron energy measurement \cite{Kozela2011}. Events for determining the $N_n$ and $R_n$ coefficients showed a ``V-track" with two reconstructed segments in one MWPC and one segment accompanied by a scintillator hit in the opposite chamber, indicating the electron was scattered on the Pb foil at one side of the neutron beam and had deposited its energy in the hodoscope at the opposite side. The ability to fully reconstruct momenta of low-energy electrons before and after the backward Mott scattering, which served as the electron polarization analyzer, was the most important feature of this setup.
The values $R_n = 0.004(12)_{\rm stat}(5)_{\rm syst}$ and $\tilde{N}_n = 0.067(11)_{\rm stat}(4)_{\rm syst}$ were obtained \cite{Kozela2012}. The final state interaction to the first amounted to $R_{FSI} \approx 0.0006$, which is below the sensitivity of the experiment whereas, for the second, the result is in agreement with the SM value $N_{SM} \approx 0.068$ \cite{Kozela2012}. The main sources of systematic error were found to be the background-subtraction procedure, the $\beta$ decay asymmetry-induced nonuniform illumination of the Mott foil, and the uncertainty in the determination of the effective Mott-analyzing power. The results can be used to constrain the relative strength of a scalar interaction and related parameters in SM extensions with leptoquark exchange, and in the MSSM with R-parity violation \cite{Kozela2012}, while also providing limits on Lorentz invariance violation \cite{Bodek2013}.

A new experiment measuring the transverse polarization of electrons emitted from polarized $^8$Li nuclei (called MTV) is currently ongoing at ISAC-TRIUMF \cite{Murata2011, Totsuka2014, Murata2016}. The 80\% polarized $^8$Li beam is stopped in an aluminum stopper foil. The transverse polarization of the decay electrons is again determined with a Mott polarimeter. In a first phase, the setup consisted of a MWDC placed between a wall of plastic scintillators and a 100~$\mu$m thick lead analyzer foil. Electrons backscattered from this foil were recognized by their typical ``V-tracks" and their transverse polarization deduced from the left-right backscattering asymmetry \cite{Murata2011, Murata2014}. Data from the 2010 run have provided a value $R = 0.065(36)$ with a modest 3.6\% precision \cite{Murata2011}. In the second phase, a new Mott polarimeter has been constructed, consisting of a cylindrical drift chamber surrounded by a Pb analyzer foil and with both being installed between two cylindrical trigger and stopping counters \cite{Tanaka2013}. Data taking is ongoing. The experiment is aiming at a statistical precision below 0.1\%, which is similar to the size of the final-state interaction of $R_{\rm FSI} \approx +0.07 \%$, while at the same time trying to limit the systematic effect to be no larger than 0.02\%  \cite{Murata2011}.

With experiments having proven that the transverse polarization of electrons emitted in the decay of cold neutrons can be accurately measured with Mott scattering polarimetry, a new experiment, called BRAND, was proposed that will measure 11 correlation coefficients ({\it a, A, B, D, H, L, N, R, S, U, V}) \cite{Bodek2011, Bodek2016}. The experiment will be optimized towards the transverse electron polarization related coefficients, i.e.\ {\it H, L, N, R, S, U}, and {\it V}, with the simultaneous measurements of the coefficients {\it a, b, A, B} and {\it D} initially being used to keep systematics under control. At a later stage, dedicated measurements of the latter coefficients are planned as well. The experiment will use electron tracking in an MWDC with cylindrical geometry and with readout of both wire ends. Both direct and Mott-scattered electrons will be detected in plastic scintillator hodoscopes. Protons from neutron decays will be accelerated to 20-30 keV and converted into bunches of electrons ejected from a thin LiF layer that will be accelerated and subsequently be detected in a plastic scintillator. Measurements with both an unpolarized as well as a longitudinally polarized cold neutron beam are planned. With a 2 m long detecting system, 10$^5$ decays per second in the fiducial volume and a three month data taking period, a sensitivity of about $5 \times 10^{-4}$ for the transverse electron polarization-related correlation coefficients is anticipated \cite{Bodek2011}. This would improve significantly the bounds on both real as well as imaginary scalar {\it versus} tensor  couplings, on the leptoquark exchange helicity projection amplitudes, and on the R-parity violating MSSM coupling constants combinations \cite{Bodek2011, Bodek2016}.

{\bf ${\cal T}$-odd momentum correlation in radiative $\beta$ decay.}
%\label{sec:Radiative}
%
The small, but now experimentally well-established radiative $\beta$ decay of the neutron \cite{Beck2002, Nico2006, Cooper2010, Bales2016} offers the opportunity of studying additional $\mathcal{T}$-odd momentum correlations, which do not appear in ordinary $\beta$ decay. Gardner and He \cite{Gardner2012,Gardner:2013aiw} considered the correlation $\xi = \bfvec{p}_\nu \cdot (\bfvec{p}_e \times \bfvec{k})$ with $\bfvec{k}$ the momentum of the radiative photon, both within and beyond the SM. Defining the $\mathcal{T}$-odd asymmetry,
$A_{\xi} = (N_+ - N_-)/(N_+ + N_-)$,
\noindent with $N_{+(-)}$ the total number of decay events with positive (negative) $\xi$, SM values for $A_{\xi}$ up to $1.9 \times 10^{-4}$ were calculated for free neutron decay (via FSI effects), with the exact value depending on the covered energy region. With  advanced detection techniques recently developed \cite{Cooper2012, Bales2016} such a measurement could become feasible in the near future. A proposal to measure this time-reversal violating 3-momentum asymmetry in the radiative beta decay of $^{38m}$K at ISAC-TRIUMF, was recently approved \cite{Behr2017}.

%%%%%%%%%%%%%%%%%%%%%%%%%%%%%%%%%%%%%%%%%%%%%%%%%%%%%%%%%%%%%%%%%%%%%%%%%%%%%%%%
\begin{table*}
\caption{Selected ongoing and planned experiments discussed in~\sref{exp}. See main text for details.
The approximate relative precision goals are given together with their reference. If the SM value is zero,
the absolute precision goal is then given. When precision goals are given as a percentage,
relative uncertainties are meant. The symbol ${\cal O}$ refers to the estimated order of
magnitude for a precision goal. 
The precisions given for $a$ are obtained setting the Fierz term $b$ to zero
(see~\sref{fit_procedure} and Ref.~\cite{Gonzalez-Alonso:2016jzm}).}
\begin{center}
\begin{tabular}{l r l l}
\hline\hline
Coefficient	&	Precision goal		   					&	Experiment (Laboratory)							&   Comments	\\
\hline
$\tau_n$		&	$ 1.0\,\rm{s};~ 0.1\,\rm{s}$ \cite{Wietfeldt2014}  &	BL2, BL3 (NIST)	~\cite{Wietfeldt2014}				&  In preparation; two phases \\
			&	$ 1.0\,\rm{s};~ 0.3\,\rm{s}$ \cite{Otono2017}  	&	LiNA (J-PARC)		~\cite{Arimoto2015, Otono2017} 	&  In preparation; two phases \\
			&	$ 0.2\,\rm{s}$	\cite{Serebrov2013}   	      	&	Gravitrap (ILL)		~\cite{Serebrov2013, Ser17}		&  Apparatus being upgraded \\
			&	$ 0.3\,\rm{s}$	\cite{Ezhov2014}              		&	Ezhov (ILL)		~\cite{Ezhov2014}				&  Under construction  \\
			&	$ 0.1\,\rm{s}$	\cite{Materne2009}      	  	&	PENeLOPE (Munich)~\cite{Materne2009}				&  Being developed \\
			&	$ \lesssim0.1\,\rm{s}$ \cite{Walstrom2009}    	&	UCN$\tau$ (LANL)	~\cite{Walstrom2009, Young2014, Morris2017, Pattie2017}	& Ongoing \\
			&	$ 0.5\,\rm{s}$	\cite{Leung2009}        	  	&	HOPE (ILL)		~\cite{Leung2009, Leung2016, Young2014}	&Proof of principle Ref.~\cite{Leung2016}  \\
			&	$ 1.0\,\rm{s};~ 0.2\,\rm{s}$ \cite{Young2014}      &	$\tau$SPECT (Mainz)~\cite{Karch2017, Young2014}		&  Taking data; two phases \\
\hline
$\beta$-spectrum  & $ {\cal O}(0.01)$ \cite{Knutson2014}			&	Supercond. spectr. (Madison)  ~\cite{Knutson2014}		&  Shape factor \eref{shapefactor}. Ongoing \\
$\beta$-spectrum  & $ {\cal O}(0.01)$ \cite{Bisch2014}			&	Si-det. spectr. (Saclay)~\cite{Bisch2014, Bisch2014a}	&  Shape factor \eref{shapefactor}. Ongoing \\
$b_{GT}$		&	$0.001$								& Calorimetry (NSCL)~\cite{Huyan:2016zfp, Huyan2018}	&  Analysis ongoing ($^6$He, $^{20}$F)\\
			&	${\cal O}(0.001)$ \cite{Perkowski2018}		&	miniBETA (Krakow-Leuven)~\cite{Lojek2009, Severijns2014, Lojek2015, Perkowski2018}	& Being commissioned \\
			&	${\cal O}(0.001)$ \cite{Hayen2018}			&	UCNA-Nab-Leuven (LANL)~\cite{Wexler2016, Plaster2012, Hayen2018}			&  Analysis ongoing ($^{45}$Ca) \\
$b_n$		&	$<0.05$ \cite{Hickerson:2013,Plaster:private}		&	UCNA (LANL)		~\cite{Plaster2008}				&  Ongoing with $A_n$ data \\
			&	$0.03$ \cite{Markisch2017}				&	PERKEO III (ILL)	~\cite{Markisch2017}				&  Possible with $A_n$ data \\
			&	$ 0.003$ \cite{Pocanic2009}				&	Nab (LANL)		~\cite{Bowman2005, Pocanic2009, Baessler2013, Young2014}	& In preparation \\
			&	$ 0.001$ \cite{Dubbers2008}				&	PERC (Munich)		~\cite{Dubbers2008, Konrad2012}	&  Planned \\
\hline
${a}_F$		&	$ 0.1$\% \cite{Behr2014}					&	TRINAT (TRIUMF)	~\cite{Behr2005, Behr2014}		&  Planned ($^{38}$K)  \\
			&	$ 0.1$\% \cite{Melconian2014}				&	TAMUTRAP (TA\&M)	~\cite{Melconian2014}			&  Superallowed $\beta\,p$ emitters \\
			&	$ 0.1$\% \cite{Severijns2017}				&	WISArD (ISOLDE)	~\cite{Blank2016, Severijns2017}	&  In preparation ($^{32}$Ar $\beta\,p$ decay)   \\
$a$			&	not stated								&	Ne-MOT (SARAF)	~\cite{Ron2014, Mardor2018}		&  In preparation ($^{18}$Ne, $^{19}$Ne, $^{23}$Ne)\\
$a_{GT}$		&	$ {\cal O}(0.1)$\% \cite{Knecht2010}			&	$^6$He-MOT (Seattle)~\cite{Knecht2011, Knecht2010}	&   Ongoing ($^6$He) \\
			&	not stated								&	EIBT (Weizmann Inst.)~\cite{Zajfman1997, Hass2011, Aviv2012}	&   In preparation ($^6$He) \\
			&	$ 0.5$\% \cite{Lienard2015}				&	LPCTrap (GANIL)	~\cite{Velten2011, Ban2013, Lienard2015, Fabian2015}&  Analysis ongoing ($^6$He, $^{35}$Ar) \\
$a_{mirror}$	&	$ 0.5$\% \cite{Brodeur2016}				&	NSL-Trap (Notre Dame)~\cite{Becchetti2003, Brodeur2016, OMalley2016}	&  Planned ($^{11}$C, $^{13}$N, $^{15}$O, $^{17}$F) \\
$\tilde{a}_n$	&	$ 1.0$\% \cite{Darius2017}				&	$a$CORN (NIST)	~\cite{Wietfeldt2005, Wietfeldt2009, Collett2017, Darius2017}	&  Data taking ongoing  \\
$a_n$		&	$ 1.0-1.5$\% \cite{Heil2017}				&	$a$SPECT (ILL)	~\cite{Zimmer2000, Gluck2005, Heil2017}					&  Analysis being finalized  \\
			&	$ 0.15$\% \cite{Baessler2013, Young2014}	&	Nab (LANL)		~\cite{Bowman2005, Pocanic2009, Baessler2013, Young2014}	&  In preparation  \\
\hline
$\tilde{A}_n$	&	$ 0.14$\% \cite{Plaster2018}			&	UCNA (LANL)		~\cite{Plaster2008}  				&  Data taking planned \\
			&	$ 0.18$\% \cite{Markisch2017}				&	PERKEO III (ILL)	~\cite{Markisch2017}				&  Analysis ongoing    \\
$\tilde{A}_{mirror}$&	$ {\cal O}(0.1)$\% \cite{Fenker2016}			&	TRINAT (TRIUMF)	~\cite{Fenker2016}				&  Planned \\
\hline
$\tilde{B}_n$	&	$ 0.01$\% \cite{Wilburn2009}		      &	UCNB (LANL)				~\cite{Wilburn2009}				&   Planned      \\
\hline
$\tilde{A}_n$ ($a_n, \tilde{B}_n, \ldots$)	&$ 0.05$\% \cite{Dubbers2008}	&    PERC (Munich)		~\cite{Dubbers2008, Konrad2012}	&  In preparation  \\
$\tilde{A}_n$ ($a_n, \tilde{B}_n, \ldots$)	& $<{\cal O}(0.1)$\% \cite{Bodek2011}&	BRAND (ILL/ESS)		~\cite{Bodek2011, Bodek2016}		&  Proposed \\
\hline
$D$			&	$ {\cal O}(10^{-4})$	\cite{Delahaye2017}		&	MORA (GANIL / JYFL)~\cite{Delahaye2017}			&  In preparation ($^{23}$Mg)  \\
$R$			&	$ {\cal O}(10^{-3})$	\cite{Murata2011}		&	MTV (TRIUMF)		~\cite{Murata2011, Totsuka2014, Murata2016}	&  Data taking ongoing ($^8$Li) \\
$D, R$		&	$ {\cal O}(0.1)$\% \cite{Bodek2011}			&	BRAND (ILL)		~\cite{Bodek2011, Bodek2016}		&  Proposal  \\
\hline\hline
\end{tabular}
\end{center}
\label{tab:expSummary}
%\small{$^\dagger$Prueba}
\end{table*}

%%%%%%%%%%%%%%%%%%%%%%%%%%%%%%%%%%%%%%%%%%%%%%%%%%%%%%%%%%%%%%%%%%%%%%%%%%%%%%%%

\newpage
%%%%%%%%%%%%%%%%%%%%%%%%%%%%%%%%%%%%%%%%%%%%%%%%%%%%%%%%%%%%%%%%%%%%%%%%%%%%%%%
\section{Numerical application}
\label{sec:fit}

The couplings that enter the 
$\beta$-decay Lagrangian, \eref{leffJTW}
have to be determined from experiments that measure spectroscopic quantities
and correlations
in allowed transitions. Several global fits have been carried out in the past
and one of the
most complete of them was performed about a decade ago
\cite{Severijns:2006dr}. Since then,
other analyses have been realized that focus either on neutron decay
data \cite{Konrad2010,Dubbers2011} or looked specifically for constraints on
tensor type interactions \cite{Pattie:2013gka,Wauters:2013loa}. The analysis in
Ref.~\cite{Vos:2015eba} showed the sensitivity from nuclear only data
compared with data from neutron and nuclear mirror transitions.

In this section we present a fit to the existing $\beta$-decay data, which
updates the constraints on exotic scalar and tensor couplings and the value
of the vector and axial-vector couplings, associated in the SM to $V_{ud}$
and $g_A$ respectively. The analysis is similar to
that described in Ref.~\cite{Severijns:2006dr}, although it contains
several significant improvements. The experimental
data is added step by step in order to illustrate the impact of each data set
on the constraints and also contour plots are given
for all parameters involved in the minimizations.
Superallowed $0^+\to 0^+$ transitions are consistently incorporated
for the first time in a global fit 
and experimental errors are rescaled using a standard prescription
to reflect internal inconsistencies.

%%%%%%%%%%%%%%%%%%%%%%%%%%%%%%%%%%%%%%%%%%%%%%%%%%%%%%%%%%%%%%%%%%%%%%%%%%%%%%%%

The transitions considered in this analysis are allowed
pure Fermi transitions, pure Gamow-Teller transitions and
neutron decay. Neutrino masses are neglected and the interaction is assumed
to be described by the Lee-Yang Lagrangian of~\eref{leffJTW}. The SM subleading
corrections (\sref{corrections}) are assumed to be properly removed for each
experimental input. This was explicitly shown in~\eref{Ft} for the total rates,
through the introduction of the corrected ${\cal F}t$ values, and will be
assumed to hold for the correlation coefficients.

%%%%%%%%%%%%%%%%%%%%%%%%%%%%%%%%%%%%%%%%%%%%%%%%%%%%%%%%%%%%%%%%%%%%%%%%%%%%%%%%
\subsection{Experimental data}
\label{sec:exp_data}

Similar to the analyses presented in Refs.~\cite{Wauters:2013loa,Vos:2015eba}
we restricted the experimental data to the most sensitive measurements, 
namely:
the ${\cal F}t$-values of super-allowed $0^+\rightarrow 0^+$
transitions, the neutron lifetime $\tau_n$, 
the correlation coefficients $a_n$, $\tilde{a}_n$, $\tilde{A}_n$,
$\tilde{B}_n$, $R_n$, $D_n$ and the combination of coefficients
$\lambda_{AB} = (A_n-B_n)/(A_n+B_n)$ in neutron decay, and selected measurements of
correlations in nuclei. These include
$a$ and $\tilde{a}$, the ratio $P_F/P_{GT}$ between the
longitudinal polarizations
of positrons emitted from pure Fermi and pure Gamow-Teller
transitions,
$\tilde{A}$ in pure Gamow-Teller transitions and the triple correlation $R$.
The exclusion of other experimental data does not affect in any
significant way the constraints on scalar and tensor couplings extracted from the
fits presented here.
We do not include the value of $C_n$ from Ref.~\cite{Schumann2008}, since
it is derived from the same data set as the parameter $B_n$ of
Ref.~\cite{Schumann2007}, which is included in our analysis.

The data used in the fits are
given in Tables~\ref{tab:ftFermi}, \ref{tab:neutron} and \ref{tab:nuclear}.
The tables give the central value, the $1\sigma$ uncertainty of the
measured property and an estimate of the value of $\langle m_e/E_e \rangle$.
Instead of using the Fierz term extracted from the
${\cal F}t$ values of super-allowed Fermi transitions \cite{Hardy:2014qxa}
we use the individual corrected ${\cal F}t$ values listed in Table IX of
Ref.~\cite{Hardy:2014qxa} and reproduced in \tref{ftFermi}. In this
way it is possible to keep track of the correlation between the bound on
the Fierz term and the CKM unitarity constraint, which is relevant if the underlying
NP model generates contributions to both quantities. 
The uncertainty of the radiative correction $\delta'_R$ is not included
in \tref{ftFermi}. This uncertainty is included in Ref.~\cite{Hardy:2014qxa}
as a systematic effect in its SM fit, by adding and subtracting one-third of
the $Z^2\alpha^3$ correction to each ${\cal F}t$ value simultaneously, and
looking at the average ${\cal F}t$ value obtained in each case. This takes
into account that the correction is expected to be a smooth function of $Z^2$
and thus to shift all ${\cal F}t$ values in the same direction. We follow
this prescription in the fits below in a statistical way, so that we can
calculate how it affects the correlations in a multidimensional BSM fit.
Explicitly we use an extra parameter, $\eta=0\pm 1/3$, gaussianly distributed
to include the uncertainty on $\delta'_R$ in the ${\cal F}t$ values in a
correlated way by generating:
${\cal F}t'_i = {\cal F}t_i ( 1 + \eta\,[\delta'_{R}]_{Z^2\alpha^3}^i )$,
where the unprimed ${\cal F}t$ values are those without the $\delta'_R$ uncertainty
listed in \tref{ftFermi}. This uncertainty is a subdominant one, but its proper
inclusion enables to fully recover the results of Ref.~\cite{Hardy:2014qxa} in
the SM limit.

\tref{neutron} contains the data from neutron decay and
\tref{nuclear} from nuclear transitions.
In all tables, the values of $\langle m_e/E_e \rangle$ are quoted only for
those parameters in which they were used as a sensitivity factor to the
Fierz term. They were calculated by averaging $m_e/E_e$ using the
statistical rate function given in \eref{fdefinition}, with the integration
limits fixed by the experimental conditions or by the selections made in
the data analysis.
For all values for which an asymmetric systematic uncertainty was quoted,
the largest absolute value of this uncertainty was adopted.

For the neutron data, we first calculated a weighted average value
for the lifetime, $a_n$, $\tilde{A}_n$ and $\tilde{B}_n$ from a set of
existing results instead of
directly including the individual results as separate inputs. 
The uncertainty on the average was then determined following the prescription of the
PDG \cite{Patrignani:2016xqp} of increasing the uncertainty obtained from
the fit by $\sqrt{\chi_{\rm{min}}^2/\nu}$ if this factor is larger than 1.
This is particularly important for the neutron lifetime and for
$\tilde{A}_n$, for which the internal consistency of
the data is poor. Because of the larger uncertainty resulting from such a
procedure, the constraints on some couplings resulting
from the fits are weaker than those obtained by previous analyses
that used directly the data from individual experiments.
We consider the procedure adopted here to be more reliable since it takes into account the
inconsistencies existing in the experimental data. 
The values of the experimental inputs and the associated averages are
given in~\tref{neut_averages}. 
We do not use the $\tilde{a}_n$ and $\tilde{B}_n$ results from Refs.~\cite{Gri68,Chr70,Ero70}, which are used by the PDG. Those results have relative uncertainties that are at least a factor of 10 larger than the most precise result and have therefore negligible impact on the associated averages. 

The use of average values for the correlation coefficients raises the
question about the proper value of $\langle m_e/E_e \rangle$ to be used
in those fits that are sensitive to the Fierz term.
This value depends on the experimental conditions or on the interval
of the $\beta$-energy spectrum adopted in the analysis. However, the
inspection of the values in~\tref{neut_averages} shows that the relative variations of
$\langle m_e/E_e \rangle$ are respectively $\pm 2.4$\% and
$\pm 3.9$\% for the measurements of $\tilde{A}_n$ and $\tilde{B}_n$.
Such small variations have a negligible impact on the constraints.
The value of $\langle m_e/E_e \rangle$ adopted for the weighted
mean of a parameter is then the weighted mean value from the
individual measurements.

%%%%%%%%%%%%%%%%%%%%%%%%%%%%%%%%%%%%%%%%%%%%%%%%%%%%%%%%%%%%%%%%%%%%%%%%%%%%%%%%
\begin{table*}
\caption{${\cal F}t$ values from superallowed Fermi
decays used in the fits. The ${\cal F}t$ values are extracted
from Table IX of Ref.~\cite{Hardy:2014qxa}, whereas the $\langle m_e/E_e \rangle$
values were calculated in this work. The table does not show the uncertainty
associated to $\delta'_R$, which is correlated between decays.}
\begin{center}
\begin{tabular}{r@{\hspace{8mm}} r@{\hspace{8mm}} c} 
\hline\hline
Parent & ${\cal F}t$ (s)~~~ & $\langle m_e/E_e \rangle$\\
\hline
$^{10}$C   & $3078.0\pm 4.5$ &  0.619  \\
$^{14}$O   & $3071.4\pm 3.2$ &  0.438  \\
$^{22}$Mg  & $3077.9\pm 7.3$ &  0.310  \\
$^{26m}$Al  & $3072.9\pm 1.0$ &  0.300  \\
$^{34}$Cl  & $3070.7\pm 1.8$ &  0.234  \\
$^{34}$Ar  & $3065.6\pm 8.4$ &  0.212  \\
$^{38m}$K   & $3071.6\pm 2.0$ &  0.213  \\
$^{38}$Ca  & $3076.4\pm 7.2$ &  0.195  \\
$^{42}$Sc  & $3072.4\pm 2.3$ &  0.201  \\
$^{46}$V   & $3074.1\pm 2.0$ &  0.183  \\
$^{50}$Mn  & $3071.2\pm 2.1$ &  0.169  \\
$^{54}$Co  & $3069.8\pm 2.6$ &  0.157  \\
$^{62}$Ga  & $3071.5\pm 6.7$ &  0.141  \\
$^{74}$Rb  & $3076.0\pm11.0$ &  0.125  \\
\hline\hline
\end{tabular}
\end{center}
\label{tab:ftFermi}
\end{table*}
%%%%%%%%%%%%%%%%%%%%%%%%%%%%%%%%%%%%%%%%%%%%%%%%%%%%%%%%%%%%%%%%%%%%%%%%%%%%%%%%

%%%%%%%%%%%%%%%%%%%%%%%%%%%%%%%%%%%%%%%%%%%%%%%%%%%%%%%%%%%%%%%%%%%%%%%%%%%%%%%%
\begin{table*}
\caption{Experimental data from neutron decay used in the fits.}
\begin{center}
\begin{tabular}{
c@{\hspace{3mm}} r@{\hspace{8mm}} r@{\hspace{5mm}}
c@{\hspace{5mm}} c }
\hline\hline
Parameter		& Value~~ 		& Rel. error	& $\langle m_e/E_e \rangle$	& Reference\\
\hline
%
% neutron lifetime
$\tau_n$~(s)	& 879.75(76)		&  0.09\,\%	& 0.655					& Average \tref{neut_averages}\\
%
% a and a-tilde in neutron
$a_n$		& $-0.1034(37)$	& 3.6	\,\%		&						& Average \tref{neut_averages}\\
$\tilde{a}_n$	& $-0.1090(41)$	& 3.8	\,\%		& 0.695					& \cite{Darius2017}\\
%
% A-tilde in neutron
$\tilde{A}_n$	&  $-0.11869(99)$	& 0.83\,\%		& 0.569					& Average \tref{neut_averages}\\
%
% B in neutron
$\tilde{B}_n$	&  0.9805(30)		& 0.31\,\%		& 0.591					& Average \tref{neut_averages}\\
%
% lambdaAB in neutron
$\lambda_{AB}$&  $-1.2686(47)$	& 0.37\,\%		& 0.581					& \cite{Mos01}\\
%
% D in neutron
$D_n$		& $-0.00012(20)$	&			&						& Average \tref{neut_averages} \\
%
% R in neutron
$R_n$		&  0.004(13)		&			&						& \cite{Kozela2012}\\
\hline\hline
\end{tabular}
\end{center}
\label{tab:neutron}
\end{table*}
%%%%%%%%%%%%%%%%%%%%%%%%%%%%%%%%%%%%%%%%%%%%%%%%%%%%%%%%%%%%%%%%%%%%%%%%%%%%%%%%

%%%%%%%%%%%%%%%%%%%%%%%%%%%%%%%%%%%%%%%%%%%%%%%%%%%%%%%%%%%%%%%%%%%%%%%%%%%%%%%%
\begin{table*}
\caption{
\label{tab:nuclear}
Data from measurements in nuclear decays used in the fits.
The columns list the parent nucleus in the transition,
the initial and final spins, the transition type,
the measured parameter, the experimental value with its $1\sigma$ uncertainty,
the relative error,
the value of $\langle m_e/E_e \rangle$,
and the reference.}
\begin{center}
\begin{tabular}{
l@{\hspace{5mm}} c@{\hspace{3mm}} c@{\hspace{3mm}} c@{\hspace{3mm}} c@{\hspace{3mm}}
r@{\hspace{8mm}} r@{\hspace{8mm}} c@{\hspace{5mm}} c
}
%
%%%%%
\hline\hline
Parent		& $J_i$	& $J_f$	& Type			& Parameter	& Value~~			& Rel. error	& $\langle m_e/E_e \rangle$	& Reference\\
\hline
$^6$He		& 0		& 1		& GT/$\beta^-$		& $a$		& $-0.3308(30)^{\rm a)}$&  0.91\,\%	&				& \cite{Johnson1963} \\
$^{32}$Ar		& 0		& 0		&  F/$\beta^+$		& $\tilde{a}$	& 0.9989(65)		& 0.65\,\%		&  0.210$^{\rm b)}$	& \cite{Adelberger1999}\\
$^{38m}$K	& 0		& 0		&  F/$\beta^+$		& $\tilde{a}$	& 0.9981(48)		& 0.48\,\%		&  0.161$^{\rm b)}$  & \cite{Gorelov2005}\\
$^{60}$Co		& 5		& 4		& GT/$\beta^-$		& $\tilde{A}$	& $-1.014(20)$		& 2.0	\,\%		& 0.704			& \cite{Wauters2010} \\
$^{67}$Cu		& 3/2		& 5/2		& GT/$\beta^-$		& $\tilde{A}$	& 0.587(14)		& 2.4	\,\%		& 0.395			& \cite{Soti2014} \\
$^{114}$In	& 1		& 0		& GT/$\beta^-$		& $\tilde{A}$	& $-0.994(14)$		& 1.4	\,\%		& 0.209			& \cite{Wauters2009} \\
$^{14}$O/$^{10}$C &	&		& F-GT/$\beta^+$	& $P_F/P_{GT}$& 0.9996(37)		& 0.37\,\%		&  0.292			&  \cite{Car91}\\
$^{26}$Al/$^{30}$P &	&		& F-GT/$\beta^+$	& $P_F/P_{GT}$& 1.0030	(40)		& 0.4	\,\%		&  0.216			& \cite{Wic87}\\
$^8$Li		& 2		& 2		& GT/$\beta^-$		& $R$		& 0.0009(22)		&			&				& \cite{Huber2003} \\
\hline\hline
\end{tabular}
\end{center}
\small{
$^{\rm a)}$ After including radiative corrections \cite{Glu98}.\\
$^{\rm b)}$
For the measurements in Refs.~\cite{Adelberger1999,Gorelov2005} we use the
values which multiply the Fierz term quoted in the original papers.
These are slightly different from the values of $\langle m_e/E_e \rangle$
that can be calculated from the experimental conditions
but they take into account the correlation between $a$ and $b$~\cite{Gonzalez-Alonso:2016jzm}. See~\sref{fit_procedure} for further details.
}
\end{table*}
%%%%%%%%%%%%%%%%%%%%%%%%%%%%%%%%%%%%%%%%%%%%%%%%%%%%%%%%%%%%%%%%%%%%%%%%%%%%%%%%

%%%%%%%%%%%%%%%%%%%%%%%%%%%%%%%%%%%%%%%%%%%%%%%%%%%%%%%%%%%%%%%%%%%%%%%%%%%%%%%%
\begin{table*}
\caption{Values of the neutron lifetime and correlation coefficients,
and their average. The fits of $\tau_n, a_n,\tilde{A}_n, \tilde{B}_n$, and $D_n$
give
$\chi_{\rm{min}}^2/\nu$ = 3.67, 0.17, 6.74, 0.83, and 0.10, respectively.
The uncertainties on the average $\tau_n$ and $\tilde{A}_n$ values have been
scaled by a factor $S=\sqrt{\chi_{\rm{min}}^2/\nu}$. 
For the $a$ parameter we note that within the SM we can also use the measurement of
Ref.~\cite{Darius2017} quoted in~\tref{neutron} since $\tilde{a}$ and $a$
coincide, resulting in an average value of $a_{SM}^{\rm exp} = -0.1060(28)$
($\chi^2_{\rm{min}}/\nu=0.63$).}
\begin{center}
\begin{tabular}{ l l l c c }
\hline\hline
Coefficient	&	Value				&	Year / Method	&	$\langle m_e/E_e \rangle$	&	Reference\\
\hline

$\tau_n$ (s)	&	$882.6 \pm 2.7$ 		&	1993 / Bottle	&		&	\cite{Mampe1993}\\
			&	$889.2 \pm 3.0 \pm 3.8$	&	1996 / Beam	&		&	\cite{Byrne1996}\\
			&	$878.5 \pm 0.7 \pm 0.3$	&	2005 / Bottle	&		&	\cite{Serebrov2005}\\
			&	$880.7 \pm 1.3 \pm 1.2$	&	2010 / Bottle 	&		&	 \cite{Pichlmaier2010}\\
			&	$882.5 \pm 1.4 \pm 1.5$	&	2012 / Bottle	&		&	 \cite{Steyerl2012}\\
			&	$887.7 \pm 1.2 \pm 1.9$	&	2013 / Beam	&		&	 \cite{Yue2013}\\
			&	$878.3 \pm 1.9$		&	2014 / Bottle	&		&	\cite{Ezhov2014}\\
			&	$880.2 \pm 1.2$		&	2015 / Bottle	&		&	 \cite{Arzumanov2015}\\
			&	$877.7 \pm 0.7 \pm 0.4$	&	2017 / Bottle	&		&	 \cite{Pattie2017}\\
			&	$881.5 \pm 0.7 \pm 0.6$	&	2017 / Bottle	&		&	 \cite{Ser17}\\
%\hline
			&	$879.75 \pm 0.76$		&				&		&	Average (S=1.9) \\
\hline
$a_n$		&	$-0.1017(51)$			&	1978			&		&	\cite{Str78}\\
			&	$-0.1054(55)$			&	2002			&		&	\cite{Byr02}\\
%\hline
			&	$-0.1034(37)$			&				&		&	Average\\
\hline
$\tilde{A}_n$	&	$-0.1146(19)$			&	1986			&  0.581	&   \cite{Bopp1986}\\
			&	$-0.1160(9)(12)$		&	1997			&  0.582	&   \cite{Liaud1997}\\
			&	$-0.1135(14)$			&	1997			&  0.558	&   \cite{Yerozolimsky1997}\\
			&	$-0.11926(31)(42)$		&	2013			&  0.559	&   \cite{Mund2013}\\
			&	$-0.12015(34)(63)$		&	2018			&  0.586	&   \cite{Bro17}\\
%\hline
			&	$-0.11869(99)$			&				& 0.569	& Average (S=2.6) \\
\hline
$\tilde{B}_n$	&	$0.9894(83)$			&	1995			&  0.554	&   \cite{Kuznetsov1995}\\
			&	$0.9801(46)$			&	1998			&  0.594	&   \cite{Serebrov1998}\\
			&	$0.9670(120)$			&	2005			&  0.600	&   \cite{Kreuz2005a}\\
			&	$0.9802(50)$			&	2007			&  0.598	&   \cite{Schumann2007}\\
%\hline
			&	$0.9805(30)$			&				& 0.591	& Average\\
\hline
$D_n\times 10^2$	&	$-0.06(12)(5)$		&	2000			&		 &  \cite{Lising2000}\\
			&	$-0.028(64)(30)$		&	2004			&		 &  \cite{Soldner2004}\\
			&	$-0.0094(189)(97)$		&	2012			&		 &  \cite{Chupp2012}\\
%\hline
			&	$-0.012(20)$			&				&		 & Average\\
\hline\hline
\end{tabular}
\end{center}
\label{tab:neut_averages}
\end{table*}
%%%%%%%%%%%%%%%%%%%%%%%%%%%%%%%%%%%%%%%%%%%%%%%%%%%%%%%%%%%%%%%%%%%%%%%%%%%%%%%%

%%%%%%%%%%%%%%%%%%%%%%%%%%%%%%%%%%%%%%%%%%%%%%%%%%%%%%%%%%%%%%%%%%%%%%%%%%%%%%%%
\subsection{Parameterization of observables and procedure}
\label{sec:fit_procedure}

The expressions of the corrected ${\cal F}t$ values and of the neutron lifetime
in terms of the couplings $C^{(\prime)}_i$ and of the nuclear matrix elements
$M_{F,GT}$ were given in~\sref{corrections}. For the correlation coefficients
the expressions can be found in Ref.~\cite{Jackson:1957auh}.
 Those expressions do not include the contribution of pseudo-scalar couplings.
For example, for $b$ this is given in Eq.~(\ref{eq:b}).

In contrast to the analyses presented in
Refs.~\cite{Severijns:2006dr,Konrad2010,Wauters:2013loa,Vos:2015eba},
we do not cast
systematically each measured correlation coefficient in an  
expression of the form $\tilde{X} = X/(1 + b \langle m_e/E_e \rangle)$ (\sref{diff-observables}). 
This concerns in particular the measurements of $a$ of Refs.~\cite{Johnson1963,Str78,Adelberger1999,Byr02,Gorelov2005,Vetter2008,Li2013}, which are obtained from differential observables, such as the recoil energy spectrum.  
For such observables, the measured distribution contains both $b$ and $a$ terms with different recoil momentum dependencies~\cite{Gonzalez-Alonso:2016jzm}.\footnote{The $\tilde{a}$ prescription does however hold if the coefficient is extracted for a fixed $\beta$ energy~\cite{Gonzalez-Alonso:2016jzm}, as in Ref.~\cite{Gri68}.} 
Most of these assume the corresponding Fierz terms to be zero or sufficiently small compared with the experimental uncertainty~\cite{Johnson1963,Str78,Byr02,Vetter2008,Li2013}, which we will show to be valid for the fits performed here. In Refs.~\cite{Adelberger1999,Gorelov2005} the extracted $a$ and $b$ combination (denoted $\hat{a}$ in Ref.~\cite{Gonzalez-Alonso:2016jzm}) happens to be very close to $\tilde{a}$. %, i.e., the values which multiply the Fierz term are only slightly different from the values of $\langle m_e/E_e \rangle$. 
For the sake of simplicity, we follow the notation of the original references and we use $\tilde{a}$, instead of $\hat{a}$, in~\tref{nuclear}.

The triple correlation parameters $D$ and $R$ are extracted from
measurements of asymmetries, usually combined in super-ratios. It is
then the parameters $\tilde{D}$ and $\tilde{R}$ that are extracted from
those measurements.
Such a distinction is however not important because the parameters
$D$ and $R$ are strongly suppressed in the SM resulting in negligible
sensitivity to the Fierz term.

We have used a $\chi^2$ minimization method in which the free
parameters are the couplings $C_i$ and $C'_i$ that enter the
expressions of the correlation coefficients or of the corrected
lifetimes. 
Combinations of the $C_i$ and $C'_i$ couplings
were also used as free parameters in the fits to illustrate specific
implications.
For some of the fits discussed below, the 2-dimensional plots of
constant $\chi^2$ are obtained by marginalizing over the remaining
parameters.
The same applies to the $1\sigma$ intervals given for each parameter. 
When the minimization results in only one minimum of the $\chi^2$, and the
distribution is approximately Gaussian in its vicinity, the values
of the coefficients are quoted with their $1\sigma$ uncertainty and the
correlation matrix is also given. 

%%%%%%%%%%%%%%%%%%%%%%%%%%%%%%%%%%%%%%%%%%%%%%%%%%%%%%%%%%%%%%%%%%%%%%%%%%%%%%%%
\subsection{Most general fit to ${\cal F}t$-values from superallowed Fermi decays}
\label{sec:fit-F}

In this section we keep all $C_i$ and $C_i$ coefficients ($i=S, V, T, A$)
as independent complex parameters. 
The fit to the ${\cal F}t$ values in \tref{ftFermi} can be performed
in a fully phenomenological parameterization following \eref{Ft-Fermi-TH}.
The two free parameters are then $\xi^{0^+ \rightarrow 0^+}$ and the
transition-independent combination $\gamma^{-1}\,b_F$, which are functions
of the vector and scalar couplings. The axial-vector and tensor couplings are
unconstrained since they do not contribute to Fermi transitions. The fit gives
\begin{alignat}{2}
\xi^{0^+ \rightarrow 0^+} & = \quad && 2.6450(18)\times10^{-10}~{\rm GeV}^{-4}~,\label{eq:fit51_chsi}\\
\gamma^{-1}\,b_F & = \quad && -0.0029(25)~, \label{eq:fit51_bF}
\end{alignat}
with a correlation coefficient $\rho=-0.94$ and $\chi_{\rm{min}}^2/\nu = 0.44$.
A sort of transition-independent ${\cal F}t_0$ value can be defined
through ${\cal F}t_i = {\cal F}t_0 / (1 + b \langle m_e/E_e \rangle )$,
with ${\cal F}t_0=3070.1\pm2.1$ s. Both this number and~\eref{fit51_bF}
are consistent with the results of Ref.~\cite{Hardy:2014qxa}. 

{\bf CKM unitarity bounds.} 
At the quark level, $\xi$ is a quadratic function of $V_{ud}$ and of the vector
and scalar BSM couplings $\epsilon_i$ and $\tilde{\epsilon}_i$. The $V_{us}$
and $V_{ub}$ values can be used to calculate $V_{ud}$ using CKM unitarity
(assuming a three-flavor structure), in such a way that the precise $\xi$
value obtained above becomes a stringent constraint on the non-standard
vector and axial couplings. Namely, at 90\% CL we find\footnote{This bound
is fully equivalent to the CKM-unitarity bound described in~\sref{CKMunitarity},
although $\Delta_{CKM}$ is not built in this case through the $\tilde{V}_{ud}$
element, but instead with $V_{ud}^{\rm exp}$ defined by
$\xi^{0^+ \rightarrow 0^+} =2\,(G_F V^{\rm exp}_{ud})^2(1+\Delta_R^V)$.}
\bea
|V_{ud}|^2 \bigg[ 2\,{\rm Re}(\eL+\eR) + |\teL+\teR|^2 + g_S^2\,|\eS|^2+
g_S^2\,|\teS|^2 - 2\,\frac{\delta G_F}{G_F} \bigg] + \delta_{us} + \delta_{ub} = 
-(0.1 \pm 1.4)\times10^{-3}~,
\label{eq:xi_ckm}
\eea
where $\delta_{us,ub}$ are possible non-standard effects affecting the
extraction of $V_{us}$ and $V_{ub}$. 
We neglected ${\cal O}\left( \epsilon_{L,R}^2, (\delta G_F)^2, \epsilon_i\,\delta G_F \right)$ terms,
which can be easily included but are expected to be very small.
We used $\tilde{V}_{us}=0.22408(87)$ obtained
when allowing for NP contributions~\cite{Gonzalez-Alonso:2016etj}, and
$\tilde{V}_{ub} =$~0.00409(39)~\cite{Patrignani:2016xqp}.
The RHS in~\eref{xi_ckm} becomes $-0.8(9)\times10^{-3}$ if $b_F=0$
and $-0.5(8)\times10^{-3}$ if in addition we use the PDG value of
$V_{us}$~\cite{Patrignani:2016xqp}, which effectively neglects non-$(V-A)$
new physics terms in kaon decays.

This result represents the strongest bound from $\beta$ decay data on
all real and imaginary couplings involved, except for Re($\eS$), which is
more strongly constrained from the $b_F$ bound in~\eref{fit51_bF}, as
shown in more detail in the next section. This illustrates the important
role of ${\cal F}t$-values from superallowed Fermi decays, not only
for the determination of the $V_{ud}$ matrix element in the SM model,
but also to constrain many types of non-standard interactions.

%%%%%%%%%%%%%%%%%%%%%%%%%%%%%%%%%%%%%%%%%%%%%%%%%%%%%%%%%%%%%%%%%%%%%%%%%%%%%%%%
\subsection{Fit without right-handed neutrinos}
\label{sec:fit-LH}

In this section we assume that there are no light right-handed neutrinos,
and thus $C'_i = C_i$ for all four couplings (i.e.\ $S, V, T, A$).
This scenario is particularly interesting for several reasons. First, it is
well motivated theoretically, including e.g. the SMEFT framework discussed
in~\sref{HEP}. Secondly, even if interactions with right-handed neutrinos
are present, the results from the fits will not change as long as the
couplings are not very large, since they enter quadratically in the observables.
Finally, due to the dominant linear dependence on the exotic couplings,
the extracted values are expected to be gaussianly distributed, what
simplifies the analysis and allows performing a general fit that can later
directly be applied to simpler scenarios, such as the SM. 
We will first assume all couplings to be real. There is then
a total of four parameters: $C_V$, $C_A$, $C_S$, and $C_T$. 
Such fits are termed ``left-handed'' for the neutrinos 
involved in any interaction. 
Under the assumptions above we have $C_A/C_V = (C_A + C'_A)/(C_V + C'_V)$,
$C_S/C_V = (C_S + C'_S)/(C_V + C'_V)$ and $C_T/C_A = (C_T + C'_T)/(C_A + C'_A)$.
To simplify the expressions of results from the fits, we use the short forms
of the ratios.
The fit including non-zero imaginary parts is presented at the end of this
section, by adding the relevant observables to the data set.

%%%%%%%%%%%%%%%%%%%%%%%%%%%%%%%%%%%%%%%%%%%%%%%%%%%%%%%%%%%%%%%%%%%%%%%%%%%%%%%
\subsubsection{${\cal F}t$-values from superallowed Fermi decays}
\label{sec:Ft_BSM}

{\bf General fit.}
The fit to the ${\cal F}t$ values in \tref{ftFermi} when all
${\cal CP}$-conserving
interactions involving left-handed neutrinos are allowed includes both the
vector and scalar couplings with $C'_V = C_V$ and $C_S = C'_S$.
In such a fit $|C_V|$ and $C_S / C_V$ were taken as free parameters. 
The expression of $C_V$ in terms of the underlying quark-level couplings is
given in \eref{CmatchingAt1loopV} and its value will
be given in units of $G_F/\sqrt{2}$. The main effect of the scalar coupling
on the ${\cal F}t$ values is through the Fierz term in~\eref{fit51_bF}, but
it also contributes quadratically through $\xi$ in~\eref{fit51_chsi}.
The results from such a fit are
\begin{alignat}{2}
|C_V| & = \quad && 0.98595(34) G_F/\sqrt{2}\label{eq:fit2_CV}~, \\
C_S/C_V & = \quad && 0.0014(12) \label{eq:fit2_CSCV}~,
\end{alignat}
with a correlation of $\rho=0.94$, and $\chi_{\rm{min}}^2/\nu = 0.44$. 
The constraint on the scalar coupling is consistent with the
result given in Ref.~\cite{Hardy:2014qxa} where the fit was performed separately
from the determination of the vector coupling. Such a procedure does
not enable to study the correlation between the parameters, which
appears to be very large in this case. 
The contours in~\fref{fit2_CVCS} around the minimum correspond
to values $\Delta \chi^2=1,4,9$,
so that the projections on the horizontal and vertical axes of the extreme
values of the three contours correspond to confidence intervals of
68\%, 95\% and 99\% for a single parameter.

\begin{figure}[!htb]
\begin{center}
\includegraphics[width=0.98\columnwidth]{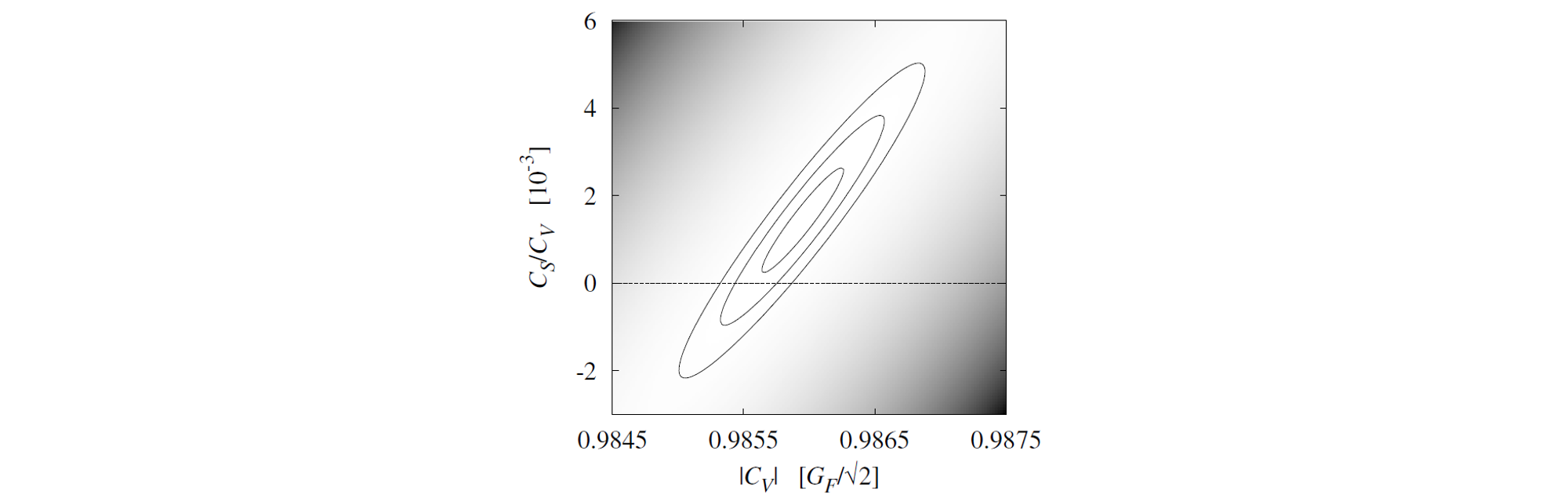}
\caption{
\label{fig:fit2_CVCS}
Results of the fit to ${\cal F}t(0^+\to 0^+)$ values assuming
${\cal CP}$-conserving interactions without right-handed neutrinos. 
The lines correspond to iso-$\chi^2$ levels with values $\Delta \chi^2=1,4,9$
around the minimum quoted in Eqs. (\ref{eq:fit2_CV}) and (\ref{eq:fit2_CSCV}).}
\end{center}
\end{figure}

%%%%%%%%%%
{\bf Standard Model fit.} 
The SM fit of the fourteen values in \tref{ftFermi}
is a particular case of the fit above for which $C_S = C'_S = 0$ (and still $C'_V = C_V$).
There is then only one free parameter, $|C_V|$,
which, under the assumptions above, corresponds to
$|V_{ud}|(1+\Delta^V_R)^{1/2}G_F/\sqrt{2}$.
The value resulting from such a fit is
\begin{equation}
|C_V|  = %0.985578(99)_{stat}(55)_{\delta'_R}\, G_F/\sqrt{2} = 
0.98558(11)\, G_F/\sqrt{2} ~,
\label{eq:fit01_CV}
\end{equation}
with $\chi_{\rm{min}}^2/\nu = 0.51$. The contribution of $\delta'_R$ to the total
uncertainty shown above is $5.5\times10^{-5}\, G_F/\sqrt{2} $. The significant reduction in the
error with respect to~\eref{fit2_CV} is a consequence of the large
correlation with the scalar coupling.   
The value of ${\cal F}t$ calculated a posteriori from
the result in \eref{fit01_CV},
is $\overline{{\cal F}t} = 3072.42(62)_{stat}(35)_{\delta'_R}$~s, consistent
with the result given in Ref.~\cite{Hardy:2014qxa}. 
Using this value for $\overline{{\cal F}t}$ in~\eref{Ft-n-num2} one obtains
the precise relation 
\bea
\tau_n\,\left( 1+  3\lambda^2 \right)	=	5172.6\pm 1.2 \,\rm{s} ~.
\label{eq:tau_n_From_Ft}
\eea
This shows that a precise determination of the neutron lifetime results in a
precise value of $|\lambda|$, as we will explicitly show in the fits below.
%
%%%%%%%%%%%%%%%%%%%%%%%%%%%%%%%%%%%%%%%%%%%%%%%%%%%%%%%%%%%%%%%%%%%%%%%%%%%%%%%
\subsubsection{Minimal left-handed fit: ${\cal F}t$-values and selected neutron data}
\label{sec:fitMin_LH}
{\bf General fit.} 
Next we consider the ${\cal F}t$ values from superallowed transitions
together with the average values of the neutron lifetime and of the
asymmetry parameter, $\tilde{A}_n$, listed in \tref{neutron},
as well as the parameter $\lambda_{AB}$.
This fit accesses two additional free parameters, $C_A/C_V$ and $C_T/C_A$, as
compared with the general fit of ${\cal F}t$ values above.
The results from such a fit, which contains 17 experimental
input data, are the following
\bea
\left(
\begin{array}{c}
 |C_V| \\
 C_A/C_V \\
 C_S/C_V \\
 C_T/C_A
\end{array}
\right)
=
\left(
\begin{array}{c}
 0.98595(34) \,G_F/\sqrt{2} \\
 -1.2728(17) \\
 0.0014(12) \\
 0.0020(22) \\
\end{array}
\right)
~~~{\rm with}~~~
\rho=
\begin{pmatrix}
1.00 \\
0.08 & 1.00 \\
0.94 & 0.08 & 1.00 \\
-0.32 & 0.85 & -0.31 & 1.00
\end{pmatrix}
~~~{\rm and}~~~
\chi_{\rm{min}}^2/\nu = 0.46~.
\label{eq:fit5}
\eea

The values of $|C_V|$ and $C_S/C_V$ and their correlation are identical to those 
obtained in the fit of the ${\cal F}t$ values alone, and thus \fref{fit2_CVCS}
still applies. 
The inclusion of the neutron data serves therefore for the determination of
$C_A/C_V$ and $C_T/C_A$. The constraint on the
tensor coupling appears to be a factor of about 2 weaker than the constraint
on the scalar coupling.
The contour plots for this fit are shown in \fref{fit5}.
The projections on the horizontal
and vertical axes of the extreme values of the three contours correspond
again to 68\%, 95\% and 99\% confidence intervals for a single
parameter, independently of the value of any other
parameter in the fit.
The covariance matrix in \eref{fit5} indicates
that the correlation between $|C_V|$ and $C_A/C_V$ is small
(left panel in \fref{fit5}) and that it is 4 times larger between $C_S/C_V$
and $C_T/C_A$ (middle panel in \fref{fit5}).
However, the matrix also shows that the correlation between
$C_A/C_V$ and $C_T/C_A$ is very large (right panel in \fref{fit5}). 
This means that, in such a general framework, improvements in the determination
of $C_A/C_V$ go along with improvements in constraining the tensor coupling.
The precise neutron lifetime fixes the width of the $C_A/C_V$-$C_T/C_A$ ellipse
along the strongly constrained direction. The asymmetry
parameter and $\lambda_{AB}$, which have larger relative uncertainties, fix the
width of the ellipse along the orthogonal direction and thus the projected bounds
for both $C_A/C_V$ and $C_T/C_A$. This also shows how the contour moves when
these inputs change, which happened in the recent past, as discussed below.

\begin{figure}[!htb]
\begin{center}
\includegraphics[width=0.98\columnwidth]{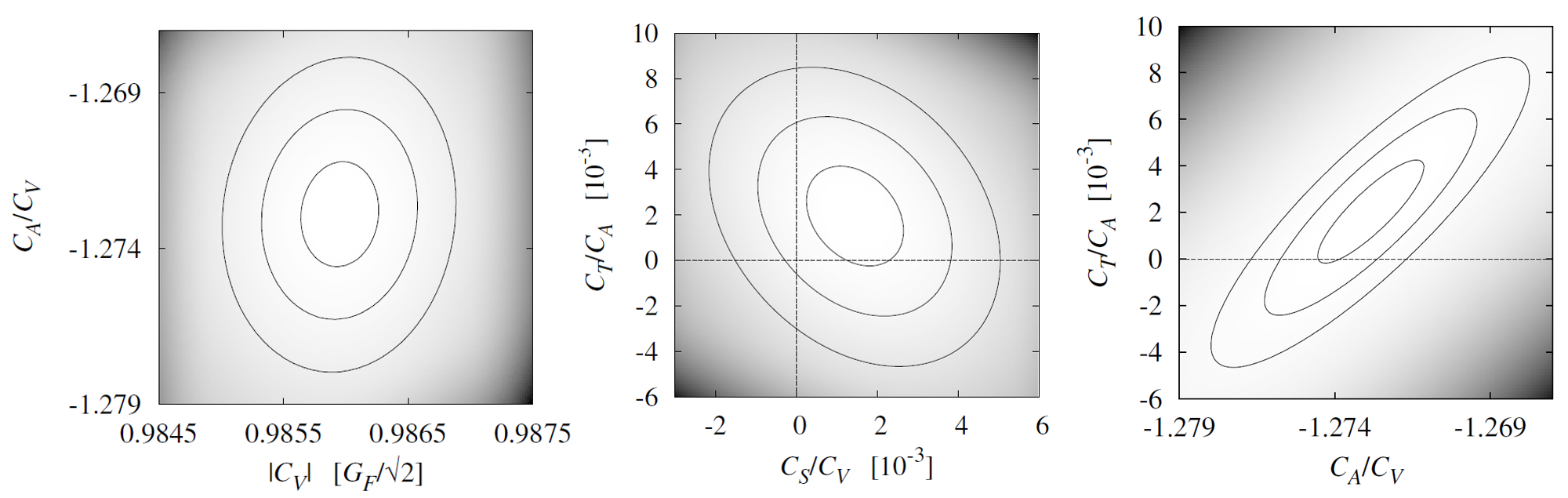}
\caption{
\label{fig:fit5}
Results from the fit to ${\cal F}t(0^+\to 0^+)$ values, $\tau_n$,
$\tilde{A}_n$ and $\lambda_{AB}$ assuming ${\cal CP}$-conserving interactions
without right-handed neutrinos. The lines correspond to iso-$\chi^2$ levels
with values $\Delta \chi^2 = 1,4,9$ around the minimum quoted in~\eref{fit5}.
Left panel: projection on the $(|C_V|,C_A/C_V)$ plane. Middle panel:
projection on the $(C_S/C_V,C_T/C_A)$ plane. Right panel: projection on
the $(C_A/C_V, C_T/C_A)$ plane.}
\end{center}
\end{figure}

\begin{figure}[!htb]
\begin{center}
\includegraphics[width=0.39\columnwidth]{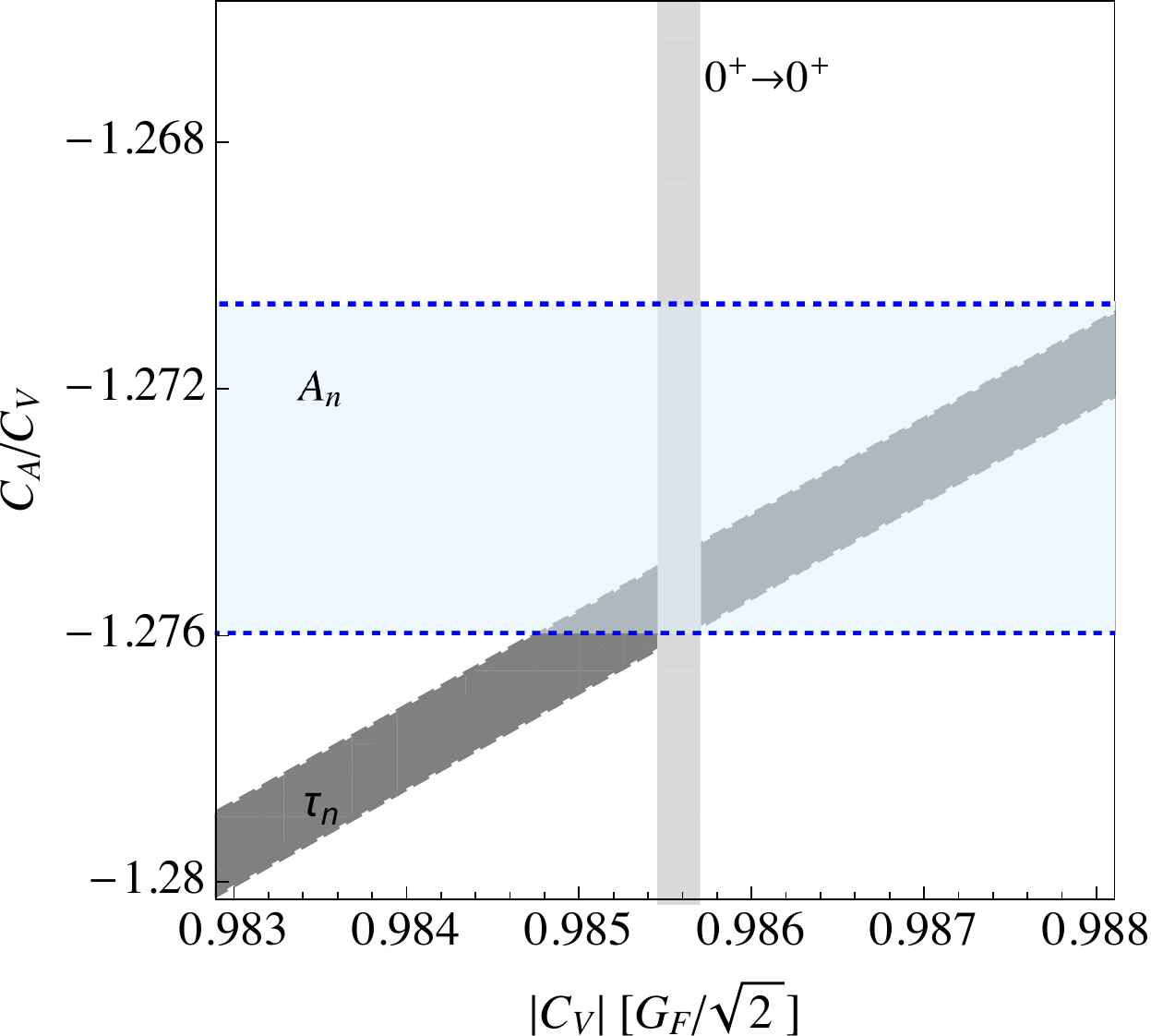}
\caption{
\label{fig:SMplot}
$1\sigma$ constraints on $|C_V|$ and $C_A/C_V $ from the ${\cal F}t$ values of $0^+\to 0^+$
decays, the neutron lifetime and the neutron $\beta$ asymmetry, when the SM is assumed.}
\end{center}
\end{figure}

%%%%%%%%%%
{\bf Standard Model fit.} 
This is a particular limit of the minimal fit above, in which
$C'_V = C_V$, $C'_A = C_A$, and $C'_S = C_S = C'_T = C_T = 0$.
There are two free parameters, $|C_V|$ and $C_A/C_V$. Such a fit leads to
\begin{alignat}{2}
|C_V| & = \quad && 0.98559(11) \,G_F/\sqrt{2} \label{eq:fit3_CV}~, \\
C_A/C_V & = \quad && -1.27510(66) \label{eq:fit3_CACV}~,
\end{alignat}
with $\rho=0.25$ and $\chi^2/\nu = 0.61$ at the minimum. 
Once again, the value of $|C_V|$ is almost identical to that obtained in the SM fit
of the ${\cal F}t$ values, Eq.~(\ref{eq:fit01_CV}), with significant reductions of
the errors relative to the general fit due to the above-discussed
large correlations. 
Figure \ref{fig:SMplot} shows the regions of parameter space probed by the dominant
inputs, and the good overall agreement among them. The region associated to the
$\lambda_{AB}$ measurement is not shown because it is less precise.

The central value of $C_A/C_V$ in \eref{fit3_CACV} has shifted about $7\sigma$
with respect to the results quoted in Ref.~\cite{Severijns:2006dr}, where the
value of the neutron lifetime obtained by Serebrov et al.~\cite{Serebrov2005} was not
included. Only the two oldest values of the neutron lifetime
in \tref{neut_averages} are common
between the present fits and those presented in Ref.~\cite{Severijns:2006dr},
and, in addition, those values have the lowest statistical weight.
The origin of the discrepancy is nothing but the change in the neutron
lifetime over the past decade. The central value was reduced by 6~s,
what represents about $7\sigma$ with the statistical uncertainty quoted
for the average value in \tref{neut_averages}.
This illustrates the fragility that has accompanied the neutron decay
data over time. 
We note that if the measurement of Serebrov et al.~\cite{Serebrov2005} would
have been included in Ref.~\cite{Severijns:2006dr} and the error would have
been scaled {\it \`a la} PDG, as done throughout the present work, it would
have produced a less precise but more reliable determination of $C_A/C_V$,
which would be in smaller tension with the current one.

%%%%%%%%%%%%%%%%%%%%%%%%%%%%%%%%%%%%%%%%%%%%%%%%%%%%%%%%%%%%%%%%%%%%%%%%%%%%%%%
\subsubsection{Full left-handed fit with ${\cal F}t$ values and all neutron and nuclear data}
\label{sec:fullFit_LH}

Under the same assumptions than the minimal fit, we now incorporate
the measurements of $a_n$, $\tilde{a}_n$ and $\tilde{B}_n$ from the neutron
(\tref{neutron})
as well as $a$, $\tilde{a}$, $\tilde{A}$ and $P_F/P_{GT}$ from nuclear
decays
(\tref{nuclear}).\footnote{
The bounds on the scalar and tensor couplings in~\eref{fit5} translate into per-mil level bounds on the Fierz terms $b_{F,GT}$, 
significantly below the sensitivity of Refs.~\cite{Johnson1963,Str78,Byr02,Vetter2008,Li2013}, which allows one to use their $a$ results (see~\sref{fit_procedure}).
} This results in a total of 28
input data. The outcome of the fit is very similar to the
minimal fit above, with a normalized $\chi^2$ at the minimum of $\chi^2/\nu = 0.65$. 
The inclusion of the three additional input data from the neutron and
the eight input data from nuclear decays reduces the statistical
uncertainty on the parameters by at most 10\%.

The origin of this moderate improvement resides in either the relatively
large experimental
uncertainties of the additional data or the weak sensitivity of the
parameters to the exotic couplings involving left-handed neutrinos.
The angular correlation parameter $a$ has a small sensitivity due to its
quadratic dependence. The parameter $\tilde{a}$ gets its sensitivity
through the Fierz term but the relative experimental uncertainty is about
an order of magnitude larger than the most precise measurement of
$\tilde{A}$. For $\tilde{B}$, which is linear in the couplings, it
has a suppressed sensitivity due an accidental cancellation with the
contribution from the Fierz term \cite{Bhattacharya:2011qm}.

However, the consistency of the results from the minimal fit with 
those derived from 
these additional less precise measurements is a
non-trivial test that should not be underestimated. First, it represents
a global consistency test of the entire theoretical framework used to
incorporate
the SM corrections and to describe NP effects in all $\beta$ decays using
a common EFT Lagrangian. It is also an important cross check from an
experimental point of view, taking into account the very different
systematic effects associated with each measurement. A clear and important
illustration of this is the agreement between the $V_{ud}$ values
derived in the
SM framework using neutron decay data, mirror transitions and superallowed
pure Fermi transitions. Despite the latter is much more precise, the former
provide non-trivial checks.

We use nevertheless the results from this full fit to calculate benchmark
values for some relevant correlation coefficients.
Using the values of $|C_V|$, $C_A/C_V$, $C_S/C_V$, and $C_T/C_A$ we proceed
then backwards and replace these in the expressions of those
coefficients and calculate the associated uncertainties,
%%% Begin Oscar
including correlations.
The results are listed in \tref{benchErrors}.
For $\tilde{A}_n$, $\tilde{a}_n$
and $\tilde{B}_n$ we have taken the same value for $\langle m_e/E_e \rangle = 0.568$.
The pure Fermi and Gamow-Teller parameters are obtained for $^{32}$Ar and
$^{6}$He with $\langle m_e/E_e \rangle =0.210$ and $0.286$ respectively.
We recall that the Fierz terms are transition independent, except for the
small Coulomb correction, see \eref{b}. 
The uncertainties listed in the second column provide then the level from
which a new measurement of the corresponding coefficient will have an impact
on the determination of some coupling involved in the fit.
The SM values (at $1\sigma$) were derived
using the results from Eqs.~(\ref{eq:fit3_CV}) and (\ref{eq:fit3_CACV}).
The value of $b_F$ is dominated by the ${\cal F}t$ values.
We see for example that a measurement of $b_{GT}$
at the $4\times10^{-3}$ level will contribute in improving
current constraints on tensor couplings. 

%%%%%%%%%%%%%%%%%%%%%%%%%%%%%%%%%%%%%%%%%%%%%%%%%%%%%%%%%%%%%%%%%%%%%%%%%%%%%%%%
\begin{table*}
\caption{ 
Current benchmark uncertainties for some correlation coefficients,
derived from the full left-handed fit of~\sref{fullFit_LH}.
For non-zero coefficients, the third column gives the relative uncertainty.
The SM values are derived using the results from
Eqs.~(\ref{eq:fit3_CV}) and (\ref{eq:fit3_CACV}).
See text for further details.}
\begin{center}
\begin{tabular}{ l@{\hspace{5mm}} c@{\hspace{5mm}} c@{\hspace{5mm}} c }
\hline\hline
Coefficient 			& Absolute uncertainty 	& Relative uncertainty	& SM value\\
\hline
$b_n$				&  $3.2\times 10^{-3}$	& 					& $0$ \\
$a_n$                           &  $4.7\times 10^{-4}$  & $4.4\times 10^{-3}$   & $-0.10648(19)$\\
$\tilde{a}_n$			&  $6.4\times 10^{-4}$	& $6.1\times 10^{-3}$	& $-0.10648(19)$\\
$A_n$                           &  $5.9\times 10^{-4}$  & $5.0\times 10^{-3}$   & $-0.11935(24)$\\
$\tilde{A}_n$                   &  $7.8\times 10^{-4}$  & $6.5\times 10^{-3}$   & $-0.11935(24)$\\
$\tilde{B}_n$			&  $1.2\times 10^{-4}$	& $1.2\times 10^{-4}$	& $0.98713(5)$\\
$b_{F}$				&  $2.3\times 10^{-3}$	& 					& $0$ \\
$b_{GT}$				&  $3.9\times 10^{-3}$	& 					& $0$ \\
$a_F$                           &  $6.4\times 10^{-6}$  & $6.4\times 10^{-6}$   & $1$\\
$\tilde{a}_F$			&  $4.7\times 10^{-4}$	& $4.7\times 10^{-4}$	& $1$\\
$a_{GT}$                        &  $4.0\times 10^{-6}$  & $1.2\times 10^{-5}$   & $-1/3$\\
$\tilde{a}_{GT}$		&  $3.7\times 10^{-4}$	& $1.1\times 10^{-3}$	& $-1/3$\\
\hline\hline
\end{tabular}
\end{center}
\label{tab:benchErrors}
\end{table*}
%%%%%%%%%%%%%%%%%%%%%%%%%%%%%%%%%%%%%%%%%%%%%%%%%%%%%%%%%%%%%%%%%%%%%%%%%%%%%%%%

\begin{figure}[!htb]
\begin{center}
\includegraphics[width=0.39\columnwidth]{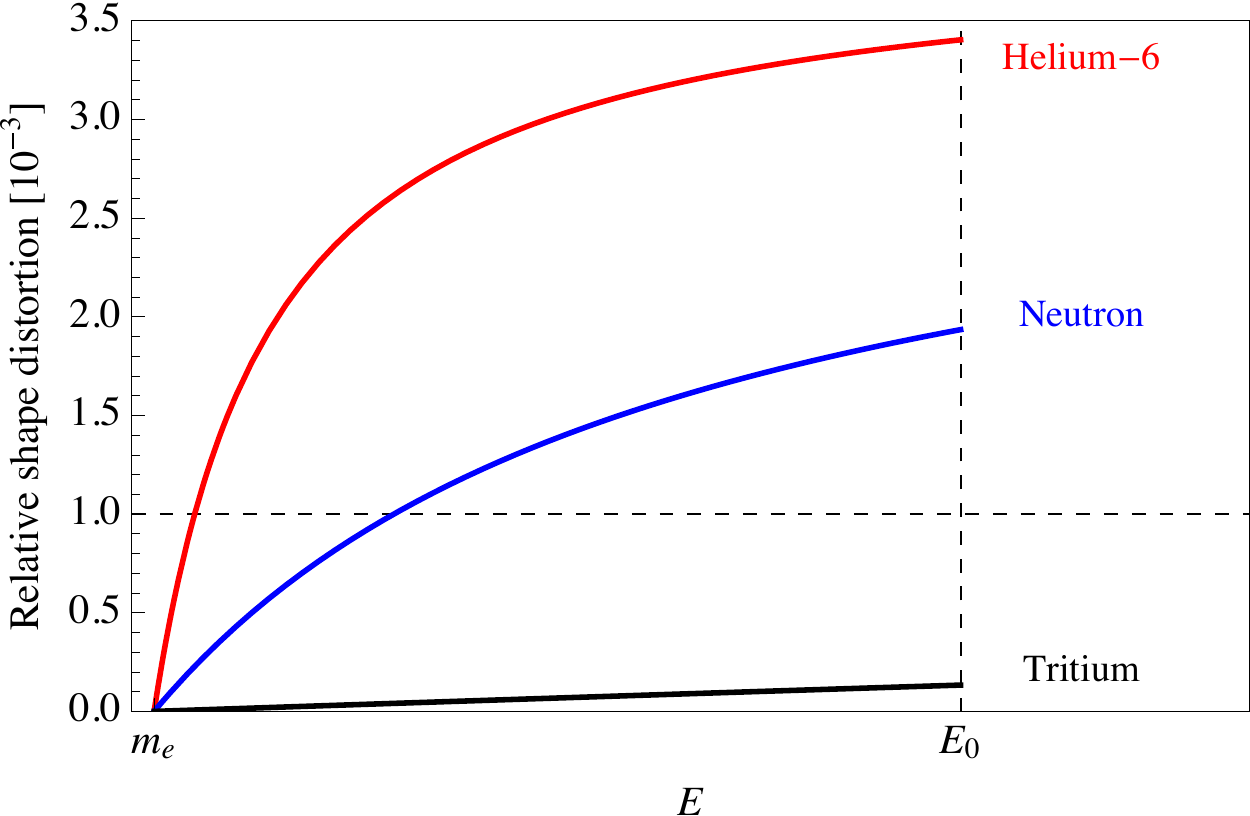}
\caption{
\label{fig:shapes}
Distortion to the $\beta$-energy spectrum in neutron, tritium and ${}^6$He decays,
associated with Fierz terms of the size of the benchmark uncertainties
quoted in \tref{benchErrors}.}
\end{center}
\end{figure}

These results can also be used to compare the precision required in
spectral shape measurements so that they provide competitive bounds.
Figure~\ref{fig:shapes} shows the distortions resulting in the $\beta$-energy
spectra, at $1\sigma$, for the neutron, tritium and ${}^6$He decays,
in absolute value.\footnote{For the tritium mixing ratio we use $\rho=-2.0951(20)$
from Ref.~\cite{Severijns:2008ep}. This value was calculated within the SM.
The mixing ratio gets however contaminated by NP which gives a subleading (quadratic)
effect in the plot.}
The effect in tritium decay is below $10^{-4}$, and thus inaccessible due to current theoretical uncertainties~\cite{Mertens:2014nha} (see also the discussion in~\sref{spectrum}).
% On the other hand, the effect in neutron and ${}^6$He is at the per-mil level, which is the precision expected in some ongoing and future experiments, as shown in~\tref{expSummary}. }

For $a_n$ and $A_n$, which are quadratic in the exotic couplings,
the uncertainty quoted in \tref{benchErrors} indicates the level of precision required
to improve the determination of $C_A/C_V$.

%%%%%%%%%%%%%%%%%%%%%%%%%%%%%%%%%%%%%%%%%%%%%%%%%%%%%%%%%%%%%%%%%%%%%%%%%%%%%%%
\subsubsection{Implications for particle physics}
\label{sec:SMFET_fit}

At the quark-level, the absence of light right-handed neutrinos
translates into $\tilde{\epsilon}_i=0$ in~\eref{leff-lowE}. As mentioned above,
this is assumed in the SMEFT framework introduced in~\sref{HEP}. Using
the matching equations in~\eref{matchLeeYang}, the hadronic charges given
in~\tref{charges} and the radiative correction $\Delta_R^V=2.361(38)\%$ introduced
in~\sref{corrections}, %we obtain
the results of the minimal fit in~\eref{fit5} can be written as
\bea
\left(
\begin{array}{c}
 |\tilde{V}_{ud}| \\
 \epsilon_R \\
 \epsilon_S \\
 \epsilon_T
\end{array}
\right)
=
\left(
\begin{array}{c}
 0.97452(34)(19) ~~~~~~~~~~~~~\\
 0.002(1)(21)~~~~~~\mbox{(90\% CL)}\\
 0.0014(20)(3) ~~~~~\mbox{(90\% CL)}\\
 -0.0007(12)(1) \,~~\mbox{(90\% CL)}\\
\end{array}
\right)
~~~{\rm with}~~~
\rho=
\begin{pmatrix}
1.00 \\
0.00 & 1.00 \\
0.83 & 0.00 & 1.00 \\
0.28 & -0.04 & 0.31 & 1.00
\end{pmatrix}
~~~{\rm and}~~~
\chi_{\rm{min}}^2/\nu = 0.46~.
\label{eq:fit08}
\eea
where the coefficients $\epsilon_i$ are evaluated at the renormalization scale $\mu=$2 GeV, 
and are approximated by $\epsilon_{R,S,T} (1+\eL+\eR)^{-1} \approx \epsilon_{R,S,T}$.
We quote here the $90\%$ CL limits (instead of the $1\sigma$ ones) to ease the comparison
with other bounds in the literature quoted in~\sref{alternativeprobes}.  
In~\eref{fit08} the first uncertainty is due to the uncertainties in the
$C_i$ coefficients whereas the second is associated with the uncertainties
in $\Delta_R^V$, $g_A$, $g_S$ and
$g_T$. For $\epsilon_{S,T}$ we neglect the uncertainty associated to two-body
corrections, which could give a similar size to the $g_{S,T}$ ones (\sref{nuclearmatrixelements}).

In the SM limit one can use instead the experimental results to extract $V_{ud}$ and
$\lambda = g_A/g_V$. The results are the following
\bea
|V_{ud}| &=& 0.97416(11)(19) = 0.97416(21) \label{eq:VudSM}~, \\
\lambda &=& 1.27510(66) \label{eq:lambda}~,
\eea
with a correlation of $\rho=-0.13$ and $\chi_{\rm{min}}^2/\nu = 0.57$.
Once again, the first uncertainty in~\eref{VudSM} comes from the uncertainty of $|C_V|$ 
whereas the second is associated with the uncertainty of $\Delta_R^V$. 
For comparison, the values of $|V_{ud}|$ obtained using only superallowed Fermi transitions
or neutron ($\tau_n, A_n, \lambda_{AB}$) decays are
\bea
|V^{0+\to 0^+}_{ud}| &=& 0.97415(21)~,\\
|V^{\rm neutron}_{ud}| &=& 0.9762(15)~.
\eea
The scale factors included in the neutron data have a very significant impact in this extraction. Indeed, if we use only bottle measurements of the neutron lifetime and the two most recent determinations of the $\beta$ asymmetry the error is two times smaller, namely $|V_{ud}| = 0.97413(82)$. We remind that the value extracted from mirror transitions is $|V^{\rm mirrors}_{ud}| = 0.9730(14)$, {\it cf.}~\eref{Vud-mirrors}.
%Let us note that using only superallowed Fermi decays we obtain $|V_{ud}| = 0.97414(21)$ whereas using only neutron data (lifetime, $\beta$ asymmetry and $\lambda_{AB}$) we find instead $|V_{ud}| = 0.9762(15)$.

The value of $\lambda$ obtained above is $3.5$ times more precise than the current value quoted in the PDG, $\lambda = 1.2723(23)$ \cite{Patrignani:2016xqp}, which was already mentioned in \sref{hadronic-charges}. The latter is obtained
from six values of the $\beta$ asymmetry parameter $A$
and the value of $\lambda_{AB}$. All these inputs or their updated
values have been included in the fit presented here. 
However, in this SM fit the absolute value of $\lambda$ is essentially determined by the precisely measured neutron
lifetime, Eq.~(\ref{eq:tau_n_From_Ft}), whereas $A$ or $\lambda_{AB}$
merely fix its sign. Removing $A$ or $\lambda_{AB}$ from
the set of input values increases the central value of $\lambda$ by about the same
amount, $\Delta\lambda = 0.00013$, with negligible impact on the statistical
uncertainty. 
We stress that the present determination of $\lambda$ does not involve any
additional assumption with respect to the PDG one. We assume here
the validity of the SM and we rescale errors to account for
the poor consistency between various measurements. Moreover, the
scaling factor of the parameter $A$ used by the PDG is larger than
that of the neutron lifetime (\tref{neut_averages}).

The values obtained in~\eref{fit08} or~\eref{VudSM} can be combined with
similar extractions of the $V_{us}$ and $V_{ub}$ elements to perform
a stringent CKM unitarity test, as discussed in~\sref{CKMunitarity}.
For illustration we show the 90\% CL result obtained using~\eref{VudSM}, i.e.
when scalar and tensor interactions are absent
\bea
\label{eq:DeltaCKMSM}
%\Delta_{CKM} = -0.46(41)_{V_{ud}}(27)_{V_{us}}\times 10^{-3} = -0.46(49)\times 10^{-3}~,
\Delta_{CKM} = -0.46(81)\times 10^{-3}~,
\eea
where we used the PDG values of $V_{us}$ and $V_{ub}$~\cite{Patrignani:2016xqp}.
The error decomposition is given by $\Delta_{CKM}= 
-0.46(68)_{V_{ud}}(45)_{V_{us}}(1)_{V_{ub}}\times 10^{-3}$.
This translates in the following 90\% CL constraint 
\bea
%\eL+\eR - 2\,\frac{\delta G_F}{G_F} = - 0.23(41)\times10^{-3}~,
2|V_{ud}|^2 \left( \eL+\eR - \frac{\delta G_F}{G_F} \right) + \delta_{us} + \delta_{ub} = - 0.46(81)\times10^{-3}~.
\label{eq:eLR_ckm}
\eea

{\bf Constraint on the pseudoscalar coupling.}
As shown in Eqs.~(\ref{eq:Plifetime})-(\ref{eq:Plifetime3}), the presence
of a pseudoscalar coupling generates an extra contribution to the neutron
lifetime. Since
this contribution enters like the Fierz term, it will also affect any
variable $\tilde{X}$ in neutron decay. 
Assuming that such a pseudoscalar contribution does not contribute
linearly to the expression of $A$ itself, the result obtained
for $C_T/C_A$ in \eref{fit08} can be translated into
\bea
\epsilon_P = - 0.08(15)~, \label{eq:fit08_CPCA}
\eea
at 90\% CL. To our knowledge, this is the first constraint on the
pseudoscalar coupling derived from neutron decay data.

{\bf Constraint on SMEFT Wilson coefficients.}
The bounds in~\eref{fit08} imply the following 90\% CL constraints on
the SMEFT coefficients at the renormalization scale of $\mu=1$ TeV
\bea
\left[w_{\varphi u d}\right]_{11}~~ &=& 0.004(41)~,\\
\left[w_{\ell equ} + 0.992\,V \, w_{\ell edq} - 0.165\,w_{\ell equ}^{(3)} \right]_{1111} &=& -0.0014(19)~,\\
\left[w_{\ell equ}^{(3)} -0.0046\, w_{\ell equ} \right]_{1111} &=& 0.0016(30)~,
\eea
where we simply used the SMEFT matching given in~\eref{matchingeqs},
as well as the running of the coefficients below and above the electroweak
scale described in~\sref{RGE}.
Finally~\eref{eLR_ckm} translates in the following 90\% CL constraint in
the flavor-universal limit described in~\sref{HEP} and~\eref{DeltaCKMsmeft}:
\bea
- w_{\varphi \ell}^{(3)} + w_{\varphi q}^{(3)}- w_{\ell q}^{(3)} + 2\,w_{\ell\ell}^{(3)} = - 0.23(41)\times10^{-3}~.
\eea

%%%%%%%%%%%%%%%%%%%%%%%%%%%%%%%%%%%%%%%%%%%%%%%%%%%%%%%%%%%%%%%%%%%%%%%%%%%%%%%%
\subsubsection{Projected sensitivities}

Given the fact that the fits carried out above are dominated by a reduced
set of measurements and considering the prospects of on-going experiments
discussed in \sref{exp}, it is interesting to estimate the impact those
measurements could have if they reach the claimed precision goals.

We have made a quantitative estimate of the impact
assuming that there will be no significant
improvements on the uncertainties of the set of ${\cal F}t$ values
and that the improvements in precision will mainly arise from the correlation
coefficients and the neutron lifetime. 
From the descriptions presented in \sref{exp} we have taken: (i) an uncertainty
of 0.1~s for the measurement of the neutron lifetime; (ii) a relative uncertainty of
0.1\% for the measurements of $\tilde{A}_n$ and $a_n$ in neutron decay
and of $\tilde{a}_F$ and $a_{GT}$ in nuclear decays; and (iii) an
absolute uncertainty of 0.001 in the measurement of $b_{GT}$ in nuclear decays. 
Once again, the pure Fermi and Gamow-Teller quantities are obtained for $^{32}$Ar 
and $^{6}$He.  
In this exercise we adopted arbitrary values for $\tau_n$, $\tilde{A}_n$ and $a_n$
that result in a consistent value of $C_A/C_V$ at the $1\sigma$ level, and SM
values for $\tilde{a}_F$, $a_{GT}$ and $b_{GT}$. As a consequence of that
choice, the central values of the fitted parameters are not relevant. 
For the left-handed fit with real couplings we obtain the following projected
uncertainties:
\bea
\left(
\begin{array}{c}
 \delta |C_V| \\
 \delta (C_A/C_V) \\
 \delta (C_S/C_V) \\
 \delta (C_T/C_A)
\end{array}
\right)
=
\left(
\begin{array}{c}
 1.9\,G_F/\sqrt{2} \\
 2.2 \\
 7.2 \\
 4.1 \\
\end{array}
\right)\times 10^{-4}~.
%
%~~~{\rm with}~~~
%
%\rho=
%\begin{pmatrix}
%1.00 \\
%0.66 & 1.00 \\
%0.84 & 0.58 & 1.00 \\
%-0.85 & -0.25 & -0.81 & 1.00
%\end{pmatrix}~.
\label{eq:fit41}
\eea

To translate these uncertainties to the quark-level parameters, we also assume
that the lattice calculation of the axial charge $g_A$ will reach the $0.5\%$
precision, which seems feasible looking at the preliminary result in
Ref.~\cite{Chang:2017oll}. For the remaining theory input
($\Delta_V^R$, $g_S$, $g_T$) we use their current values. We obtain
\bea
\left(
\begin{array}{c}
 \delta |\tilde{V}_{ud}| \\
 \delta \epsilon_R \\
 \delta \epsilon_S \\
 \delta \epsilon_T
\end{array}
\right)
=
\left(
\begin{array}{c}
 2.6 ~~~~~~~~~~~~~~~~\\
 41~~~~\mbox{(90\% CL)}\\
 12~~~~\mbox{(90\% CL)}\\
 2.2~~~\mbox{(90\% CL)}\\
\end{array}
\right)\times 10^{-4}
%
%~~~{\rm with}~~~
%
%\rho=
%\begin{pmatrix}
% 1.00 \\
% 0.02  &  1.00 \\
% 0.62  & 0.02 &  1.00 \\
% 0.61  & 0.01 & 0.78 & 1.00 \\
%\end{pmatrix}~.
\label{eq:fit42}~.
\eea
Finally, in the SM limit we find that $\lambda$ would be determined with
an absolute precision of $15\times 10^{-5}$, whereas the total error
of $V_{ud}$ would not change as it is dominated by the radiative
correction uncertainty. 
In this projected scenario, the $V_{ud}$ uncertainty using only neutron
data would be only about 15\% larger than the one from nuclear data.

Such an optimistic scenario is not necessarily unrealistic in view of
on-going experiments described in \sref{exp}. We see that the measurements
of those six parameters in neutron and nuclear decays, at the level assumed above,
would have an impressive impact in constraining exotic couplings at a level
not easily reachable by high-energy experiments, {\it cf.}~\sref{LHC searches}.

%%%%%%%%%%%%%%%%%%%%%%%%%%%%%%%%%%%%%%%%%%%%%%%%%%%%%%%%%%%%%%%%%%%%%%%%%%%%%%%%
\subsubsection{Minimal left-handed fit with real and imaginary couplings}
For clarity in the discussion, we have assumed so far all couplings
to be real. The extraction of the imaginary parts can be performed
simultaneously with the real parts but requires the inclusion of the
triple correlation coefficients $D$ and $R$.
Here we present a fit that assumes $C^\prime_i = C_i$ but where all couplings 
can have an imaginary part. The fit contains seven free parameters:
$|C_V|$, $\mbox{Re}(C_A/C_V)$, $\mbox{Im}(C_A/C_V)$, $\mbox{Re}(C_S/C_V)$,
$\mbox{Im}(C_S/C_V)$, $\mbox{Re}(C_T/C_A)$, and $\mbox{Im}(C_T/C_A)$.
The input data are those of the minimal fit, namely the ${\cal F}t$
values, $\tau_n$, $\tilde{A}$, $\lambda_{AB}$, with the addition of the
triple-correlation coefficients $R_n$ and $D_n$ from neutron decay and
$R$ from $^8$Li. 
The values obtained for the parameters are:
\bea
%\label{eq:}
\begin{aligned}
|C_V| & = 0.98592(39)\,G_F/\sqrt{2} \label{eq:fit31_CV} ~,\\
\mbox{Re}(C_A/C_V) & =  -1.2729(17) \label{eq:fit31_reCACV}~, \\
\mbox{Re}(C_S/C_V) & =  0.0014(12) \label{eq:fit31_reCSCV}~, \\
\mbox{Re}(C_T/C_A) & =  0.0020(22) \label{eq:fit31_reCTCA}~, \\
\end{aligned}
\qquad\qquad
\begin{aligned}
\\
\mbox{Im}(C_A/C_V) & = -0.00034(59) \label{eq:fit31_imCACV}~, \\
\mbox{Im}(C_S/C_V) & = -0.007(30) \label{eq:fit31_imCSCV}~, \\
\mbox{Im}(C_T/C_A) & =  0.0004(33) \label{eq:fit31_imCTCA}~,
\end{aligned}
\eea
with $\chi_{\rm{min}}^2/\nu = 0.46$. 
The constraint on $\mbox{Im}(C_A/C_V)$ improved by a factor of 3
compared to that quoted in the global fit of
Ref.~\cite{Severijns:2006dr},
owing to the improvement on the value of $D_n$ (\tref{neutron}).
The constraint on $\mbox{Im}(C_T/C_A)$ did
not improve since it is dominated by the measurement of $R$ in
$^8$Li decay. The main contribution of the measurement of $R_n$
is to provide a constraint on $\mbox{Im}(C_S/C_V)$ that was not available in
previous global fits. This bound is however not strong enough to neglect
its quadratic contribution to the ${\cal F}t$ values, which results
in a non-gaussian $\chi^2$ distribution.
A slightly tighter bound on $\mbox{Im}(C_S/C_V)$ can be obtained using
the CKM unitarity, {\it cf.}~\eref{xi_ckm}.
%
%%%%%%%%%%%%%%%%%%%%%%%%%%%%%%%%%%%%%%%%%%%%%%%%%%%%%%%%%%%%%%%%%%%%%%%%%%%%%%%%
\subsection{Fit with right-handed neutrinos}
\label{sec:fit-RH}

Fits involving left-handed neutrinos for the vector and axial-vector
interactions
and right-handed neutrinos for the scalar and tensor interactions are generally
termed ``right-handed'' fits. They correspond to $C'_V = C_V$, $C'_A = C_A$,
$C'_S = -C_S$ and $C'_T = -C_T$. The contribution of the Fierz term vanishes so that
the distinction of parameters between $X$ and $\tilde{X}$ becomes irrelevant. 
We will nevertheless maintain this distinction to indicate which
experimental data was used from the tables. 
Under the present assumptions we have $C_A/C_V = (C_A + C'_A)/(C_V + C'_V)$,
$C_S/C_V = (C_S - C'_S)/(C_V + C'_V)$ and $C_T/C_A = (C_T - C'_T)/(C_A + C'_A)$.
Again, to simplify the notation we use the short forms
of the ratios when quoting the results from the fits.
It has already been observed \cite{Severijns:2006dr} that such fits result in two
equivalent minima and that it is not possible to provide values and uncertainties for
parameters at a minimum in closed form, and to quote the covariance
matrix.

%%%%%%%%%%%%%%%%%%%%%%%%%%%%%%%%%%%%%%%%%%%%%%%%%%%%%%%%%%%%%%%%%%%%%%%%%%%%%%%
%\subsubsection{Minimal right-handed fit: ${\cal F}t$ values and selected neutron data}
%\label{sec:fit_RH_Min}

As a minimal right-handed fit, we consider first the set of 18 input data that
includes the ${\cal F}t$ values and
the neutron parameters $\tau_n$, $\tilde{A}_n$, $\lambda_{AB}$ and $\tilde{B}_n$.
The iso-$\chi^2$ contours of minimum $\chi^2$ obtained for this fit
on the $(|C_V|,C_A/C_V)$ and $(C_S/C_V,C_T/C_A)$ planes
are shown in the upper panels of \fref{fitsRH}. The extreme values
of the $(\chi^2_{\rm min} + 4)$ contours, which correspond to the $2\sigma$
level, are:
\begin{alignat}{2}
0.9799\,G_F/\sqrt{2} <& ~~~~ |C_V|  && < 0.9857\,G_F/\sqrt{2}  \label{eq:fit14_CV} ~,\\
-1.281 <& ~~C_A/C_V  && < -1.271 \label{eq:fit14_CACV} ~,\\
& ~~|C_S/C_V|  && < 0.107 \label{eq:fit14_CSCV} ~,\\
& ~~|C_T/C_A|  && < 0.081 \label{eq:fit14_CTCA} ~,
\end{alignat}
with $\chi^2/\nu = 0.76$ at the minimum.
It is seen (\fref{fitsRH} upper right panel) that for the selected data set,
the $(\chi^2_{\rm min} + 4)$ contour encloses both minima. 

\begin{figure}[!htbp]
\begin{center}
\includegraphics[width=0.95\textwidth,height=0.95\textheight,keepaspectratio]{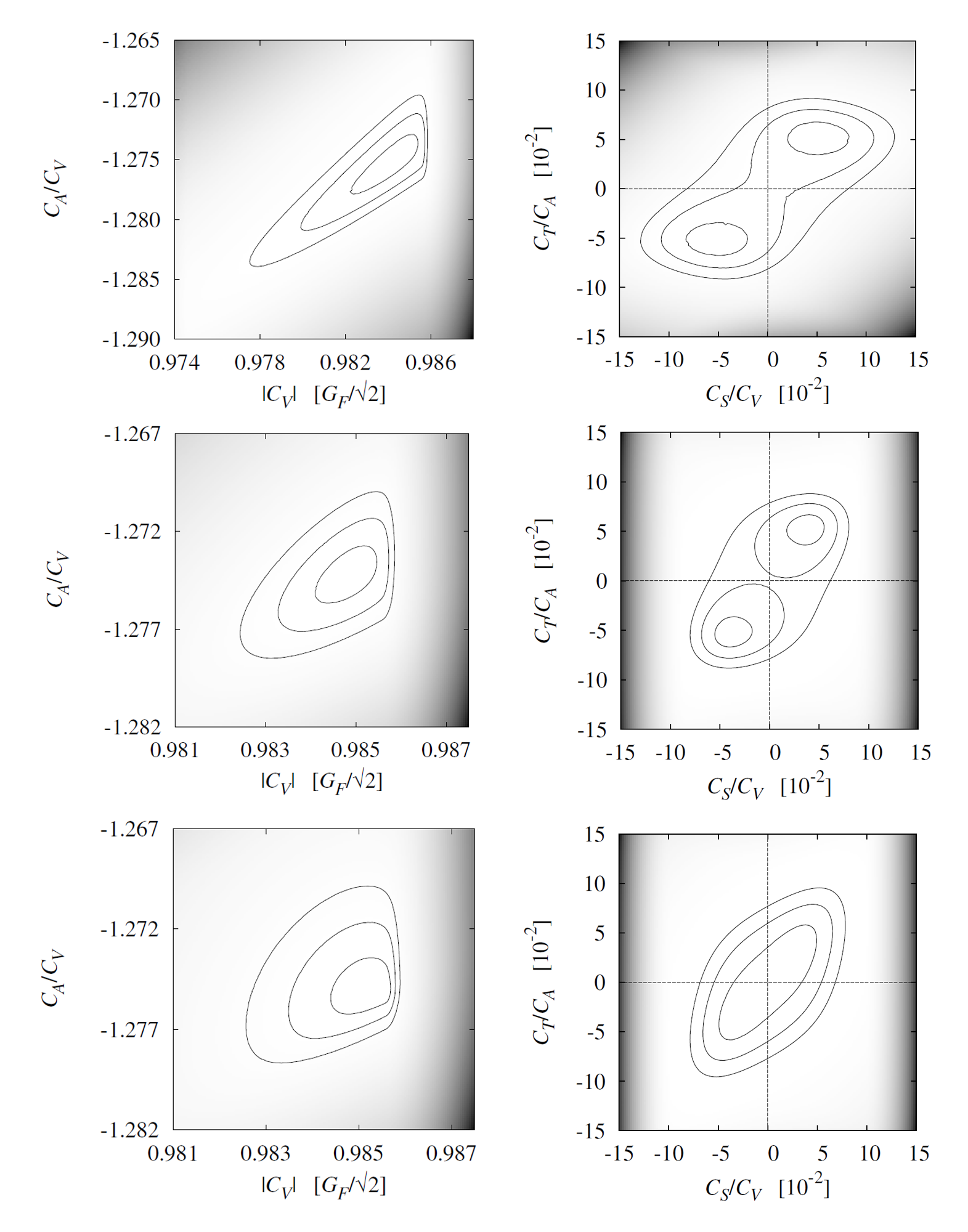}
\caption{
\label{fig:fitsRH}
Results from the fits discussed in \sref{fit-RH} with
scalar and tensor couplings that involve right-handed neutrinos.
The contours of constant $\chi^2$ correspond to iso-$\chi^2$ levels with values
$\Delta \chi^2=1,4,9$. The data used for the three fits are:
the ${\cal F}t(0^+\to 0^+)$ values and the neutron parameters $\tau_n$, $\tilde{A}_n$,
$\lambda_{AB}$ and
$\tilde{B}_n$ (upper panels); all nuclear and neutron data (middle panels);
all neutron and nuclear data except $\tilde{B}_n$ (lower panels).}
\end{center}
\end{figure}

%%%%%%%%%%%%%%%%%%%%%%%%%%%%%%%%%%%%%%%%%%%%%%%%%%%%%%%%%%%%%%%%%%%%%%%%%%%%%%%
%\subsubsection{Full fit with ${\cal F}t$ values and all neutron and nuclear data}
%\label{sec:fullFit_RH}

Next we consider a fit where, in addition to the data considered %in \sref{fit_RH_Min}, 
above, one includes also the other neutron parameters,
$a_n$ and $\tilde{a}_n$ and the nuclear inputs $a$, $\tilde{a}$, $\tilde{A}$
and $P_F/P_{GT}$, corresponding to a total of 28 input data. 
The iso-$\chi^2$ contours
are shown in the middle panels of \fref{fitsRH}. The extreme values
of the $(\chi^2_{\rm min} + 4)$ contours are:
\begin{alignat}{2}
0.9833\,G_F/\sqrt{2} <& ~~~~ |C_V|  && < 0.9857\,G_F/\sqrt{2}  \label{eq:fit9_CV} ~,\\
-1.2771 <& ~~C_A/C_V  && < -1.2713 \label{eq:fit9_CACV} ~,\\
& ~~|C_S/C_V|  && < 0.068 \label{eq:fit9_CSCV} ~,\\
0.003 <& ~~|C_T/C_A|  && < 0.078 \label{eq:fit9_CTCA}~,
\end{alignat}
with $\chi^2/\nu = 0.59$ at the minimum. 
The constraint on the scalar coupling, Eq.~(\ref{eq:fit9_CSCV}),
is reduced by a factor 1.6 compared to the fit with only the neutron data,
Eq.~(\ref{eq:fit14_CSCV}), whereas the upper limit on tensor couplings
changes slightly.
It is seen (middle panels of \fref{fitsRH}) that the projection of the
($\chi^2_{\rm min} + 4$) contour on the vertical axis is
marginally close to zero.

Under the assumptions of this fit, namely $C^\prime_S = -C_S$ and
$C^\prime_T = -C_T$, the asymmetry parameter in Gamow-Teller transitions
becomes totally insensitive to tensor couplings.
The scalar coupling is mainly improved due to the contribution of $a_F$
in $^{32}$Ar and $^{38m}$K decays. The comparison between the contours
on the left upper and middle panels of \fref{fitsRH} shows that
the additional input data has a significant impact in narrowing the
determination of $C_A/C_V$.
We stress here this point that went unnoticed in previous global fits.
Due to the correlation between $C_S/C_V$ and $C_A/C_V$, experiments that 
measure $a$ in pure Fermi transitions, and that so far were motivated to
search for the presence of scalar couplings, are also sensitive to
$C_A/C_V$ in such extended models.

The bounds on scalar and tensor interactions in Eqs.~(\ref{eq:fit9_CSCV})
and~(\ref{eq:fit9_CTCA}) can be rewritten as:
\bea
&|\teS|&< 0.063 ~(\mbox{90\% CL}) \label{eq:fit9_eS}~, \\
0.006< &|\teT|&< 0.024 ~(\mbox{90\% CL}) \label{eq:fit9_eT}~.
\eea%
An about two times stronger bound on $\teS$ is
obtained using CKM unitarity, {\it cf.}~\eref{xi_ckm}. This has been
unnoticed in the literature so far.

%
%%%%%%%%%%%%%%%%%%%%%%%%%%%%%%%%%%%%%%%%%%%%%%%%%%%%%%%%%%%%%%%%%%%%%%%%%%%%%%%
%\subsubsection{Fit without the neutrino asymmetry parameter}
%\label{sec:fit_RH_noB}

It is interesting to investigate which input data leads
to the two minima in the projection of the hyper-surfaces onto
the $(C_S/C_V,C_T/C_A)$ planes, observed in the right upper and
middle panels of
Fig.~\ref{fig:fitsRH}. The two minima are well
separated by the ($\chi^2_{\rm min} + 1$) contours and, depending on
the inclusion of other data, can also remain separated by the
($\chi^2_{\rm min} + 4$) contours (right middle panel in \fref{fitsRH}).
After some
inspection, it was noticed that the neutrino-asymmetry parameter
$\tilde{B}_n$ is the sole culprit behind the tension between
these two minima. Removing only this single input from the data
used in the last fit %of \sref{fullFit_RH} 
results in the
iso-$\chi^2$ contours shown in the lower panels of \fref{fitsRH}.
Despite the difference in shapes, the extreme
values of the $(\chi^2_{\rm min} + 4)$ contours are
very similar to those of Eqs.~(\ref{eq:fit9_CV})-(\ref{eq:fit9_CTCA}),
except that the tensor coupling has here no lower bound.

Even if the internal consistency among the values of $\tilde{B}_n$ in
\tref{neut_averages} is satisfactory, its consistency with the other
neutron decay parameters within the SM is poor, as briefly discussed
in~\sref{recoil}. 
Although this parameter is less sensitive to the extraction of
$\lambda$, the PDG does not include any of the experimental inputs
to this end.

We note that improving the above-given bounds on scalar and tensor
couplings is much harder than for the LH fit, since the dependence
of the observables on them is quadratic.

%%%%%%%%%%%%%%%%%%%%%%%%%%%%%%%%%%%%%%%%%%%%%%%%%%%%%%%%%%%%%%%%%%%%%%%%%%%%%%%

\section{Alternative probes}
\label{sec:alternativeprobes}

We discuss in this section other observables that are expected to be sensitive to nonstandard interactions present in $\beta$ decay. For pion decay we can use the quark-level effective Lagrangian in~\eref{leff-lowE} without involving further assumptions. For the other observables, the SMEFT framework is used to analyze the degree of complementarity.
\subsection{Pion decay processes}
\subsubsection{Leptonic pion decay}
\label{sec:leptonicpiondecay}
The $\pi_{e2}$ process, $\pi\to e\nu_e(\gamma)$, is sensitive at tree-level to axial-vector and especially to pseudo-scalar interactions, which receive a large chiral enhancement \cite{Gonzalez-Alonso:2016etj}
\bea
\label{eq:pie2}
\Gamma_{\pi_{e2}} \propto f_\pi^2\, G^2_F \,\tilde{V}^2_{ud}  \left[ 1 - 4\,{\rm Re}\,\eR - \frac{2m_\pi^2}{m_e(m_u+m_d)}{\rm Re}\,\eP + {\cal O}(\epsilon^2) \right]~,
\eea
where we omitted various SM factors that are not relevant for the discussion below.
In a two-body decay, the rate represents the only available observable, and thus it is clear that only this particular combination of Wilson coefficients can be extracted from $\pi_{e2}$, once it is combined with another $d\to u$ transition so that the $G_F\,\tilde{V}_{ud}$ dependence cancels. $\beta$ decay provides the additional input to separate the various coefficients, via its unique handle on $\eR$ through the determination of $g_A$, {\it cf.} \eref{gA} \cite{Gonzalez-Alonso:2016etj}.

The large chiral enhancement combined with a cancellation of theoretical and experimental effects, makes the ratio $\pi_{e2}/\pi_{\mu2}$ a particularly powerful probe, sensitive to a pseudo-scalar coupling of order $10^{-7}$, which corresponds to an effective NP scale of hundreds of TeV. However, this ratio involves also the muonic couplings, and possible cancellations between the contributions in the numerator and the denominator relax this bound.\footnote{These cancellations happen naturally in models where the $\eP$ coupling follows the hierarchy of the Yukawa matrices, e.g. in models with an extended Higgs sector~\cite{Haber:1978jt,Herczeg:2001vk}. However, for this very same reason, in these models the BSM effect is extremely suppressed in $\beta$ decays, as discussed in~\sref{specificmodels}.} Allowing for such cancellations, a global fit found \cite{Gonzalez-Alonso:2016etj}
\bea
{\rm Re}\,\eP=(0.4\pm 1.3)\times 10^{-5}\hspace{0.5cm}\text{(90\% C.L.)}~,
\eea
which is highly correlated with other NP parameters in that fit. 
This bound can be further relaxed introducing non-negligible quadratic NP terms that could cancel the linear contributions shown in \eref{pie2}. In the simplest scenario this role is played by the $\eP^2$ term, which displays a large helicity-enhancement too. This allows for a non-zero $\eP$ value compatible with pion decay bounds, namely $\eP\approx 2\,m_e(m_u+m_d)/m_\pi^2\sim4\times 10^{-4}$. In order to discard this possibility one would need an independent $\eP$ probe such as $\beta$ decay~\cite{Gonzalez-Alonso:2013ura}. 
The current bound in~\eref{fit08_CPCA} is however three orders of magnitudes too weak for this. 
If we allow for a complex $\eP$ coefficient the two-fold degeneracy becomes circular in the $\eP$ complex plane~\cite{Herczeg:1994ur,Bhattacharya:2011qm}. The same holds for any chiral-enhanced not-interfering term,
e.g. operators with light right-handed ($\teP$) or ``wrong-flavor'' neutrinos ($\eP^{e\mu}, \eP^{e\tau}$)~\cite{Herczeg:1994ur,Bhattacharya:2011qm,Abada:2012mc,deGouvea:2015euy}.
The bounds on them are~\cite{Bhattacharya:2011qm}
\bea
\label{eq:teP}
{\rm Im}\,\eP, |\teP|, |\eP^{e\alpha}| < 2.8\times 10^{-4}~ (\alpha\ne e)\hspace{0.5cm}\text{(90\% C.L.)}.
\eea

As discussed in~\sref{specificmodels}, NP models generating scalar and tensor interactions can easily generate at the same time pseudoscalar interactions of similar magnitude. Moreover, even if the NP theory generates at tree-level only the pseudoscalar interaction, the other chirality-violating interactions will be generated radiatively. Thus, the strong bounds on the $\eP,\teP$ couplings generate nontrivial bounds on tensor and scalar interactions through operator mixing (\sref{RGE}). The larger the running performed, the more stringent is the resulting bound. In specific scenarios where the above-discussed cancellations do not occur, this provides the strongest bounds on $\epsilon_{S,T}$ and $\tilde{\epsilon}_{S,T}$, but in a general setup %where the cancellations are allowed,
the bounds are significantly relaxed. For a running up to $\Lambda\sim{\cal O}(10\,\rm{TeV})$ the following 90\% CL bounds are approximately found~\cite{Bhattacharya:2011qm,Cirigliano:2012ab}
\bea
|{\rm Re}\,\eT-0.009\,{\rm Re}\,\eS| &\lesssim& 0.9\times 10^{-3}~,\\
|{\rm Re}\,\teT+0.014\,{\rm Re}\,\teS| &\lesssim& 0.7\times 10^{-3}~,
\eea
where the larger sensitivity to tensor interactions is due to the large mixing, especially in the electroweak running above $\mu=m_Z$, {\it cf.}, \eref{RGEsmeft}. Likewise, strong bounds can be obtained for the imaginary parts.

Concerning the $\eR$ coupling, its contribution cancels in the $e/\mu$ ratio because $\eR$ is lepton-independent in the SMEFT framework, {\it cf.}~\eref{matchingeqs}. There is however a sub-percent level sensitivity to $\eR$ through the comparison of the individual (axial-mediated) $\pi\to \ell\nu_\ell$ rate and (vector-mediated) nuclear superallowed decays, due to recent precise LQCD calculations of the pion decay constant and other radiative corrections.\footnote{More precisely, this sub-percent level $\eR$ sensitivity comes mainly from the ratio $\Gamma(\pi\to\mu\nu)/\Gamma(K\to\mu\nu)$ due to a more accurate SM prediction. Once again this requires the use of $V_{ud,us}$ from vector-mediated processes, and it involves many other Wilson coefficients.} However, as mentioned above, in a global fit one needs an extra handle in order to disentangle its effect from the pseudoscalar one, {\it cf.}~\eref{pie2}. 

Finally, the $\tilde{V}_{ud}$ value extracted from $\pi\to e\nu$ provides an important sensitivity to $\eL$ through a CKM unitarity test. This will however not provide a competitive bound, since it is well known that the sensitivity to $V_{ud}$ of this process is much smaller than in superallowed nuclear decays.

\subsubsection{Radiative pion decay}
The kinematic distribution of radiative pion decay is sensitive to the same nonstandard tensor interaction that is probed in $\beta$ decay~\cite{Poblaguev:1990tv,Chizhov:1995wp}. For this process the hadronization is driven, at very low momentum transfer, by the $f_T$ constant defined by~\cite{Mateu:2007tr}
\bea
\langle \gamma|\bar u\sigma^{\mu\nu}\gamma_5d|\pi^-\rangle = \frac{e}{2}\,f_T \,\left(\epsilon^\mu k^\nu-\epsilon^\nu k^\mu\right)~, \label{eq:radff}
\eea
where $k$ and $\epsilon$ are the photon momentum and polarization respectively, and $e$ is the electric elementary charge.
Analyzing the Dalitz plot of this decay, the PIBETA collaboration obtained a stringent constraint on the real part of the product $\eT f_T$~\cite{Bychkov:2008ws}, which combined with the $f_T$ estimate obtained in Ref.~\cite{Mateu:2007tr} gives~\cite{Bhattacharya:2011qm}
\begin{align}
-1.2\times10^{-3}\leq {\rm Re}\,\eT\leq1.4\times10^{-3}\hspace{0.5cm}\text{(90\% C.L.)},\label{eq:eTboundrad}
\end{align}
which is comparable to the bound from $\beta$ decay of~\eref{fit08}.

\subsubsection{Pion $\beta$ decay}
An obvious hadronic process governed by the same underlying quark-level interaction as nuclear $\beta$ decay is the so-called pion $\beta$ decay, $\pi^{+} \rightarrow \pi^{0} e^{+} \nu$. This is a pure vector transition between two spin-zero members of the pion isospin triplet. It is analogous to superallowed Fermi transitions in nuclear $\beta$ decay. The difficulty resides in the extremely low branching ratio, i.e.\ $BR(\pi_\beta) \sim 10^{-8}$. 
In 2004 PIBETA published the latest and most precise measurement of the ratio $BR(\pi_\beta)/BR(\pi_{e2})$, carried out at the Paul Scherrer Institute~\cite{Frlez:2003vg,Pocanic:2003pf}. Multiplied by the experimental value of $BR(\pi_{e2})$~\cite{Patrignani:2016xqp}, they found $BR(\pi_\beta) = 1.036(4)_{\rm stat}(4)_{\rm syst}(3)_{\pi_{\rm e2}} \times 10^{-8}$~\cite{Pocanic:2003pf}. To further reduce the error one can calculate $BR(\pi_{e2})$ as the product of the experimental value of $BR(\pi_{\mu 2})$ multiplied by the precise theoretical prediction of the $\pi_{e2}/\pi_{\mu2}$ ratio~\cite{Cirigliano:2007xi}, finding $BR(\pi_\beta) = 1.040(4)_{\rm stat}(4)_{\rm syst} \times 10^{-8}$. This branching ratio can be used as another (currently less precise) method to determine $V_{ud}$ neglecting nonstandard contributions, finding $|V_{ud}| = 0.9748(25)$. 
This value is consistent with the one obtained in the fit of \sref{SMFET_fit} (dominated by pure Fermi transitions), $|V_{ud}| = 0.97416(21)$, {\it cf.}~\eref{VudSM}, but a factor about 13 less precise, and also a factor of 2 less precise than the value from mirror $\beta$ transitions or neutron decay. %, {\it cf.}~\eref{Vud-mirrors}.
In view of the extremely small value of the branching ratio $BR(\pi_\beta)$ it will not be an easy task to improve on this result.

From a somewhat different perspective, the process described above is equivalent to use the PIBETA measurement to extract the value of the pion decay constant $f_\pi$ with better precision than current lattice QCD calculations, which can then be used to extract $V_{ud}$ from $BR(\pi_{e2})$.

One might wonder about the BSM effects that pion $\beta$ decay is sensitive to. First, the linear contribution from the tensor term $\eT$ is suppressed by the $m_e/m_\pi$ ratio, and the scalar term $\eS$ is hidden inside the SM scalar form factor, which is itself negligible in this process. This can be seen, e.g., in the $K_{\ell 3}$ expressions of Ref.~\cite{Gonzalez-Alonso:2016etj} doing the proper replacement to describe pion $\beta$ decay. All in all, this removes any sensitivity to chirality-changing operators. On the other hand, contribution of the vector coupling, $\eL+\eR$, to pion $\beta$ decay can be re-absorbed in $\tilde{V}_{ud}$ as in the superallowed transitions, {\it cf.}~\eref{Vtilde}. Since the latter provide a much more precise $\tilde{V}_{ud}$ determination, the situation is the same as in the SM case, and pion $\beta$ decay measurements cannot provide competitive constraints.

\subsection{LEP and other electroweak precision measurements}

An important feature of $\beta$ decay is its very low energy, which makes the observables sensitive to the interference between the SM chirality-conserving and nonstandard chirality-violating interactions, which scales with the ratio $m_e/E$. For experiments with much larger energies, such as LEP, this is not the case, and the linear sensitivity to chirality-violating interactions is lost, as well as to right-handed vector ones. On the other hand they are still sensitive to the interference of the SM term with chirality-conserving operators, which makes possible to set strong bounds on them, or at least on certain linear combinations. This was explicitly shown in the first global SMEFT analysis of LEP and other precision electroweak observables, assuming a $U(3)^5$-invariant structure \cite{Han:2004az}. This analysis was recently extended to the flavor-general case~\cite{Efrati:2015eaa,Falkowski:2015krw,Falkowski:2017pss}.

For the sake of simplicity and due to the insensitivity of LEP and other precision measurements to chirality-violating operators, it is useful to work in the $U(3)^5$-invariant SMEFT, where only chirality-conserving operators are present.
As discussed in \sref{HEP}, in this case the only nonstandard effect in $\beta$ decay is an apparent violation of CKM unitarity induced by the following combination of operators:
\bea
%\label{eq:DeltaCKMsmeft}
\Delta_{\rm CKM} = 2~\left( - w_{\varphi \ell}^{(3)} + w_{\varphi q}^{(3)}- w_{\ell q}^{(3)} + 2\,w_{\ell\ell}^{(3)}        \right)~.\nonumber
\eea
In order to assess the competitiveness of this CKM unitarity test, the bounds on these operators were studied \cite{Cirigliano:2009wk}
from a long list of electroweak precision observables, including most importantly LEP and atomic parity violation data. Their sensitivity is limited by the $w_{\ell q}^{(3)}$ coefficient, which cannot be probed by the precise $Z$-pole measurements. If the corresponding operator is the only one present one finds
\bea
w_{\ell q}^{(3)} = +2.5(1.7)\times 10^{-3}~~\to~~\Delta_{\rm CKM} =-5.0(3.4)\times 10^{-3}\hspace{0.5cm}\text{(90\% C.L.)}~,
\eea
which arises mainly from  $e^+ e^- \to q \bar{q}$ cross-section measurements at LEP2. This indirect bound is %, as we will see in \sref{},
approximately 4 times weaker than the direct bound on $\Delta_{CKM}$ from muon, $\beta$ and kaon decay data, {\it cf.}~\eref{DeltaCKMSM}. 
%from muon and $\beta$ decay data. 
%
This conclusion is of course enhanced in more general scenarios with all operators present at the same time~\cite{Cirigliano:2009wk}, with a general flavor structure~\cite{Falkowski:2017pss} or if a non-linear EFT framework is used~\cite{Brivio:2016fzo}, since more operators have to be considered, which relaxes significantly the indirect bound on $\Delta_{\rm CKM}$. The opposite can also happen in specific BSM scenarios, where the sensitivity to nonstandard effects of the CKM-unitarity test can be much smaller than that of other probes. In the SMEFT framework this means that the four operators in~\eref{DeltaCKMsmeft} are either suppressed in those models or their contributions cancel to some extent.

\subsection{LHC searches}
\label{sec:LHC searches}

As mentioned in the previous section, the SMEFT Wilson coefficients contributing to $\eR,\eS,\eP$, and $\eT$ are not accessible by LEP searches at order $v^2/\Lambda^2$ in the SMEFT expansion due to their chiral structure. On the other hand, the LHC can still provide stringent limits, due to the ${\cal O}(s^2/v^4)$ enhancement of the quadratic term, related to their contact-interaction nature~\cite{Bhattacharya:2011qm,Cirigliano:2012ab,Alioli:2017ces}. The same holds for $\eLc$ which does interfere with the SM, $\teR,\teS,\teP$ and $\teT$~\cite{Cirigliano:2012ab}.

The natural channel to study at the LHC is $pp\to e+ {\rm MET}+X$ (where MET denotes missing transverse energy), since the underlying partonic process is the same as in $\beta$ decay. % ($\bar{u}d\to e\bar{\nu}$). % and so we expect it to be sensitive to the same kind of NP.
This process was studied within the SMEFT in Refs.~\cite{Bhattacharya:2011qm,Cirigliano:2012ab}, and later updated in Ref.~\cite{Gonzalez-Alonso:2013uqa}.\footnote{
The results in Fig.~28 (left panel) of Ref.~\cite{Bhattacharya:2016zcn} are not actual constraints, but the projections calculated in Ref.~\cite{Cirigliano:2012ab} for 8 TeV data. We note also that the disagreement between the LHC projections shown in Fig.~28 (right panel) of Ref.~\cite{Bhattacharya:2016zcn} and Fig.~11 in Ref.~\cite{Bhattacharya:2011qm} are simply due to a numerical error in the production of the former figure~\cite{Cirigliano:private}.}
The extracted bounds on the SMEFT coefficients $w_i$ can then be translated into bounds on the low-energy coefficients $\epsilon_i$ and $\tilde{\epsilon}$ using the matching conditions in~\eref{matchingeqs}. Equivalently one can directly express the collider observables in terms of the coefficients of the low-energy theory to make the connection more explicit, but keeping in mind that the SMEFT is the valid theory at LHC scales. The cross-section $\sigma(pp\to e+{\rm MET}+X)$ with transverse mass higher than $\overline{m}_T$ takes the following form:
\bea
\label{eq:sigmamt}
\sigma(m_T \!\!>\! \overline{m}_T) &=& \sigma_W \Big[ \Big| 1 +   \eLv \Big|^2+  |\teL|^2  + | \eR|^2 \Big] - 2 \, \sigma_{WL}\, \mbox{Re} \left( \eLc  +  \eLc \eLv^* \right) +~ \sigma_R  \Big[ |\teR|^2 \!+ |\eLc|^2 \Big]
\nonumber\\
&& + \,\sigma_S \Big[ |\eS|^2  \!+  |\teS|^2  \!+ |\eP|^2 \!+  |\teP|^2   \Big] +  \sigma_T  \Big[ |\eT|^2  \!+  |\teT|^2     \Big]~,
\eea
where $\sigma_W (\overline{m}_T)$ represents the SM contribution and $\sigma_{WL, R, S,T}(\overline{m}_T)$ are new functions, explicitly given in Ref.~\cite{Cirigliano:2012ab}. They contain the ${\cal O}(s/v^2,s^2/v^4)$ enhancement that makes them several orders of magnitudes larger than the SM contribution, which compensates for the smallness of the NP couplings and makes it possible to set tight bounds on them from these searches. On the other hand, the lack of such an enhancement makes this search not so sensitive to $\eLv$, $\teL$ and $\eR$. It is worth noting also that high-energy searches probe separately the vertex correction $\eLv$ and contact $\eLc$ contributions to the coupling $\eL$.

The latest bounds using this channel were obtained in Ref.~\cite{Gonzalez-Alonso:2013uqa} from 20 fb$^{-1}$ of data recorded at $\sqrt{s}=$ 8 TeV by the CMS collaboration ~\cite{Khachatryan:2014tva}. One single event was found with a transverse mass above $1.5$ TeV, to be compared with the SM expectation of $2.02(26)$ events. This absence of an excess of high-$m_T$ events in this channel translates into the following 90\% C.L. bounds~\cite{Gonzalez-Alonso:2013uqa}
\bea
\label{eq:CCbounds1}
&&|\epsilon_{S,P}|, |\tilde{\epsilon}_{S,P}| < 5.8\times 10^{-3} ~,\\
\label{eq:CCbounds2}
&&|\eT|, |\teT| < 1.3\times 10^{-3}~,\\
&&|\teR|,|\mbox{Im}~\eLc| < 2.2\times 10^{-3}~,\\
&& -1.1\times 10^{-3} < \mbox{Re}~\eLc < 4.5\times 10^{-3}~,
\eea
assuming that only one operator at a time is present. 
This analysis reinterprets an experimental search of $W'$ bosons~\cite{Khachatryan:2014tva}, which looks for a peak in the differential distribution and treats the SM tail as a background. In the SMEFT context the NP effect is not a peak but a modification of the tail, and thus it would benefit from a specific experimental analysis %(not available yet)
where the SM is actually the signal.
The projected future bounds using this channel are $\eSP\lesssim 1.9\times10^{-3}$ and $\eT \lesssim 0.5\times10^{-3}$ at 90\% C.L. with 300 fb$^{-1}$ at 14 TeV \cite{Bhattacharya:2011qm}.

SMEFT four-fermion operators can also be probed via the $pp \to e^+ e^-X$ process at the LHC, as first studied in Refs.~\cite{Cirigliano:2012ab,deBlas:2013qqa}, and more recently used in Refs.~\cite{Farina:2016rws,Greljo:2017vvb,Falkowski:2017pss}. In this channel there is an ATLAS measurement of the differential $e^+ e^-$ cross section during the LHC run-1 ($\sqrt{s}=$8 TeV, 20.3 fb$^{-1}$), which measures very precisely the SM high-$m_{ee}$ tail~\cite{Aad:2016zzw}.
The constraints on various SMEFT operators translate in the following $90\%$~C.L. bounds \cite{Falkowski:2017pss}
\bea
\label{eq:NCbounds1}
\mbox{Re}~\eLc &\approx& +0.7(1.2)\times10^{-3}~,\\
\label{eq:NCbounds2}
|\eSP| &\lesssim& 4.4 \times10^{-3}~,\\
\label{eq:NCbounds3}
|\eT| &\lesssim& 0.6 \times10^{-3}~,
\eea
which are stronger than those obtained using the charge-current channel. These bounds were obtained using three bins in the range  $0.5 \leq m_{ee} \leq 1.5$~TeV of the $e^+e^-$ differential distribution~\cite{Aad:2016zzw}, and assuming only one SMEFT operator is present at a time, although large cancellations do not seem to appear in a global fit~\cite{Greljo:2017vvb}. The neutral-current channel is not sensitive to operators with right-handed neutrinos, which generate the $\tilde{\epsilon}_i$ couplings at low-energies. The running of the limits in~Eqs.~(\ref{eq:CCbounds1})-(\ref{eq:NCbounds3}) from 1 TeV to 2 GeV was taken into account.
These constraints will also improve in the near future, with the availability of new LHC data at 13 TeV. % energies.
In fact, the recent recast \cite{Greljo:2017vvb} of 13-TeV ATLAS $pp \to e^+ e^-X$ data~\cite{ATLAS:2017wce} shows this is already the case with the currently available luminosity (36.1 fb$^{-1}$). This analysis, which focuses on chirality-conserving operators, calculates also the expected bounds with 3000 fb$^{-1}$ of data at 13 TeV, finding a bound at the $10^{-5}$ level on $\mbox{Re}~\eLc$.

However the LHC bounds derived above require extra assumptions compared with the low-energy bounds. First, the SMEFT analysis of LHC data assumes a larger gap between the electroweak scale and the new states. A BSM setup with new particles at, {\it e.g.}, 1 TeV requires a model-dependent analysis of the LHC data, whereas it can still benefit from the EFT analysis of the low-energy searches. Furthermore, even when new states are heavy enough, the sensitivity to $\cO(\Lambda^{-4})$ terms (from quadratic dimension-6 operators) is very different for these two classes of searches. Low-energy observables are very insensitive to these terms, because the factor ${\cal O}(s/v^2)$ is not large, and their strength comes directly from their accuracy. In contrast, LHC bounds present usually a similar sensitivity to linear and quadratic terms, and for chirality-flipping operators there is no interference at all. It is thus unclear whether the effect of higher dimensional operators is negligible, as assumed by construction, except for BSM scenarios where one expects them to be suppressed. This is the case of strongly coupled UV completions, where the enhancement of dimension-6 squared terms with respect to dimension-8 contributions comes from a large new physics coupling~\cite{Contino:2016jqw}.
For the sake of applicability to a wider class of BSM models, it is interesting to decrease the $\sqrt{s}$ range of the used LHC data at some $M_{\rm cut}$ above which the SMEFT is no longer valid. It has been shown that the LHC constraints are relaxed by a factor 2-3 when lower cuts ($M_{\rm cut} \sim 0.7$~TeV) are used~\cite{Falkowski:2017pss}.

The associated production of a Higgs and a $W^\pm$ boson 
can provide access to the right-handed current operator $\cO_{\varphi ud}$ and thus to the low-energy coefficient $\eR$ \cite{Alioli:2017ces}. Once again the crucial point is to have a contact interaction ($udW\!H$ in this case), to benefit from an ${\cal O}(s^2/v^4)$ enhancement with respect to the non-contact SM background. This was not possible for the $\cO_{\varphi ud}$ operator in the $pp\to e\nu X$ channel, but it occurs in $W\!H$ production due to the presence of a Higgs field. The sensitivity of the cross section to $\eR$ is quadratic, due to its chiral structure. The following $90 \%$ C.L. bound was found \cite{Alioli:2017ces} using LHC run 1 data~\cite{Khachatryan:2016vau},
\bea
%\label{eq:}
|\eR| &\lesssim& 0.04~.
\eea
The same bound is found using 13 TeV ATLAS data~\cite{ATLAS:2016pkl}, whereas a factor of 2 improvement is expected with future data~\cite{Alioli:2017ces}.

The comparison between the above-given LHC bounds and those obtained from $\beta$ decay in~\sref{fit} shows an interesting complementarity, which will continue in the near future with significant progress in both ends. This is illustrated in~\fref{LHC}. The interplay becomes much more interesting if a non-zero result is obtained for one of the Wilson coefficients, either in the low-energy experiments or in collider searches.
It was shown \cite{Bhattacharya:2011qm} that a hypothetical scalar resonance found at the LHC in the $pp\to e^\pm+{\rm MET}+X$ channel would imply a lower bound in the value of $|\eS|$ that should then be confirmed in nuclear and neutron decay experiments.
\begin{figure}[!htb]
\centering
\includegraphics[width=0.4\columnwidth]{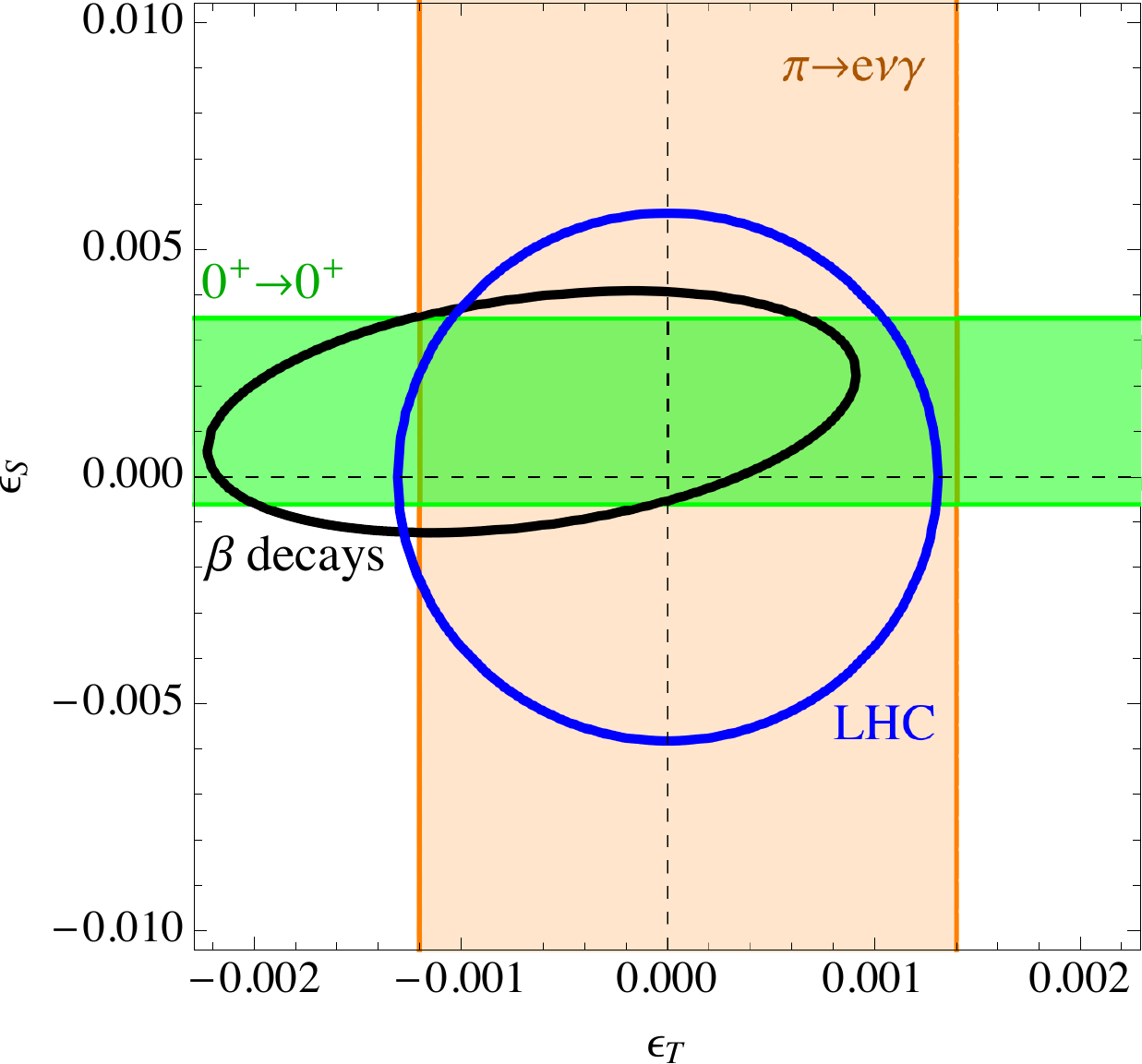}
\hspace{1cm}
\includegraphics[width=0.4\columnwidth]{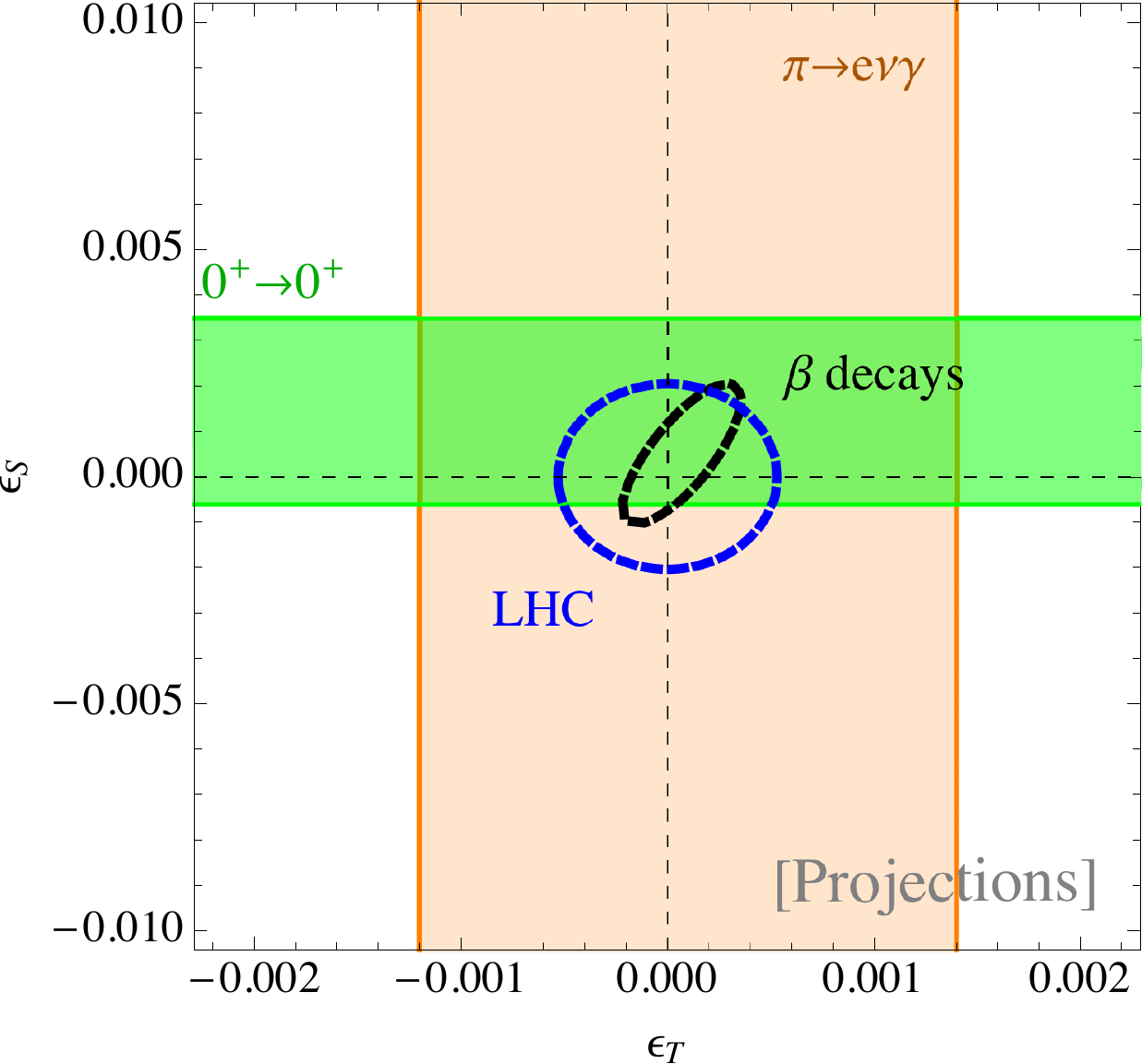}
\caption{
(Left) 90\% CL constraints on $\epsilon_{S,T}$ at $\mu= 2$ GeV from $\beta$-decay data, {\it cf.}~\eref{fit08}, with $\Delta\chi^2=4.61$, (black ellipse), from the analysis of $pp\to e+ {\rm MET}+X$ at the 8-TeV LHC (20 fb$^{-1}$)~\cite{Gonzalez-Alonso:2013uqa} (blue ellipse), and from radiative pion decay, {\it cf.}~\eref{eTboundrad}~\cite{Bhattacharya:2011qm} (orange band). The green band shows the 90\% CL bound ($\Delta\chi^2=2.71$) using only superallowed Fermi decays. 
(Right) Same figure but using projected $\beta$-decay data, {\it cf.}~\eref{fit42} (black) and projected LHC bounds from $pp\to e+ {\rm MET}+X$ searches with 14 TeV and 300 fb$^{-1}$~\cite{Bhattacharya:2011qm} (blue).}
\label{fig:LHC}
\end{figure}

\subsection{Neutrino mass bounds}
Effective operators involving right-handed neutrinos in~\eref{leff-lowE} generate a contribution to the neutrino mass at loop level, which can be used to set bounds on them assuming they are not canceled by some other contributions~\cite{Prezeau:2004md,Ito:2004sh}. Requiring that the leading logarithmic part of the 2-loop correction is not larger than current bounds on the neutrino mass, the following bounds were found \cite{Vos:2015eba}
\bea
|\teL|& \lesssim&10^{-2}~,\\
|\teS\pm\teP|&\lesssim&2\times10^{-3}~,\\
|\teT|&\lesssim&0.5\times10^{-3}~,
\eea
where $\mu=1$ TeV was used as the initial running scale. 
The bounds on scalar and tensor interactions are about 3 times stronger than those derived from LHC data in Eqs.~(\ref{eq:CCbounds1})-(\ref{eq:CCbounds2}) and orders of magnitude stronger than those from $\beta$ decay, {\it cf.}~\sref{fit-RH}. The bound on the pseudoscalar coupling is also 3 times stronger than the LHC one, but still weaker than that from pion decay, {\it cf.}~\eref{teP}. Finally, the neutrino-mass considerations above offer a valuable alternative probe for the $\teL$ coupling, which can also be accessed through CKM unitarity, but with slightly less accuracy, {\it cf.}~\eref{xi_ckm}.

\subsection{Electric dipole moments}
It can be shown that in the SMEFT framework, the same dimension-6 effective operators generating $\mathcal{CP}$-violating effects in $\beta$ decay would also generate at tree- or one-loop-level a non-zero nuclear and neutron Electric Dipole Moment (EDM)~\cite{Ng:2011ui}. As a result one can translate the stringent EDM bounds~\cite{Chupp:2017rkp} in indirect limits on the $\beta$-decay $\mathcal{CP}$-violating coefficients, such as $D$ or $R$, which are two orders of magnitudes stronger than their direct limits from $\beta$-decay measurements~\cite{Vos:2015eba}. This takes into account the calculation of Ref.~\cite{Seng:2014pba} that relaxed the EDM bound by an order of magnitude with respect to Ref.~\cite{Ng:2011ui}.

In principle, these indirect bounds can be avoided through a fine-tuned cancellation with additional dimension-6 operators contributing to the EDMs, or using dimension-8 operators. The precise realization in specific models is however nontrivial, as shown for instance for leptoquark models, where the connection with EDMs is still present, although the indirect bounds can be relaxed in this case~\cite{Ng:2011ui}. Finally, the EDM bounds can be avoided abandoning altogether the SMEFT framework, introducing for example light new particles.
Thus, current measurements of $\mathcal{CP}$-violating coefficients in $\beta$ decay can be considered as probes of the SMEFT framework itself, or at least its simpler realizations where large fine-tunings are not considered.  A recent and detailed review of the connection between EDMs and $\beta$-decay measurements is presented in Ref.~\cite{Vos:2015eba}.

%%%%%%%%%%%%%%%%%%%%%%%%%%%%%%%%%%%%%%%%%%%%%%%%%%%%%%%%%%%%%%%%%%%%%%%%%%%%%%%%
\section{Conclusions}
\label{sec:conclusion}

We have reviewed the role of precision measurements in nuclear and neutron $\beta$ decay, as useful tools to improve our understanding of fundamental interactions. 
Transitions with small nuclear-structure uncertainties (or none in neutron decay) are used to learn about QCD, to extract the values of fundamental SM parameters such as $V_{ud}$, and to search for new physics.

First, we have introduced the theoretical formalism that describes $\beta$ decay
at the elementary level
with special attention to the latest developments, such as the precise
calculations of the hadronic charges in the lattice, or the SMEFT framework
that enables the connection with other experimental searches. 
We have discussed
also recent measurements, as well as ongoing and planned experiments, which are summarized in~\tref{expSummary}.
The present inconsistencies of various observables in neutron decay are
expected to be clarified in the near future, considering the number of
experiments with improved precision that are ongoing. An important activity
is also taking place to measure the Fierz term from the differential energy
distributions both in nuclear and neutron decays.

We have performed global fits to the available $\beta$ decay data including, for the first time in a consistent way,
the ${\cal F}t$ values of $0^+\to 0^+$ transitions, neutron decay data and
correlations coefficients in nuclear decays. This allows the extraction of
correlations between the SM parameters ($V_{ud}$ and $g_A$) as well as those
for exotic couplings. 
This is shown for example in~\eref{fit5} in the so-called left-handed fit, where only interactions involving LH neutrinos are included. These correlated bounds represent one of our main numerical results. It allows the calculation of any given $\beta$-decay observable including those correlations, as shown in~\tref{benchErrors}. In the SM limit it gives a determination of the axial charge 3.5 times more precise than the current PDG average, {\it cf.}~\eref{lambda}. We have moreover identified the observables that dominate the results in each scenario. For instance, in the left-handed fit the result is completely dominated by the ${\cal F}t$ values of $0^+\to 0^+$ transitions, the neutron lifetime and the $\beta$ asymmetry parameter $A_n$. We have tentatively calculated the future expected bounds provided the ongoing experiments reach their claimed goals, finding a significant improvement on the bound on scalar and tensor interactions. The uncertainty on $V_{ud}$ is not expected to change significantly since it is dominated by the uncertainty on the inner radiative corrections. A modest improvement in the uncertainty of the CKM unitarity test is nevertheless expected mainly due to the reduction of the error on $V_{us}$. Improved lattice determinations of $g_A$ will directly translate into more stringent constraints over the right-handed coupling, $\epsilon_R$, which is expected to reach the per-mil level in the near future.

We have also presented results from fits with scalar and tensor currents
involving right-handed neutrinos. 
We stress that the most stringent constraint on such scalar interactions comes actually from CKM unitarity, which was not noticed before. 
We found also that the neutrino asymmetry
coefficient $B$ plays currently a crucial role in this fit. Within the SM, its current
value is in $\sim2\sigma$ tension with other neutron data for the extraction
of $\lambda$, which translates into a non-zero value for the RH tensor current.
New measurements of $B$ are desirable to clarify this situation.
This right-handed
fit illustrates how parameters that are not relevant for the left-handed fit
can have a significant impact in a different or more general framework. Moreover,
the agreement between different measurements is always a non-trivial cross
check since the systematic and theory errors are usually very different.

Finally we have analyzed the complementarity of $\beta$-decay searches with
other probes, such as pion decay or LHC searches. 
The CKM unitarity test is an extremely strong constraint on the vector
interaction,
which clearly provides a non-redundant constraint in the EFT approach and which
is known to constrain many BSM extensions. $\beta$ decay remains a unique
model-independent probe of RH vector interactions ($\eR$). Scalar and tensor
interactions are harder to embed in a realistic model without generating
at the same time pseudo-scalar interactions, which are extremely constrained
by the leptonic pion decay. However, in the most general framework, bounds
are relaxed and the three currents must be kept separated. Current bounds on
LH scalar and tensor interactions from $\beta$ decay are comparable to 
those obtained
from radiative pion decay or LHC searches. The latter will improve
with the arrival of new data and higher energies, but $\beta$ decay will remain
competitive if the expected precision goals of the various ongoing experiments
are reached. 

In summary, the intense activity in precision measurements in nuclear and
neutron $\beta$ decay, along with a number of theoretical developments, is
expected to contribute in paving the way toward the discovery of the new
Standard Model.

%%%%%%%%%%%%%%%%%%%%%%%%%%%%%%%%%%%%%%%%%%%%%%%%%%%%%%%%%%%%%%%%%%%%%%%%%%%%%%%%

\section*{Note Added in Proof}
Several significant results appeared after this review was submitted to the journal,
such as those from Refs.~\cite{Chang:2018uxx,Gupta:2018qil,Seng2018}.
The literature survey ended by March 2018 and
we have not updated the discussion and conclusions presented in this work using
later results.

\section*{Acknowledgments}
This review is based on collaborations and exchanges with many colleagues over many years, to whom we are deeply indebted. This work has been supported by a Marie Sk\l{}odowska-Curie Individual Fellowship of the European Commission's Horizon 2020 Programme under contract number 745954 Tau-SYNERGIES, by the US National Science Foundation under Grant No. PHY-1565546 and by the FWO-Fund for Scientific Research Flanders.

%\section*{References}
%%%%%%%%%%%%%%%%%%%%%%%%%%%%%%%%%%%%%%%%%%%%%%%%%%%%%%%%%%%%%%%%%%%%%%%%%%%%%%%%
\providecommand{\href}[2]{#2}\begingroup\raggedright

\endgroup

%%%%%%%%%%%%%%%%%%%%%%%%%%%%%%%%%%%%%%%%%%%%%%%%%%%%%%%%%%%%%%%%%%%%%%%%%%%%%%%%
\end{document}